\documentclass[aps,prx,twocolumn,floatfix,longbibliography]{revtex4-2}
\usepackage{tabularx}
\usepackage{bm}
\usepackage{epsfig,subfigure}
\usepackage{multirow}
\usepackage{graphicx}
\usepackage{color}
\usepackage{amsfonts}
\usepackage{exscale}
\usepackage{amsbsy}
\usepackage[colorlinks=true,urlcolor=blue,linkcolor=blue,citecolor=red]{hyperref}

\usepackage{amsmath}
\usepackage{amssymb}
\usepackage{braket}

\begin{document}
\title{Generalized $\eta$-pairing approach to interacting non-Hermitian systems in arbitrary dimensions}
\author{Kai Lieta}
\email{kaili@zzu.edu.cn}
\affiliation{School of Physics, Zhengzhou University, Zhengzhou 450001, China}

\begin{abstract}
Developing a general and rigorous analytical approach to non-Hermitian many-body systems is a challenging task. Here, we generalize the eta-pairing theory to very general non-Hermitian Hubbard models and find many novel phenomena without Hermitian analogs. For instance, the Hermitian conjugate of an eta-pairing eigenoperator may not be an eigenoperator, eta-pairing eigenoperators can be spatially modulated, and the $SU(2)$ pseudospin symmetry may not be possible even if $H$ commutes with the eta-pairing operators.
Remarkably, these novel non-Hermitian phenomena are closely related to each other by several theorems we establish and can lead to, for example, new types of eta-pairing operators (e.g., the notion of \emph{non-Hermitian angular-momentum operators}) and the Fock space localization of many-body eta-pairing eigenstates. Some issues on the $SO(4)$ and particle-hole symmetries are clarified. Our general eta-pairing theory also reveals a previously unnoticed unification of these symmetries of the Hubbard model.
These general results can be illustrated with concrete examples such as the generalized Hatano-Nelson-Hubbard models and a general two-sublattice model. In particular, the general two-sublattice model can reveal the eta-pairing structure [e.g., the $SO(4)$ symmetry] in systems with Hermitian hoppings, including the original eta-pairing theory for square lattice, the extension to triangular lattice, and some topological systems.
Our results establish a new and rigorous theoretical framework for studying interacting non-Hermitian systems in arbitrary spatial dimensions, even without bulk translation symmetry.

\end{abstract}
\date{\today}

\maketitle
\section{Introduction}

Non-Hermitian quantum physics has recently attracted growing interest \cite{Ashida2020,Bergholtz2021,Okuma2023}.  In particular, non-Hermitian quantum systems host many interesting phenomena without Hermitian analogs \cite{Gong2018,Kawabata2019,Kawabata2021}. One such phenomenon is the non-Hermitian skin effect originating from the non-Hermiticity of the Hamiltonian \cite{Torres2018,Yao2018}, where many eigenstates are localized at the boundaries of the system. Note that the conventional non-Hermitian skin effect is a phenomenon of the single-particle eigenstates of noninteracting systems with bulk translation symmetry, which has been extensively studied and is well understood in the framework of non-Bloch band theory \cite{Slager2020,Okuma2020,Zhang2020,Kawabata2020,Amoeba2024}.
Then, a natural question arises: Can the non-Hermitian localization survive in the presence of many-body interactions or in the absence of bulk translation symmetry? The answer to this question clearly goes beyond the framework of band theory. Recently, there have been various numerical studies of the fate of skin effects under interactions  \cite{Lee2020,Liu2020,Ryu2022,Alsallom2022,Neupert2022,Yoshida2022,Longhi2023,YZNK2024,Rangi2025}, which focus on the case of one-dimensional systems.

However, in contrast to the single-particle case, even a clear and proper characterization of localization for non-Hermitian many-body systems is elusive. A recent work \cite{Shimomura-Sato2024} gives a very general and precise definition of localization, which generalizes the notion of localization from the conventional real space to the abstract Hilbert space. Based on this definition, Ref. \cite{Shimomura-Sato2024} reveals a localization phenomenon intrinsic to non-Hermitian systems (i.e., a general criterion for non-Hermitian skin effects) and establishes relations between the skin effect, the spectral sensitivity to perturbations, and the non-normality.

Despite the above studies of non-Hermitian systems, it is challenging to establish rigorous analytical results for interacting many-body systems in arbitrary spatial dimensions (with or without the bulk translation symmetry). Note that all the previous exact results on the interacting non-Hermitian systems are based on the Bethe-ansatz method and are limited to one-dimensional systems with bulk translation symmetry \cite{F-K1998,N-K-U2021,M-H-P2023,K-P-A2023,WLSW2023,ZQWCC2024,WZYW2026}.

An important rigorous theory for interacting Hermitian many-body systems is the so-called eta-pairing theory established in the seminal work of C. N. Yang \cite{Yang1989}, which was very influential and has wide applications in many branches of condensed matter and quantum physics, ranging from superconductivity, quantum phase transitions, quantum thermalization, to quantum many-body scars \cite{Essler1992,Essler1993,Boer1,Zhang1997,Roy2018,Nandkishore2015,Garrison2017,Veness2017,Clark2018,Bernevig2020,Motrunich2020,Kai2020,Divic2025}. This theory provides a mechanism for analytically constructing exact and nontrivial many-particle eigenstates of interacting fermionic systems in arbitrary spatial dimensions (even without the bulk translation symmetry) \cite{Yang1991}. Specifically, for Hermitian Hubbard models defined on a lattice $\Lambda$, the eta-pairing operator takes the general form
\begin{equation}
\eta^\dag=\sum_{j\in\Lambda} e^{{\rm i}\theta_j} c_{j\uparrow}^\dag c_{j\downarrow}^\dag,
\label{eq:hermitian-eta}
\end{equation}
which is a linear combination of on-site singlet-pairing operators $c_{j\uparrow}^\dag c_{j\downarrow}^\dag$ with the coefficients (e.g., phase factors) $e^{{\rm i}\theta_j}$. Note that the phase parameter $\theta_j\in\mathbb{R}$ may depend on site $j$; however, the absolute value of each coefficient (dubbed the \emph{on-site pairing amplitude}), say $|e^{{\rm i}\theta_j}|$, does not depend on $j$, e.g., $|e^{{\rm i}\theta_j}|\equiv1$. Thus, the eta-pairing states $\eta^\dag |0\rangle, (\eta^\dag)^2 |0\rangle, \cdots$ in Hermitian systems are always \emph{spatially extended}. Moreover, the eta-pairing states are nontrivial many-body states in the sense that they possess off-diagonal long-range order \cite{Yang1989} and quantum entanglement \cite{Fan2005,Vafek2017,Clark2018}. Importantly, eta-pairing states can be exact eigenstates of the Hubbard model \cite{Yang1989} and even become the ground states of some extended Hubbard models \cite{Essler1992,Essler1993,Boer1,Kai2020}.

The goal of this article is to develop a generalized eta-pairing theory for very general non-Hermitian Hubbard models [see Eq.~\eqref{eq:nH-Hubbard}], which would provide a new and rigorous framework for understanding interacting non-Hermitian systems in arbitrary spatial dimensions (even without the bulk translation symmetry). Our general non-Hermitian eta-pairing theory consists of several theorems and conclusions. To be specific, Theorems 1 and 2 give the sufficient and necessary condition for the eta-pairing operator to be an eigenoperator of the non-Hermitian Hubbard Hamiltonian. Based on Theorems 1 and 2, we derive many unusual properties of the non-Hermitian eta-pairing theory, which have no counterparts in Hermitian systems. For example, Theorem 5 and Corollary 6 say that the Hermitian conjugate of an eta-pairing eigenoperator may not be an eigenoperator; and in the main text we refer to this phenomenon as the property $(a)$. Note that the precise meaning of the property $(a)$ is twofold, as discussed below Corollary 6. The second exotic non-Hermitian phenomenon, namely the property $(b)$, revealed by Theorem 7 and Corollary 8 is that the on-site pairing amplitude of an eta-pairing eigenoperator can be spatially modulated (i.e., site dependent). The property $(b)$ directly implies that the eta-pairing eigenstates can exhibit localization in the Hilbert space \cite{Shimomura-Sato2024}, namely the Fock space localization; see Eqs.~\eqref{eq:m-eta-state} and \eqref{eq:Fock-space}. This property can also lead to the notion of \emph{non-Hermitian angular-momentum operators} [see, e.g., Eq.~\eqref{eq:nHcomm}], as explained in this work. Moreover, Corollary 9 indicates that the so-called $SU(2)$ pseudospin symmetry of Hubbard model (see below) does not make sense in the presence of asymmetric hoppings, even when the Hamiltonian commutes with the eta-pairing operators; this phenomenon is referred to as the property $(c)$.

Interestingly, the above three properties $(a)$, $(b)$, and $(c)$ are equivalent to each other; see, e.g., Eq.~\eqref{eq:relations}. We also find other exotic non-Hermitian phenomena such as the existence of models with \emph{non-unique} eta-raising or -lowering eigenoperators [namely the property $(e)$]. Actually, the property $(e)$ implies the three properties $(a)$, $(b)$, and $(c)$, as summarized in Eq.~\eqref{eq:relations}.

To exemplify the above general results, we study several concrete model systems, which include the generalized Hatano-Nelson-Hubbard models in one and two dimensions and a general two-sublattice model defined on an arbitrary lattice. In particular, the two-sublattice model can reveal the eta-pairing structure [e.g., the $SO(4)$ symmetry discussed below] in systems with Hermitian hoppings, including the original eta-pairing theory for square lattice, the extension to triangular lattice, and some well-known topological systems.

Another important implication of eta-pairing theory for Hubbard model is the hidden $SU(2)$ pseudospin symmetry \cite{Zhang1990}, which is generated by the eta-pairing operators. Now, together with the usual $SU(2)$ spin symmetry, what is the full symmetry of the Hubbard model? A naive guess is the $SU(2)\times SU(2)$ symmetry, but the true symmetry turns out to be $SO(4)=\frac{SU(2)\times SU(2)}{\mathbb{Z}_2}$ in the Majorana representation \cite{Yang1991}. However, in the literature \cite{Zhang1991,Yang1991,Arovas2022,Moudgalya2023,Pakrouski2023}, there are still some ambiguities in identifying the symmetry group of the Hubbard model in the usual electron representation, e.g., the Hamiltonian has the $SU(2)\times SU(2)$ symmetry when $N_\Lambda$ is odd and has the $SO(4)$ symmetry when $N_\Lambda$ is even, where $N_\Lambda$ denotes the total number of lattice sites. Thus, to address the above issue, a more careful analysis of the symmetry group of the Hubbard model is needed.

In this work, we reexamine the group $G(N_\Lambda)$, which is generated by the pseudospin (e.g., eta-pairing) and spin operators, in more detail. As mentioned above, the group structure turns out to be $G(N_\Lambda)=SU(2)\times SU(2)$ for odd $N_\Lambda$ or $G(N_\Lambda)=SO(4)$ for even $N_\Lambda$.
We notice that $G(N_\Lambda)$ with odd $N_\Lambda$ contains a $\mathbb{Z}_2$ gauge transformation and $G(N_\Lambda)$ with even $N_\Lambda$ does not contain gauge transformations. So, the physical symmetry group is $G(N_\Lambda)/\mathbb{Z}_2=\frac{SU(2)\times SU(2)}{\mathbb{Z}_2}=SO(4)$ for odd $N_\Lambda$ and is simply $G(N_\Lambda)=SO(4)$ for even $N_\Lambda$. Thus, the true symmetry is always $SO(4)$ that does not rely on $N_\Lambda$, which is, of course, consistent with the result in the Majorana representation.

The Hubbard model can also possess discrete symmetries, e.g., the well-known $\mathbb{Z}_2$ particle-hole (PH) transformation: $c_{j\sigma}^\dag \leftrightarrow c_{j\sigma}$, up to some phase factors. Given the above $SO(4)$ symmetry and the $\mathbb{Z}_2$ PH symmetry, an interesting issue that is seldom discussed in the literature is the relations between these symmetries. As shown in previous works \cite{Hermele2007,You2017,Kai2020}, the $SO(4)$ symmetry or, more precisely, the $SU(2)$ pseudospin symmetry contains a $\mathbb{Z}_4$ PH symmetry. [Note that the $SO(4)$ symmetry contains the $SU(2)$ pseudospin symmetry as a subgroup.] Here, we revisit the usual $\mathbb{Z}_2$ PH symmetry and show that it actually belongs to the $SO(4)$ symmetry (e.g., as a combination of the $\mathbb{Z}_4$ PH symmetry and a particular spin rotation). Interestingly, we also find that the combination of the $\mathbb{Z}_4$ PH transformation and an arbitrary spin rotation can realize many different kinds of PH symmetries, namely a general $\mathbb{Z}_{2k}$ PH symmetry for any positive integer $k$.

Therefore, it is clear that if the Hubbard model has the $SU(2)$ pseudospin symmetry, then all the above discrete PH symmetries that belong to the $SO(4)$ symmetry are also respected. This leads us to wonder whether the converse is true; i.e., if the Hubbard model has the discrete PH symmetry, then the $SU(2)$ pseudospin symmetry is also respected. Based on our general eta-pairing theory, we will prove that the converse is true, which is a nontrivial result. In other words, we show that in the context of Hubbard model, all these different symmetries, including the $\mathbb{Z}_2$ Shiba transformation which does not belong to the $SO(4)$ symmetry, share the same underlying algebraic equations and hence are equivalent to each other.

The rest of this paper is organized as follows. In Sec.~\ref{sec:general-theory}, we briefly review the Hermitian eta-pairing theory, and then generalize it to non-Hermitian systems. Our general non-Hermitian eta-pairing theory reveals many new eta-pairing phenomena that are impossible in Hermitian systems, and these phenomena are closely related to each other [as shown in Eq.~\eqref{eq:relations}]. In Sec.~\ref{sec:symmetries}, we discuss the various symmetry properties of the (non-Hermitian) Hubbard model and uncover a previously unnoticed unification of these symmetries in the framework of our general eta-pairing theory. Finally, in Sec.~\ref{sec:examples}, we study several concrete models to illustrate the various non-Hermitian eta-pairing phenomena proposed in Sec.~\ref{sec:general-theory}.

\section{General non-Hermitian eta-pairing theory}\label{sec:general-theory}

We consider a generic non-Hermitian Hubbard model on  an arbitrary lattice $\Lambda$, defined by the Hamiltonian
\begin{eqnarray}
H&=&T+V,
\nonumber\\
T&=&\sum_{i\neq j}t_{ij} (c_{i\uparrow}^\dag c_{j\uparrow}+ c_{i\downarrow}^\dag c_{j\downarrow}),
\nonumber\\
V&=&\sum_{i\in\Lambda} U_in_{i\uparrow}n_{i\downarrow}-\mu_i(n_{i\uparrow}+n_{i\downarrow}).
\label{eq:nH-Hubbard}
\end{eqnarray}
Here, all the complex-valued model parameters, namely $t_{ij},U_i,\mu_i\in\mathbb{C}$, are allowed to depend on  lattice sites, and  the  hopping $t_{ij}$ may be nonzero between two arbitrary sites $i$ and $j$.
The fermionic operators $c_{i\sigma}^\dag$ and $c_{i\sigma}$ ($\sigma=\uparrow,\downarrow$) satisfy the usual anticommutation relations, and $U_i$ ($\mu_i$) with number operators $n_{i\sigma}=c_{i\sigma}^\dag c_{i\sigma}$  represents the on-site Hubbard interaction (chemical potential). Notice that the Hamiltonian is Hermitian (i.e., $H^\dag=H$) \emph{if and only if} all the off-site hoppings satisfy $t_{ij}=t_{ji}^*$ (i.e., $T^\dag=T$) and all the on-site $U_i$ and $\mu_i$ are real (i.e.,  $V^\dag=V$).

The lattice $\Lambda$ (typically but not necessarily a regular lattice) is simply defined as a collection of sites in any spatial dimension and hence can be arbitrary, including the case of irregular lattices without periodic structures \cite{Lieb1989}. $N_\Lambda$ denotes the total number of sites in $\Lambda$. There is said to be a bond between sites $i$ and $j$ ($i\neq j$), denoted by $\{i,j\}=\{j,i\}$, if either $t_{ij}$ or $t_{ji}$ is nonzero. We further call the bond $\{i,j\}$ a \emph{strong} bond if both $t_{ij}$ and $t_{ji}$ are nonzero. [Thus, Hermiticity (e.g., $t_{ij}=t_{ji}^*$) implies that every bond is a strong bond.] Throughout our paper, the lattice $\Lambda$ is assumed to be connected; i.e., there exists a connected path of bonds  between two arbitrary sites. (No extra assumption about $\Lambda$ is made. Importantly, our general theory and results in this article do not rely on a bipartite lattice structure or an even $N_\Lambda$.)

\subsection{Review of Hermitian eta-pairing theory}

Our primary aim in this work is to obtain a general eta-pairing theory for non-Hermitian Hubbard models \eqref{eq:nH-Hubbard}. Before doing this, let us review the conventional eta-pairing theory for Hermitian systems. In the Hermitian case, the eta-pairing operator can be written as Eq.~\eqref{eq:hermitian-eta}. Importantly, the phase parameter $\theta_j$ and the model parameters can be suitably chosen so that $\eta^\dag$ is an eigenoperator of the Hamiltonian $H$, in the sense that $[H,\eta^\dag]=\lambda\eta^\dag$  with a constant $\lambda\in\mathbb{R}$; then, one can construct many eta-pairing eigenstates, i.e., $H(\eta^\dag)^m |0\rangle=m\lambda(\eta^\dag)^m |0\rangle$ with $m=0,1,\ldots,N_\Lambda$.

Now, we highlight some basic properties of the Hermitian eta-pairing theory, which were not clarified in previous works. First, due to the Hermiticity (i.e., $H^\dag=H$), the equation $[H,\eta^\dag]=\lambda\eta^\dag$ is equivalent to $[H,\eta]=-\lambda\eta$. This means that the Hermitian conjugate of an eta-pairing eigenoperator is still an eigenoperator and the constants $\pm\lambda$ in the above two equations are equal (up to a minus sign). Second, the eta-raising or -lowering eigenoperator (if one exists) must be unique (up to a trivial factor, i.e., multiplication by an arbitrary nonzero constant). [In this paper, we call a linear combination of the operators $c_{j\uparrow}^\dag c_{j\downarrow}^\dag$ ($c_{j\downarrow}c_{j\uparrow}$) an eta-raising (-lowering) operator; see, e.g., Eqs.~\eqref{eq:eta-raising} and \eqref{eq:eta-lowering}.]
Third, as mentioned below Eq.~\eqref{eq:hermitian-eta}, the on-site pairing amplitude of eta-pairing eigenoperators is site independent, e.g., $|e^{{\rm i}\theta_j}|\equiv1$. Last, the Hamiltonian $H$ possesses the $SU(2)$ pseudospin symmetry when $\lambda=0$. (The reasons for the second and third properties will be clear later.)

\subsection{Non-Hermitian eta-pairing theory}

Non-Hermiticity can break all of the above basic properties of the conventional eta-pairing theory, leading to unusual eta-pairing phenomena. To see this, we start by writing the eta-raising operator as
\begin{equation}
\eta^\dag=\sum_{j\in\Lambda} \omega_j c_{j\uparrow}^\dag c_{j\downarrow}^\dag,  \qquad  \omega_j=|\omega_j|e^{{\rm i}\theta_j}\in\mathbb{C},
\label{eq:eta-raising}
\end{equation}
where the on-site pairing amplitude $|\omega_j|$ is allowed to depend on $j$ (see Corollary 8). Note that the spatially modulated $|\omega_j|$ is a fundamentally new feature.
Now, the sufficient and necessary condition for $\eta^\dag$  in Eq.~\eqref{eq:eta-raising} to be an eigenoperator of $H$  in Eq.~\eqref{eq:nH-Hubbard} can be summarized in the following theorem (see Appendix \ref{subsec:proofTheorem1-2} for a proof):

\textbf{Theorem 1:} The eta-raising eigenoperator-equation $[H,\eta^\dag]=\lambda\eta^\dag$ is equivalent to the following set of coupled equations:
\begin{subequations}
\begin{align}
t_{ij}\omega_j+t_{ji}\omega_i=0     \qquad     \text{for each bond $\{i,j\}$},      \label{eq:constraints1a}\\
(U_j-2\mu_j)\omega_j=\lambda\omega_j   \qquad      \text{for each site $j\in\Lambda$} ,    \label{eq:constraints1b}
\end{align}
\label{eq:constraints1}
\end{subequations}
where $\lambda\in\mathbb{C}$ is a constant.  In fact, Eqs.~\eqref{eq:constraints1a} and \eqref{eq:constraints1b} are equivalent to $[T,\eta^\dag]=0$ and $[V,\eta^\dag]=\lambda\eta^\dag$, respectively, where $T$ and $V$ are defined in Eq.~\eqref{eq:nH-Hubbard}.

For non-Hermitian systems, the equation $[H,\eta^\dag]=\lambda\eta^\dag$ is not equivalent to $[H,\eta]=-\lambda\eta$. Thus, as we will see from Corollary 6, the eta-lowering eigenoperator (if one exists),  written as
\begin{equation}
\eta'=\sum_{j\in\Lambda} (\omega'_j)^* c_{j\downarrow}c_{j\uparrow},
\label{eq:eta-lowering}
\end{equation}
may be distinct from the Hermitian conjugate of the eta-raising eigenoperator. Similar to Theorem 1, our second theorem  is stated as follows (see Appendix \ref{subsec:proofTheorem1-2} for a proof):

\textbf{Theorem 2:} The eta-lowering eigenoperator-equation $[H,\eta']=-\lambda'\eta'$ is equivalent to the following set of coupled equations:
\begin{subequations}
\begin{align}
t_{ij}(\omega'_i)^*+t_{ji}(\omega'_j)^*=0      \qquad   \text{for each bond $\{i,j\}$},      \label{eq:constraints2a}\\
(U_j-2\mu_j)(\omega'_j)^*=\lambda'(\omega'_j)^*     \qquad   \text{for each site $j\in\Lambda$} ,    \label{eq:constraints2b}
\end{align}
\label{eq:constraints2}
\end{subequations}
where $\lambda'\in\mathbb{C}$ is a constant.  In fact, Eqs.~\eqref{eq:constraints2a} and \eqref{eq:constraints2b} are equivalent to $[T,\eta']=0$ and $[V,\eta']=-\lambda'\eta'$, respectively.

Before proceeding, let us consider a simple example of Theorems 1 and 2: Assuming that the hoppings satisfy $t_{ij}=-t_{ji}$ (or equivalently, $t_{ij}+t_{ji}=0$) for every bond, Eqs.~\eqref{eq:constraints1a} and \eqref{eq:constraints2a} will give a site-independent solution, namely $\omega_j=\omega'_j=1$ (up to a trivial factor). If we further suppose that $U_j-2\mu_j$ does not depend on $j$, then according to Theorems 1 and 2,
$\eta^\dag$ in Eq.~\eqref{eq:eta-raising} with $\omega_j=1$ and $\eta'=(\eta^\dag)^\dag$ are the eta-raising and -lowering eigenoperators, respectively. Thus, when $\mu_j=U_j/2$ ($\forall j\in\Lambda$), the Hubbard model \eqref{eq:nH-Hubbard} possesses the $SU(2)$ pseudospin symmetry (see Sec.~\ref{symmetry-pseudospin} for more details). [Note that in the above discussion we do not assume Hermiticity for the hoppings, and the fact that $\eta^\dag$ and $\eta'$ are Hermitian conjugate of each other is due to Theorem 4 (see below).]

As a concrete model with hoppings satisfying $t_{ij}=-t_{ji}$, the $SU(2)$ pseudospin symmetry generated by the eta-pairing operators with $\omega_j=1$ plays an important role in the study of the triangular lattice Hofstadter-Hubbard model \cite{Divic2025}. Interestingly, as shown in our previous work \cite{Kai2020}, the triangular lattice Kitaev-Hubbard model \cite{Kai2016} can also possess eta-pairing eigenoperators with $\omega_j=1$ and hence the $SU(2)$ pseudospin symmetry; moreover, this model hosts a weak-coupling integer quantum Hall phase, just as in the triangular lattice Hofstadter-Hubbard model. [Note that one important difference between these two triangular lattice models is that the Kitaev-Hubbard model is a spin-orbit coupled system which explicitly breaks the $SU(2)$ spin symmetry.]

As shown in Appendix \ref{subsec:RL-eta}, using the eigenoperator equations $[H,\eta^\dag]=\lambda\eta^\dag$ and $[H,\eta']=-\lambda'\eta'$, one can construct many right and left eta-pairing eigenstates; and we also discuss the general relations between these eta-pairing eigenstates (see Remarks 1 and 2 in Appendix \ref{subsec:RL-eta}).
Moreover, from Eq.~\eqref{eq:constraints1a} [Eq.~\eqref{eq:constraints2a}],  $\omega_j$ ($\omega'_j$) and hence $\eta^\dag$ ($\eta'$) are completely determined by the off-site hoppings $t_{ij}$ or simply $T$, provided that Eq.~\eqref{eq:constraints1b} [Eq.~\eqref{eq:constraints2b}] holds (see the discussion below Theorem 3). Therefore, the exact eta-pairing eigenstates are insensitive to the Hubbard interaction or the on-site part $V$.

Now, a natural question arises: Given a Hamiltonian $H$, is the eta-raising (-lowering) eigenoperator $\eta^\dag$ ($\eta'$), if one exists, unique (up to a trivial factor)? The answer depends on the details of $H$. Nevertheless, we have the following conclusion (see Appendix \ref{subsec:proofTheorem3} for a proof):

\textbf{Theorem 3:} For the Hamiltonian \eqref{eq:nH-Hubbard}, if there exists a connected path of strong bonds (i.e., each bond within this path is a strong bond) between two arbitrary sites, then (a) the eta-raising (-lowering) eigenoperator $\eta^\dag$ ($\eta'$) given by Eq.~\eqref{eq:eta-raising} [Eq.~\eqref{eq:eta-lowering}], if one exists, must be unique (up to a trivial factor); (b) $\forall j\in\Lambda$, $\omega_j\neq0$ ($\omega'_j\neq0$).

Note that Theorem 3 provides a sufficient, but not necessary, condition for the uniqueness of eta-pairing eigenoperator; see the Remarks in Appendixes \ref{subsec:abcd-e} and \ref{subsec:onlyone-exists} and also the concrete examples in Sec.~\ref{sec:examples}. In Appendixes \ref{subsec:abcde} and \ref{subsec:abce-d}, we also present models with non-unique eta-raising or -lowering eigenoperators. Conclusion (b) in Theorem 3 implies that Eq.~\eqref{eq:constraints1b} or \eqref{eq:constraints2b} can reduce to $U_j-2\mu_j=\lambda$ or $\lambda'$ $(j\in\Lambda)$, i.e., a site-independent constant (note that $U_j$ or $\mu_j$ may still depend on $j$). However, as explained later (see Theorem 12), if not all the bonds are strong bonds, then $\omega_j$ or $\omega'_j$ must be zero on some site(s), and Eq.~\eqref{eq:constraints1b} or \eqref{eq:constraints2b} is not equivalent to the stronger constraint $U_j-2\mu_j=\lambda$ or $\lambda'$ $(j\in\Lambda)$ but is actually equivalent to $U_j-2\mu_j=\lambda$ $(j\in\Lambda_\omega)$ or $\lambda'$ $(j\in\Lambda_{\omega'})$, where $\Lambda_\omega=\{j\in\Lambda\mid\omega_j\neq0\}\subseteq\Lambda$ denotes the set of all the nonzero points of $\omega_j$; in other words, $U_j-2\mu_j$ is required to be a constant equal to $\lambda$ ($\lambda'$) in sublattice $\Lambda_\omega$ ($\Lambda_{\omega'}$) rather than the whole lattice $\Lambda$.

According to Theorem 3, the Hermiticity of $H$ (implying that all the bonds are strong bonds), together with the lattice connectivity of $\Lambda$, leads to the uniqueness of eta-raising or -lowering eigenoperator in Hermitian systems. Moreover, Hermiticity implies that for each bond $\{i,j\}$, we have $|t_{ij}|=|t_{ji}|$ and hence $|\omega_i|=|\omega_j|$ due to Eq.~\eqref{eq:constraints1a}. This, together with the lattice connectivity, explains the site-independent nature of $|\omega_j|$ in Hermitian systems. In fact, we have a more general result in the non-Hermitian case (see Appendix \ref{subsec:proofTheorem4} for a proof):

\textbf{Theorem 4:}
For the Hamiltonian \eqref{eq:nH-Hubbard}, if the hopping amplitudes are symmetric (i.e., $|t_{ij}|=|t_{ji}|$) for every bond $\{i,j\}$ and $\eta^\dag$ ($\eta'$) given by Eq.~\eqref{eq:eta-raising} [Eq.~\eqref{eq:eta-lowering}] is an eta-raising (-lowering) eigenoperator, then (a) $\eta^\dag$ ($\eta'$) is the unique eta-raising (-lowering) eigenoperator, and there also exists a unique eta-lowering (-raising) eigenoperator that is simply the Hermitian conjugate of $\eta^\dag$ ($\eta'$), up to a trivial factor; (b) $|\omega_j|$ ($|\omega'_j|$) does not depend on $j\in\Lambda$; (c) $U_j-2\mu_j$ does not depend on $j\in\Lambda$ and $\lambda=\lambda'=U_j-2\mu_j$.

Thus, Theorem 4 tells us that when the hopping amplitudes are symmetric for every bond, all the basic properties of the Hermitian eta-pairing theory are preserved even in non-Hermitian systems (see also the Remark in Appendix \ref{subsec:proofTheorem4}). On the other hand, when the hopping amplitudes are asymmetric (e.g., $|t_{ij}|\neq|t_{ji}|$) for some bond(s), we expect that some new eta-pairing phenomena arise: for example, the Hermitian conjugate of an eta-pairing eigenoperator is not an eigenoperator, due to the following theorem (see Appendix \ref{subsec:proofTheorem5} for a proof).

\textbf{Theorem 5:}
If an eta-pairing operator and its Hermitian conjugate are both eigenoperators of the Hamiltonian \eqref{eq:nH-Hubbard}, then the hopping amplitudes must be symmetric for every bond.

According to Theorem 4, the above symmetric hoppings in turn imply the same conclusions as in Theorem 4, which are omitted from Theorem 5.
Now, combining Theorems 4 and 5, we arrive at

\textbf{Corollary 6:}
The Hermitian conjugate of an eta-pairing eigenoperator is not an eigenoperator \emph{if and only if} the hopping amplitudes are asymmetric for some bond(s) (but not necessarily for every bond).

Therefore, in the presence of asymmetric hoppings, the eta-raising and -lowering eigenoperators are not Hermitian conjugate of each other, provided that they both exist. This can lead to the distinction between right and left eta-pairing eigenstates (see Appendix \ref{subsec:RL-eta} for detailed discussions), as will be illustrated with concrete models in Sec.~\ref{sec:examples}. Moreover, when not all the bonds are strong bonds, the existence of eta-raising (-lowering) eigenoperator does not necessarily imply the existence of eta-lowering (-raising) eigenoperator, in contrast to the conclusion (a) in Theorem 4 or its more general version Theorem 11. Indeed, we find that there are models in which one of the eta-raising and -lowering eigenoperators exists but the other does not exist (see Sec.~\ref{examples-twosub}).

So, when we say that the Hermitian conjugate of an eta-pairing eigenoperator is not an eigenoperator, there are two different situations, as discussed above.

Moreover, asymmetric hoppings inevitably lead to a site-dependent $|\omega_j|$, due to the following theorem:

\textbf{Theorem 7:}
If $\eta^\dag$ ($\eta'$) defined by Eq.~\eqref{eq:eta-raising} [Eq.~\eqref{eq:eta-lowering}] is an eta-raising (-lowering) eigenoperator and $|\omega_j|$ ($|\omega'_j|$) does not depend on $j\in\Lambda$, then the hopping amplitudes must be symmetric for every bond.

This theorem follows directly from Eq.~\eqref{eq:constraints1a} [Eq.~\eqref{eq:constraints2a}]. Note that according to Theorem 4, the above symmetric hoppings in turn imply the same conclusions as in Theorem 4. Now, combining Theorems 4 and 7, we get

\textbf{Corollary 8:}
The on-site pairing amplitude of an eta-pairing eigenoperator is site dependent \emph{if and only if} the hopping amplitudes are asymmetric for some bond(s) (but not necessarily for every bond).

Explicit examples of the spatially modulated eta-pairing eigenoperators in Corollary 8 can be found in Sec.~\ref{sec:examples}.
Remarkably, the properties of the eta-pairing operators and states with site-dependent $|\omega_j|$ can be very different from the case of site-independent $|\omega_j|$.
As a direct consequence of site-dependent $|\omega_j|$, the eta-pairing states can exhibit localization in the Fock space rather than the usual real space \cite{Shimomura-Sato2024}: Let us consider the general eta-pairing state $(\eta^\dag)^m |0\rangle$ with $0\leq m\leq N_\Lambda$; using Eq.~\eqref{eq:eta-raising}, we have
\begin{equation}
(\eta^\dag)^m|0\rangle=m!\sum_{j_1,\ldots,j_m\in\Lambda} \omega_{j_1}\cdots \omega_{j_m}\eta_{j_1}^\dag\cdots\eta_{j_m}^\dag|0\rangle,
\label{eq:m-eta-state}
\end{equation}
where $\eta_j^\dag=c_{j\uparrow}^\dag c_{j\downarrow}^\dag$ ($j\in\Lambda$), and $j_1,\ldots,j_m$ represent $m$ arbitrary and different sites of $\Lambda$; see also Eq.~\eqref{eq:eta-pairing-state4-sm}.
For a fixed $m$, we introduce the Fock space $\mathcal{F}_m$ spanned by the orthonormal states $\eta_{j_1}^\dag\cdots\eta_{j_m}^\dag|0\rangle$ in Eq.~\eqref{eq:m-eta-state}
\begin{equation}
\mathcal{F}_m=\{\sum_{j_1,\ldots,j_m\in\Lambda} \chi_{j_1,\ldots,j_m} \eta_{j_1}^\dag\cdots\eta_{j_m}^\dag|0\rangle\mid\chi_{j_1,\ldots,j_m}\in\mathbb{C}\}.
\label{eq:Fock-space}
\end{equation}
Thus, $\mathcal{F}_m$ has the dimension dim$\mathcal{F}_m=\binom{N_\Lambda}{m}=\frac{N_\Lambda!}{m!(N_\Lambda-m)!}$. Now, since $|\omega_j|$ is site dependent, the amplitude $|\omega_{j_1}\cdots \omega_{j_m}|$ varies with $\{j_1,\ldots,j_m\}$ (i.e., the set of $m$ different sites). Then, we arrange the order of the Fock basis $\eta_{j_1}^\dag\cdots\eta_{j_m}^\dag|0\rangle$ such that $|\omega_{j_1}\cdots \omega_{j_m}|$ decreases with respect to the order. So, from Eq.~\eqref{eq:m-eta-state}, we can say that the many-body eta-pairing state $(\eta^\dag)^m|0\rangle$ is localized in the Fock space $\mathcal{F}_m$ \cite{Shimomura-Sato2024}. In Sec.~\ref{sec:examples},
the Fock space localization of Eq.~\eqref{eq:m-eta-state} will be further illustrated with several explicit examples. Moreover,  the two eta-pairing states $(\eta^\dag)^m |0\rangle$ and $\eta^{N_\Lambda-m} |2\rangle$  are distinct from each other when $|\omega_j|$ depends on $j$, where $|2\rangle=\prod_{j\in\Lambda}c_{j\uparrow}^\dag c_{j\downarrow}^\dag|0\rangle$; see Remark 1 in Appendix \ref{subsec:proofLemma10} for more details on this distinction.

As another consequence of site-dependent $|\omega_j|$, the eta-pairing operator and its Hermitian conjugate may not be able to form an angular momentum operator. In fact, we have the following \textbf{proposition} (see Appendix \ref{subsec:eta-angular} for a proof): \emph{An eta-pairing operator, together with its Hermitian conjugate, can form an angular momentum operator if and only if the nonzero $|\omega_j|$ does not depend on $j\in\Lambda_\omega(\subseteq\Lambda)$}.

Now, the precise meaning of the $SU(2)$ pseudospin symmetry of the Hamiltonian \eqref{eq:nH-Hubbard} is twofold: (i) the eta-pairing operator and its Hermitian conjugate can form an angular momentum operator, and (ii) they both commute with the Hamiltonian. Note that the latter (ii) actually implies the former (i), according to Theorem 5 (see also the discussion below it) and the above proposition [now $\Lambda_\omega=\Lambda$ and $|\omega_j|$ is a nonzero constant, due to conclusion (b) in Theorem 4]. Taking these statements together, we get the conclusion that \emph{it is possible for the Hamiltonian \eqref{eq:nH-Hubbard} to have the $SU(2)$ pseudospin symmetry if and only if the hopping amplitudes are symmetric for every bond} (see Sec.~\ref{symmetry-pseudospin} for further results). In other words, we have

\textbf{Corollary 9:}
Even when the Hamiltonian \eqref{eq:nH-Hubbard} commutes with the eta-pairing operators, the $SU(2)$ pseudospin symmetry is not possible \emph{if and only if} the hopping amplitudes are asymmetric for some bond(s) (but not necessarily for every bond).

Despite the above unusual properties of eta-pairing operators and states, we have a general result (see Appendix \ref{subsec:proofLemma10} for a proof):

\textbf{Lemma 10:}
Consider the eta-pairing operators $\eta^\dag$ and $\eta'$ defined by Eqs.~\eqref{eq:eta-raising} and \eqref{eq:eta-lowering}, respectively, and the operator $J_z=\frac{1}{2}\sum_{i\in\Lambda}(n_{i\uparrow}+n_{i\downarrow}-1)$. If the on-site product $\omega_j (\omega'_j)^*$ ($ j\in\Lambda$) is a nonzero constant $C\in\mathbb{C}$, then (a) $(\eta^\dag)^m |0\rangle$ and $(\eta')^{N_\Lambda-m} |2\rangle$ ($0\leq m\leq N_\Lambda$) represent the same eta-pairing state; (b) we have the commutation relations
\begin{equation}
[\eta^\dag,\eta']= 2 CJ_z,  \quad [J_z,\eta^\dag]=\eta^\dag,  \quad \text{and} \quad [J_z,\eta']=-\eta'.
\label{eq:nHcomm}
\end{equation}

Several remarks about this lemma are in order. First, Lemma 10 does not rely on any specific Hamiltonian: $\eta^\dag$ and $\eta'$ [or, $\eta=(\eta^\dag)^\dag$ and $(\eta')^\dag$] are not necessarily eigenoperators of any specific Hamiltonian. Second, $\eta^\dag$ and $\eta'$ are not assumed to be Hermitian conjugate of each other, but are related to each other by the site-independent product $\omega_j (\omega'_j)^*$; thus, the operators $\eta^\dag$ and $\eta'$ obeying Eq.~\eqref{eq:nHcomm} may be called \emph{non-Hermitian angular-momentum operators} (see Remark 2 in Appendix \ref{subsec:proofLemma10} for an explanation of this terminology).
Third, from the assumption of site-independent $\omega_j (\omega'_j)^*$, it follows that $\eta'=\eta$ (up to a trivial factor) if and only if $|\omega_j|$ (or equivalently, $|\omega'_j|$) does not depend on $j\in\Lambda$.
So, when $|\omega_j|$ does not depend on $j$, conclusion (a) means that $(\eta^\dag)^m |0\rangle$ and $\eta^{N_\Lambda-m} |2\rangle$ represent the same state (see the discussion below Corollary 8 for comparison), and conclusion (b) [i.e., Eq.~\eqref{eq:nHcomm}] means that the operators $\eta^\dag$, $\eta$, and $J_z$ form the usual angular momentum operator, which is consistent with the preceding proposition.

In the following, we show that the site-independent product $\omega_j (\omega'_j)^*$ is actually a universal property of our non-Hermitian eta-pairing theory.

\textbf{Theorem 11:}
Assume that for the Hamiltonian \eqref{eq:nH-Hubbard}, all the bonds are strong bonds (i.e., $t_{ij}t_{ji}\neq0$ for every bond $\{i,j\}$). If $\eta^\dag$ ($\eta'$) given by Eq.~\eqref{eq:eta-raising} [Eq.~\eqref{eq:eta-lowering}] is an eta-raising (-lowering) eigenoperator, then (a) $\eta^\dag$ ($\eta'$) is the unique eta-raising (-lowering) eigenoperator, and there also exists a unique eta-lowering (-raising) eigenoperator $\eta'$ ($\eta^\dag$) written as Eq.~\eqref{eq:eta-lowering} [Eq.~\eqref{eq:eta-raising}]; (b) the product $\omega_j (\omega'_j)^*$ is a nonzero constant, and $\eta^\dag$ and $\eta'$ obey Eq.~\eqref{eq:nHcomm}; (c) $U_j-2\mu_j$ does not depend on $j\in\Lambda$ and $\lambda=\lambda'=U_j-2\mu_j $.

This theorem (see Appendix \ref{subsec:proofTheorem11} for a proof) includes Theorem 4 as a special case (see Remark 1 in Appendix \ref{subsec:proofTheorem11}). Note that according to Lemma 10, conclusion (b) in Theorem 11 in turn implies the identities $(\eta^\dag)^m \ket{0}=(\eta')^{N_\Lambda-m} \ket{2}$ and $[(\eta')^\dag]^m \ket{0}=\eta^{N_\Lambda-m} \ket{2}$, up to trivial factors, which represent the right and left eta-pairing eigenstates with eigenvalues $m\lambda$ and $m\lambda^*$, respectively (see Appendix \ref{subsec:RL-eta}).
Furthermore, Theorem 7 is a special case of

\textbf{Theorem 12:}
If $\eta^\dag$ ($\eta'$) in Eq.~\eqref{eq:eta-raising} [Eq.~\eqref{eq:eta-lowering}] is an eta-raising (-lowering) eigenoperator and $\forall j\in\Lambda$, $\omega_j\neq0$ ($\omega'_j\neq0$), then all the bonds must be strong bonds.

This theorem follows directly from Eq.~\eqref{eq:constraints1a} [Eq.~\eqref{eq:constraints2a}]. Note that according to Theorem 11, the above strong bonds in turn imply the same conclusions as in Theorem 11.
From Theorems 11 and 12: \emph{If $\eta^\dag$ ($\eta'$) in Eq.~\eqref{eq:eta-raising} [Eq.~\eqref{eq:eta-lowering}] is an eta-raising (-lowering) eigenoperator. Then, $\forall j\in\Lambda$, $\omega_j\neq0$ ($\omega'_j\neq0$) if and only if all the bonds are strong bonds}.

A more general version of Theorem 5 is (see Appendix \ref{subsec:proofTheorem13})

\textbf{Theorem 13:}
If $\eta^\dag$ and $\eta'$ in Eqs.~\eqref{eq:eta-raising} and \eqref{eq:eta-lowering}, respectively, are both eigenoperators, and there is a site $k$ such that $\omega_k\omega'_k\neq0$. Then, $\forall j\in\Lambda$, $\omega_j\omega'_j\neq0$, or equivalently, all the bonds are strong bonds.

According to Theorem 11, the above strong bonds in turn imply the same conclusions as in Theorem 11. In particular, the product $\omega_j (\omega'_j)^*$ in Theorem 13 is a nonzero constant. There is actually a more general result (see Appendix \ref{subsec:proofCorollary14} for a proof):

\textbf{Corollary 14:}
If $\eta^\dag$ and $\eta'$ in Eqs.~\eqref{eq:eta-raising} and \eqref{eq:eta-lowering}, respectively, are both eigenoperators. Then, the product $\omega_j (\omega'_j)^*$ is a nonzero (zero) constant, when all the bonds are strong bonds (not all the bonds are strong bonds).

For a generic system, the constant $\omega_j (\omega'_j)^*$ in Corollary 14 imposes a universal constraint on $\omega_j$ and $\omega'_j$, which relates the eta-raising with the eta-lowering eigenoperators (see Appendix \ref{subsec:proofCorollary14}). In the presence of asymmetric hoppings, this constraint implies that the right and left eta-pairing eigenstates have distinct spatial distributions  (see Remark 2 in Appendix \ref{subsec:proofCorollary14} and also the concrete models studied in Sec.~\ref{sec:examples}).

Based on the above results, we can prove the following theorem (see Appendix \ref{subsec:proofTheorem15}) which contains the last unusual non-Hermitian eta-pairing phenomenon studied in this paper.

\textbf{Theorem 15:}
If $\eta^\dag$ and $\eta'$ in Eqs. \eqref{eq:eta-raising} and \eqref{eq:eta-lowering}, respectively, are both eigenoperators, i.e., $[H,\eta^\dag]=\lambda\eta^\dag$ and $[H,\eta']=-\lambda'\eta'$. Then, the two constants $\lambda$ and $\lambda'$ can be independent of each other (i.e., they are not necessarily equal) \emph{if and only if} not all the bonds are strong bonds.

This is in sharp contrast to the Hermitian case, where all the bonds are strong bonds and we always have $\lambda=\lambda'$.

Before closing this section, we summarize the relations between the various non-Hermitian eta-pairing phenomena discussed above, which include $(a)$ the Hermitian conjugate of an eta-pairing eigenoperator is not an eigenoperator; $(b)$ the on-site pairing amplitude of an eta-pairing eigenoperator is site dependent; $(c)$ the $SU(2)$ pseudospin symmetry is absent even in the presence of eta-pairing eigenoperators; $(d)$ the two constants $\lambda$ and $\lambda'$ can be independent of each other; $(e)$ the eta-raising or -lowering eigenoperator is not unique. All of these five properties are impossible in Hermitian systems. Remarkably, we have the following relations (see Appendix \ref{subsec:proof-edabc} for a proof):
\begin{equation}
(e)\Rightarrow (d) \Rightarrow(a)\Leftrightarrow(b) \Leftrightarrow (c).
\label{eq:relations}
\end{equation}
Note that when one of the eta-raising and -lowering eigenoperators exists but the other does not exist [and hence the property $(a)$ holds], the property $(d)$ does not make sense. Thus, in this case, we have
$(e) \Rightarrow(a)\Leftrightarrow(b) \Leftrightarrow (c)$.

\section{Symmetries of the Hubbard model}\label{sec:symmetries}

We now turn to discussing the symmetries of the general non-Hermitian Hubbard model Eq.~\eqref{eq:nH-Hubbard} on an arbitrary lattice (which, of course, includes the Hermitian case).

\subsection{$SU(2)$ pseudospin symmetry}\label{symmetry-pseudospin}

The Hamiltonian $H$ in Eq.~\eqref{eq:nH-Hubbard}  under consideration has the $SU(2)$ spin symmetry (see Appendix \ref{subsec:physicalSO4} for some details).
What about the $SU(2)$ pseudospin symmetry? From Theorem 1 (or Theorem 2) and the discussion above Corollary 9, it follows that $H$ has the $SU(2)$ pseudospin symmetry if and only if
\begin{subequations}
\begin{align}
t_{ij}e^{{\rm i}\theta_j}+t_{ji}e^{{\rm i}\theta_i}=0   \qquad     \text{for each bond $\{i,j\}$},      \label{eq:pseudospin-symmetry-a}\\
U_j-2\mu_j=0   \qquad      \text{for each site $j\in\Lambda$} ,    \label{eq:pseudospin-symmetry-b}
\end{align}
\label{eq:pseudospin-symmetry}
\end{subequations}
where $\theta_j$ is the phase of $\omega_j=|\omega_j|e^{{\rm i}\theta_j}$ [see, e.g., Eq.~\eqref{eq:eta-raising}] and $|\omega_j|$ does not depend on $j$ [note that $\omega'_j=\omega_j$, i.e., $\eta'=\eta=(\eta^\dag)^\dag$, up to a trivial factor]; see also Appendixes \ref{subsec:f-fermion} and \ref{subsec:physicalSO4} for more details on the pseudospin symmetry.
From Eq.~\eqref{eq:pseudospin-symmetry-a}, it follows that the hopping amplitudes are symmetric for every bond (see also Corollary 9 or the conclusion above it); note that this does not imply Hermiticity for the Hubbard model \eqref{eq:nH-Hubbard}.

\subsection{$SO(4)$ symmetry}

When both the $SU(2)$ spin and $SU(2)$ pseudospin symmetries are respected, what is the full symmetry group of the Hubbard model? To describe the combination of these $SU(2)$ symmetries, consider the unitary operator $R$ in Eq.~\eqref{eq:so4-8}, which acts on the electron operators as Eq.~\eqref{eq:so4-9} and commutes with the Hamiltonian. Now, let $G(N_\Lambda)$ be the group of the unitary operators $R$ [see Eq.~\eqref{eq:so4-11}]. Then, as shown in Eq.~\eqref{eq:so4-16}, we have  \cite{YangZhang1990,Yang1991}:
\begin{equation}
  G(N_\Lambda) = \left\{
  \begin{array}{ll}
    SU(2)\times SU(2) & \text{for odd  $N_\Lambda$} \\
   SO(4) & \text{for even  $N_\Lambda$}
  \end{array}
  \right..
  \label{eq:group-GN}
\end{equation}
However, as discussed in Appendix \ref{subsec:physicalSO4}, the group $G(N_\Lambda)$ with odd $N_\Lambda$ does not represent the true physical symmetry because it contains a $\mathbb{Z}_2$ gauge transformation (i.e., $-1$), while $G(N_\Lambda)$ with even $N_\Lambda$ does not contain gauge transformations. So, the physical symmetry group is $G(N_\Lambda)/\mathbb{Z}_2=\frac{SU(2)\times SU(2)}{\mathbb{Z}_2}=SO(4)$ for odd $N_\Lambda$ and is simply $G(N_\Lambda)=SO(4)$ for even $N_\Lambda$. Thus, the true symmetry of the system is always $SO(4)$, which does not rely on $N_\Lambda$. This clarifies some ambiguities in the understanding of $SO(4)$ symmetry in previous works \cite{Zhang1991,Yang1991,Arovas2022,Moudgalya2023,Pakrouski2023}. In fact, the global $SO(4)$ symmetry becomes explicit in the Majorana representation \cite{Yang1991}; see Appendix \ref{subsec:Majorana-rep} for details.

\subsection{Particle-hole (PH) symmetries}

The group $SO(4)$ has many interesting subgroups, such as the continuous $SU(2)$ spin and $SU(2)$ pseudospin symmetries discussed above. There are also discrete subgroups generated by, e.g., the PH symmetries.
For example, as shown in Eqs.~\eqref{eq:ph-1} and \eqref{eq:ph-2}, the $SU(2)$ pseudospin rotation contains a $\mathbb{Z}_4$ PH transformation \cite{Kai2020}: $c_{j\uparrow}\rightarrow -e^{{\rm i}\theta_j}c_{j\downarrow}^\dag$ and $c_{j\downarrow}\rightarrow e^{{\rm i}\theta_j}c_{j\uparrow}^\dag$, which is realized by the unitary operator $e^{{\rm i}\pi J_y}$ [with $J_y$ being the pseudospin operator defined in Eq.~\eqref{eq:f4}]. The combination of this $\mathbb{Z}_4$ transformation and an arbitrary spin rotation $R_S$, described by the unitary operator [see Eq.~\eqref{eq:ph-3}]
\begin{equation}
\mathcal{C}_S=e^{{\rm i}\pi J_y}R_S,
\label{eq:ph-general}
\end{equation}
can realize a very general PH transformation [see Eq.~\eqref{eq:ph-4}]; on the other hand, $\mathcal{C}_S$ also represents a charge conjugation, which flips the sign of the charge at every site [see Eq.~\eqref{eq:ph-4-1}]. Notably, when $R_S=e^{-{\rm i}\pi S_y}$ [with $S_y$ being the spin operator defined in Eq.~\eqref{eq:f6}], $\mathcal{C}_S$ in Eq.~\eqref{eq:ph-general} describes the usual $\mathbb{Z}_2$ PH transformation: $c_{j\uparrow}\rightarrow e^{{\rm i}\theta_j}c_{j\uparrow}^\dag$ and $c_{j\downarrow}\rightarrow e^{{\rm i}\theta_j}c_{j\downarrow}^\dag$ [see also Eq.~\eqref{eq:ph-5}].

In addition to the above $\mathbb{Z}_4$ and $\mathbb{Z}_2$ PH symmetries, Eq.~\eqref{eq:ph-general} can also realize a $\mathbb{Z}_{2k}$ PH transformation for any positive integer $k$. To see this, we consider $\mathcal{C}_S$ in Eq.~\eqref{eq:ph-general} with the $U(1)$ spin rotation $R_S=e^{{\rm i}\phi S_z}$, where $0\leq \phi<4\pi$ and $S_z$ is the spin operator defined in Eq.~\eqref{eq:f6}, i.e.,  the following unitary operator [see Eq.~\eqref{eq:ph-6-1}]
\begin{equation}
\mathcal{C}_\phi=e^{{\rm i}\pi J_y}e^{{\rm i}\phi S_z}.
\label{eq:ph-phi}
\end{equation}
As shown in Appendix \ref{subsec:PHsymmetries}, when the spin rotation angle $\phi$ satisfies [see also Eq.~\eqref{eq:ph-7}]
\begin{equation}
\frac{\phi+\pi}{2\pi}=\frac{l}{k} \quad (k>0),
\label{eq:phi-rational}
\end{equation}
$\mathcal{C}_\phi$ in Eq.~\eqref{eq:ph-phi} realizes a $\mathbb{Z}_{2k}$ PH transformation, where $l$ and $k$ are coprime integers and the fraction $l/k$ represents a rational number. Note that $k$ in Eq.~\eqref{eq:phi-rational} can be any positive integer, as discussed in Appendix \ref{subsec:PHsymmetries}. As two concrete examples, let us consider the case of $\phi=0$ and $\phi=\pi$. When $\phi=0$, the rational number in Eq.~\eqref{eq:phi-rational} is $l/k=1/2$ and hence $k=2$, meaning that $\mathcal{C}_{\phi=0}=e^{{\rm i}\pi J_y}$ in Eq.~\eqref{eq:ph-phi} describes a $\mathbb{Z}_{4}$ PH transformation as discussed at the beginning of this subsection. When $\phi=\pi$, Eq.~\eqref{eq:phi-rational} simply gives $k=1$; thus, $\mathcal{C}_{\phi=\pi}=e^{{\rm i}\pi J_y}e^{{\rm i}\pi S_z}$ in Eq.~\eqref{eq:ph-phi} describes a $\mathbb{Z}_{2}$ PH transformation: $c_{j\uparrow}\rightarrow {\rm i}e^{{\rm i}\theta_j}c_{j\downarrow}^\dag$ and $c_{j\downarrow}\rightarrow {\rm i}e^{{\rm i}\theta_j}c_{j\uparrow}^\dag$ (see Appendix \ref{subsec:PHsymmetries}). Comparing this with the usual $\mathbb{Z}_2$ PH transformation discussed above, we see that different PH transformations may share the same symmetry group $\mathbb{Z}_{2k}$.

On the other hand, in contrast to Eq.~\eqref{eq:phi-rational}, when $(\phi+\pi)/(2\pi)$ is an irrational number, the unitary operator $\mathcal{C}_\phi$ in Eq.~\eqref{eq:ph-phi} generates a $\mathbb{Z}$ symmetry (namely a $\mathbb{Z}$ PH transformation), where $\mathbb{Z}$ denotes the additive group of the integers; see Appendix \ref{subsec:PHsymmetries} for more details on this statement. Note that the infinite cyclic group $\mathbb{Z}$ may be viewed as the above group $\mathbb{Z}_{2k}$ with $k=\infty$.

For the Hubbard model \eqref{eq:nH-Hubbard}, all the PH symmetries described by Eq.~\eqref{eq:ph-general} (with arbitrary spin rotations $R_S$) are equivalent to each other, which is actually a consequence of the $SU(2)$ spin symmetry of the Hamiltonian $H$ (see Theorem 21 and its proof in Appendix \ref{subsec:PHsymmetries}). Furthermore, for any given $R_S$, $H$ has the PH symmetry described by Eq.~\eqref{eq:ph-general} (i.e., $\mathcal{C}_S H \mathcal{C}_S^{-1}=H$) if and only if Eqs.~\eqref{eq:pseudospin-symmetry-a} and \eqref{eq:pseudospin-symmetry-b} hold; see the proof of Theorem 21 in Appendix \ref{subsec:PHsymmetries}.

\subsection{$\mathbb{Z}_2$ Shiba transformation}

In addition to the above various discrete PH transformations, the so-called $\mathbb{Z}_2$ Shiba transformation plays an important role in understanding the symmetry properties of Hubbard model. It is a partial PH transformation defined as follows [see Eq.~\eqref{eq:shiba1}]:
\begin{equation}
c_{j\uparrow}\rightarrow c_{j\uparrow}, \qquad  c_{j\downarrow}\rightarrow e^{{\rm i}\theta_j}c_{j\downarrow}^\dag,
\label{eq:shiba-ph}
\end{equation}
which is realized by the unitary operator $U_\downarrow$ given by Eq.~\eqref{eq:shiba2}. However, as stated in Appendix \ref{subsec:physicalSO4}, the $\mathbb{Z}_2$ Shiba transformation does not belong to the $SO(4)$ symmetry; i.e., it cannot be written as the combination of the $SU(2)$ spin and $SU(2)$  pseudospin rotations. This also becomes clear from the Majorana representation [see Eq.~\eqref{eq:mr-22} and the related discussion].

Moreover, the $\mathbb{Z}_2$ Shiba transformation  cannot be a symmetry of the Hubbard model. Actually, the $\mathbb{Z}_2$ Shiba symmetry can be respected by the off-site part $T$ of $H$ in Eq.~\eqref{eq:nH-Hubbard}, but is explicitly broken by the on-site part $V$ of $H$ [see Eq.~\eqref{eq:shiba5}]. More precisely, $T$ has the $\mathbb{Z}_2$ Shiba symmetry (i.e., $U_\downarrow T U^{-1}_\downarrow=T$) if and only if Eq.~\eqref{eq:pseudospin-symmetry-a} holds, as shown in Appendix \ref{subsec:Shiba}; in this case, together with the $SO(4)$ symmetry, the full symmetry of $T$ becomes $O(4)=\mathbb{Z}_2\ltimes SO(4)$, with $\mathbb{Z}_2$ representing the Shiba symmetry [see Eq.~\eqref{eq:mr-26}]. By contrast, from Eq.~\eqref{eq:shiba5}, it is easy to see that the $\mathbb{Z}_2$ Shiba transformation simply flips the sign of $V$ (i.e., $U_\downarrow V U^{-1}_\downarrow=-V$, up to addition of some constant) if and only if Eq.~\eqref{eq:pseudospin-symmetry-b} holds; in this case, $V$ has a local $SO(4)$ symmetry, which becomes explicit in the Majorana representation [see Eq.~\eqref{eq:mr-27} and the related discussion].

Based on the above observation, we see that the $\mathbb{Z}_2$ Shiba transformation \eqref{eq:shiba-ph} changes the total Hamiltonian $H=T+V$ in Eq.~\eqref{eq:nH-Hubbard} as
\begin{equation}
U_\downarrow (T+V) U^{-1}_\downarrow=T-V
\label{eq:shiba-hamiltonian}
\end{equation}
(up to addition of some constant) if and only if Eqs.~\eqref{eq:pseudospin-symmetry-a} and \eqref{eq:pseudospin-symmetry-b} hold.

\subsection{Symmetry unification}\label{symmetry-unification}

From the preceding subsections, we see that the Hubbard model \eqref{eq:nH-Hubbard} can have many different symmetries, including the $SU(2)$ pseudospin [or $SO(4)$] symmetry, the various PH symmetries belonging to the $SO(4)$ symmetry, and the $\mathbb{Z}_2$ Shiba symmetry [in the sense of Eq.~\eqref{eq:shiba-hamiltonian}].

Interestingly, all these different symmetry properties of the Hubbard model share the same underlying algebraic equations, namely Eqs.~\eqref{eq:pseudospin-symmetry-a} and \eqref{eq:pseudospin-symmetry-b}. This means that all these symmetries are equivalent to each other; i.e., if the Hamiltonian $H$ in Eq.~\eqref{eq:nH-Hubbard} possesses any one of these symmetries, then $H$ must possess all the other symmetries (see also Theorems 17 and 21 and the related discussion in Appendix \ref{app:symmetries}).

\section{examples}\label{sec:examples}

We now study several explicit examples of non-Hermitian Hubbard models, which illustrate the various exotic non-Hermitian eta-pairing phenomena [namely the properties $(a)$, $(b)$, $(c)$, $(d)$, and $(e)$] uncovered by our general non-Hermitian eta-pairing theory [see, e.g., Eq.~\eqref{eq:relations} and the related discussion and also the Remark in Appendix \ref{subsec:proof-edabc}].

In particular, for these concrete models and based on our general theory in Sec.~\ref{sec:general-theory}, we give the explicit expressions for both the spatially modulated eta-pairing eigenoperators [e.g., non-Hermitian angular-momentum operators in the sense of Eq.~\eqref{eq:nHcomm}] and the exact many-body eta-pairing eigenstates exhibiting the Fock space localization \cite{Shimomura-Sato2024} (which reduces to the conventional real space localization for special cases).

Finally, in Appendix \ref{app:ODLRO}, we give a brief discussion on the off-diagonal long-range order of the eta-pairing eigenstates in non-Hermitian systems.

\subsection{One-dimensional Hatano-Nelson-Hubbard (HNH) model}
We start with a generalized HNH model \cite{HN1996,HN1997} in one dimension, which is described by Eq.~\eqref{eq:nH-Hubbard} with the ansatz
\begin{equation}
t_{j,j+1}=\alpha_j(1-\gamma),  \qquad  t_{j+1,j}=\alpha_j(1+\gamma)
\label{eq:1dHNH}
\end{equation}
with $\alpha_j\in\mathbb{C}$, $\gamma\in\mathbb{R}$, and $1\leq j\leq N_\Lambda-1$; all other hoppings are zero (implying the open boundary conditions). Here, the nonzero factor $\alpha_j$ is allowed to depend on bond $\{j,j+1\}$. So, either a bond-dependent $\alpha_j$ or a site-dependent $U_j$ (or $\mu_j$) would break the bulk translation symmetry. [Note that when $\alpha_j\equiv1$ and $U_j\equiv0$, the corresponding Hamiltonian reduces to the original noninteracting Hatano-Nelson model (with the spin degree of freedom)].

We first consider the case when $\gamma(\gamma^{2}-1)\neq0$  (e.g., all the bonds are strong bonds and the hopping amplitudes are asymmetric for every bond) and assume that $U_j-2\mu_j$ does not depend on site according to Theorem 11. The unique eta-raising and -lowering eigenoperators of our HNH model, namely $\eta^\dag$ and $\eta'$, can be readily obtained by solving Eqs.~\eqref{eq:constraints1} and \eqref{eq:constraints2}, respectively:
\begin{equation}
\eta^\dag=\sum_{j=1}^{N_\Lambda} q^{-j} c_{j\uparrow}^\dag c_{j\downarrow}^\dag,  \qquad  \eta'=\sum_{j=1}^{N_\Lambda} q^j c_{j\downarrow}c_{j\uparrow}
\label{eq:1d-eta-pairing}
\end{equation}
with $q=(\gamma-1)/(\gamma+1)$. From Eq.~\eqref{eq:1d-eta-pairing}, the product $\omega_j (\omega'_j)^*=q^{-j}q^{j}\equiv 1$ is a nonzero constant, which is consistent with Theorem 11 (or Corollary 14); and $\eta^\dag$ and $\eta'$ obey Eq.~\eqref{eq:nHcomm} (with $C=1$), according to Lemma 10 (or Theorem 11).
Moreover, both $|\omega_j|=|q|^{-j}$ and $|\omega'_j|=|q|^j$  as a function of $j$ take the exponential form, where $|q|\neq 1$ and $0<|q|<\infty$ since $\gamma(\gamma^2-1)\neq0$; thus, the right and left two-particle eta-pairing eigenstates, i.e., $\eta^\dag \ket{0}$ and $(\eta')^\dag \ket{0}$, respectively, are exponentially localized at opposite boundaries of the chain. (It is also interesting to see that the right and left two-hole eta-pairing eigenstates, i.e., $\eta' \ket{2}$ and $\eta \ket{2}$, respectively, are exponentially localized at opposite boundaries of the chain.) Obviously, the three equivalent properties $(a)$, $(b)$, and $(c)$ [see, e.g., Eq. \eqref{eq:relations}] hold.

The above localization discussion focuses on the two-particle states. Now, let us consider the general case for the many-body eta-pairing eigenstates $(\eta^\dag)^m|0\rangle$ with arbitrary $m$. From Eqs.~\eqref{eq:m-eta-state} and \eqref{eq:1d-eta-pairing}, we have
\begin{equation}
(\eta^\dag)^m|0\rangle=m!\sum_{k=k_{\texttt{min}}}^{k_{\texttt{max}}} {q}^{-k} \sum_{k_1+\cdots+k_m=k} \eta_{k_1}^\dag\cdots\eta_{k_m}^\dag|0\rangle,
\label{eq:1d-eta-state}
\end{equation}
where $k_1,\ldots,k_m$ represent $m$ arbitrary and different lattice sites (e.g., $1\leq k_1<\cdots<k_m\leq N_\Lambda$), and we denote the sum of these integers as $k=k_1+\cdots+k_m$, whose minimal and maximal values are respectively $k_{\texttt{min}}=m(1+m)/2$ and $k_{\texttt{max}}=m(2N_\Lambda-m+1)/2$. So the exponential ${q}^{-k}$ in Eq.~\eqref{eq:1d-eta-state} spreads over a wide range of $k$, namely $k_{\texttt{max}}-k_{\texttt{min}}=m(N_\Lambda-m)$, which is proportional to $N_\Lambda^2$ if $m$ is proportional to $N_\Lambda$.
Now, as discussed below Eq.~\eqref{eq:Fock-space}, if we arrange the order of the Fock basis such that the amplitude $|{q}^{-k}|$ decreases with respect to the order (for example, when $|q|>1$, the index $k$ increases with respect to the order), then the many-body eta-pairing eigenstate $(\eta^\dag)^m|0\rangle$ in Eq.~\eqref{eq:1d-eta-state} is exponentially localized in the Fock space Eq.~\eqref{eq:Fock-space}.

According to Theorem 15, the property $(d)$ arises when not all the bonds are strong bonds. For example, consider Eq.~\eqref{eq:1dHNH} with $\gamma=1$; and hence there is no strong bond. In this case, there is no constraint on $U_j$ and $\mu_j$, indicating that $U_j-2\mu_j$ at different sites can be independent of each other. Now, solving Eqs.~\eqref{eq:constraints1} and \eqref{eq:constraints2} gives the unique eta-raising and -lowering eigenoperators
\begin{equation}
\eta^\dag=c_{N_\Lambda\uparrow}^\dag c_{N_\Lambda\downarrow}^\dag,  \qquad  \eta'=c_{1\downarrow}c_{1\uparrow}
\label{eq:1d-eta-pairing-0}
\end{equation}
with $\lambda=U_{N_\Lambda}-2\mu_{N_\Lambda}$ and $\lambda'=U_1-2\mu_1$, respectively. [Note that the product $\omega_j (\omega'_j)^*\equiv 0$ is a zero constant, which is consistent with Corollary 14.] Thus, the property $(d)$ occurs. Clearly, the properties $(a)$, $(b)$, and $(c)$ also occur. From Eq.~\eqref{eq:1d-eta-pairing-0}, we see that the right and left eta-pairing eigenstates $\eta^\dag \ket{0}$ and $(\eta')^\dag \ket{0}$ are exactly localized at the two opposite ends of the chain.

Finally, when $\gamma=0$ and from Eq.~\eqref{eq:1dHNH}, the hopping amplitudes are symmetric for every bond (but the Hamiltonian is not necessarily Hermitian, since $\alpha_j$, $U_j$, and $\mu_j$ can be complex valued). Now, the eta-pairing eigenoperators are $\eta^\dag=\sum_{j=1}^{N_\Lambda} (-1)^j c_{j\uparrow}^\dag c_{j\downarrow}^\dag$ and $\eta'=\eta$; and the Hamiltonian possesses the $SU(2)$ pseudospin symmetry when $U_j\equiv2\mu_j$ (see Sec.~\ref{symmetry-pseudospin}). Thus, the right and left eta-pairing eigenstates coincide with each other and are spatially extended states. These results are the same as the Hermitian case (see Theorem 4 and the discussion below it).

\subsection{Two-dimensional HNH model}\label{2d-HNH}
We now extend the above HNH model \eqref{eq:1dHNH} to two dimensions (e.g., a square lattice of size $N_\Lambda=N_x\times N_y$), which is described by Eq.~\eqref{eq:nH-Hubbard} with the $x$-bond ansatz
\begin{equation}
t_{j,j+\vec{x}}=\alpha_j(1-\gamma_x),  \qquad  t_{j+\vec{x},j}=\alpha_j(1+\gamma_x)
\label{eq:2dHNHx}
\end{equation}
with $\alpha_j\in\mathbb{C}$, $\gamma_x\in\mathbb{R}$, $\vec{x}=(1,0)$, $j=(j_x, j_y)$, $1\leq j_x\leq N_x-1$, and $1\leq j_y\leq N_y$; and the $y$-bond ansatz
\begin{equation}
t_{j,j+\vec{y}}=\beta_j(1-\gamma_y),  \qquad  t_{j+\vec{y},j}=\beta_j(1+\gamma_y)
\label{eq:2dHNHy}
\end{equation}
with $\beta_j\in\mathbb{C}$, $\gamma_y\in\mathbb{R}$, $\vec{y}=(0,1)$, $j=(j_x, j_y)$, $1\leq j_x\leq N_x$, and $1\leq j_y\leq N_y-1$.
All other hoppings are zero (implying the open boundary conditions along both the $x$ and $y$ directions). Here, the nonzero factor $\alpha_j$ ($\beta_j$) may depend on the $x$ ($y$) bonds. So, either a bond-dependent $\alpha_j$ (or $\beta_j$) or a site-dependent $U_j$ (or $\mu_j$) would break the bulk translation symmetry.

Let us now focus on the case when $(\gamma_x^{2}-1)(\gamma_y^{2}-1)\neq0$  (e.g., all the bonds are strong bonds) and assume that $U_j-2\mu_j$ does not depend on site according to Theorem 11. The unique eta-raising and -lowering eigenoperators, namely $\eta^\dag$ and $\eta'$, can be obtained by solving Eqs.~\eqref{eq:constraints1} and \eqref{eq:constraints2}, respectively:
\begin{equation}
\eta^\dag=\sum_{j\in\Lambda} q_x^{-j_x}q_y^{-j_y} c_{j\uparrow}^\dag c_{j\downarrow}^\dag,  \quad  \eta'=\sum_{j\in\Lambda} q_x^{j_x}q_y^{j_y} c_{j\downarrow}c_{j\uparrow}
\label{eq:2d-eta-pairing}
\end{equation}
with $q_{x,y}=(\gamma_{x,y}-1)/(\gamma_{x,y}+1)$, $j=(j_x, j_y)$, $1\leq j_x\leq N_x$, and $1\leq j_y\leq N_y$. From Eq.~\eqref{eq:2d-eta-pairing}, the product $\omega_j (\omega'_j)^*\equiv 1$ is a nonzero constant, which is consistent with Theorem 11 (or Corollary 14); and $\eta^\dag$ and $\eta'$ obey Eq.~\eqref{eq:nHcomm} (with $C=1$), according to Lemma 10 (or Theorem 11).

Now, we discuss the localization behavior of eta-pairing eigenstates. As can be seen from Eq.~\eqref{eq:2d-eta-pairing}, the two-particle eta-pairing eigenstates can exhibit the usual real space localization: When one of $\gamma_x$ and $\gamma_y$ is zero but the other is nonzero, the right and left eta-pairing eigenstates $\eta^\dag \ket{0}$ and $(\eta')^\dag \ket{0}$ are localized at opposite edges of the square lattice; when both $\gamma_x$ and $\gamma_y$ are nonzero, the right and left eta-pairing eigenstates $\eta^\dag \ket{0}$ and $(\eta')^\dag \ket{0}$ are localized at opposite corners.

For the general many-body eta-pairing eigenstates $(\eta^\dag)^m|0\rangle$ with arbitrary $m$, using Eqs.~\eqref{eq:m-eta-state} and \eqref{eq:2d-eta-pairing}, we have
\begin{equation}
(\eta^\dag)^m|0\rangle=m!\sum_{j_1,\ldots,j_m\in\Lambda} {q_x}^{-k} {q_y}^{-l} \eta_{j_1}^\dag\cdots\eta_{j_m}^\dag|0\rangle,
\label{eq:2d-eta-state}
\end{equation}
where $j_1=(k_1,l_1), \ldots, j_m=(k_m,l_m)$ represent $m$ arbitrary and different sites of $\Lambda$, and $k=k_1+\cdots+k_m$ and $l=l_1+\cdots+l_m$.
Now, as discussed below Eq.~\eqref{eq:Fock-space}, if we arrange the order of the Fock basis such that the amplitude $|{q_x}^{-k}{q_y}^{-l}|$ decreases with respect to the order (for example, when $|q_x|=|q_y|>1$, the index $k+l$ increases with respect to the order), then the many-body eta-pairing eigenstate $(\eta^\dag)^m|0\rangle$ in Eq.~\eqref{eq:2d-eta-state} is exponentially localized in the Fock space Eq.~\eqref{eq:Fock-space}.

\textbf{Remark:} In the above one- and two-dimensional HNH models, the right and left two-particle eta-pairing eigenstates are always localized at opposite boundaries (or corners) of the system, which clearly shows the distinction between right and left eta-pairing eigenstates. This opposite localization behavior can be understood from Corollary 14.

\subsection{A general two-sublattice model}\label{examples-twosub}

As discussed above, the one- and two-dimensional HNH models have manifested most of the exotic non-Hermitian eta-pairing phenomena. However, two are left untouched: (1) one of the eta-raising and -lowering eigenoperators exists but the other does not exist, which belongs to the property $(a)$; and (2) the property $(e)$.

To realize all of the exotic eta-pairing phenomena, now we construct a general two-sublattice model, which is defined on an arbitrary lattice $\Lambda$ that is divided into two disjoint sublattices $A$ and $B$. (There are totally $2^{N_\Lambda-1}-1$ distinct partitions of a given $\Lambda$ into two disjoint and nonempty subsets.) The model is described by Eq.~\eqref{eq:nH-Hubbard} with the hoppings between different sublattices
\begin{equation}
  t_{ij} = \left\{
  \begin{array}{ll}
    \omega_A \lambda_{ij} & \text{if  $i\in A$ and $j\in B$} \\
   -\omega_B \lambda_{ij} & \text{if  $i\in B$ and $j\in A$}
  \end{array}
  \right.,
  \label{eq:twosublattice-1}
\end{equation}
where $\lambda_{ij}=\lambda_{ji}$ and $\omega_A, \omega_B, \lambda_{ij}\in\mathbb{C}$; and the hoppings within the same sublattice
\begin{equation}
  t_{ij} =-t_{ji} \quad \text{if  $i,j\in A$ or $B$},
  \label{eq:twosublattice-2}
\end{equation}
where $i\neq j$ and $t_{ij}\in\mathbb{C}$. Note that $\Lambda$ is bipartite if all the hoppings in Eq.~\eqref{eq:twosublattice-2} vanish; otherwise, $\Lambda$ may be nonbipartite (see Appendix \ref{app:two-sublattice-model} for an explanation of this statement). So, the bipartite condition is not necessary for our two-sublattice model.

Now, solving Eqs.~\eqref{eq:constraints1} and \eqref{eq:constraints2} for the above model, one can verify that the eta-pairing operators
\begin{eqnarray}
\eta^\dag &=& \omega_A \sum_{j\in A} c_{j\uparrow}^\dag c_{j\downarrow}^\dag+\omega_B \sum_{j\in B} c_{j\uparrow}^\dag c_{j\downarrow}^\dag,
\nonumber\\
\eta' &=& \omega_B \sum_{j\in A} c_{j\downarrow}c_{j\uparrow}+\omega_A \sum_{j\in B} c_{j\downarrow}c_{j\uparrow}
\label{eq:twosublattice-3}
\end{eqnarray}
are eigenoperators, provided that: $U_j-2\mu_j$ does not depend on $j\in\Lambda$ if $\omega_A \omega_B\neq0$ (i.e., all the bonds are strong bonds); and $U_j-2\mu_j$ within the sublattices $A$ and $B$ are two independent constants, respectively, if $\omega_A \omega_B=0$ (i.e., not all the bonds are strong bonds). From Eq.~\eqref{eq:twosublattice-3}, the product $\omega_j (\omega'_j)^*\equiv \omega_A \omega_B$ is a constant being either nonzero or zero, which is fully consistent with Corollary 14; and $\eta^\dag$ and $\eta'$ obey Eq.~\eqref{eq:nHcomm} (with $C=\omega_A \omega_B$) when $\omega_A \omega_B\neq0$ , according to Lemma 10 (or Theorem 11). Also note that $\eta=\eta'$ (up to a trivial factor) if and only if $|\omega_A|=|\omega_B|$ (i.e., the case of symmetric hoppings for every bond), which is consistent with Corollary 6.

According to our general eta-pairing theory, the unusual non-Hermitian eta-pairing phenomena arise from asymmetric hoppings. Here, for our two-sublattice model, the properties $(a)$, $(b)$, and $(c)$ certainly occur when $|\omega_A|\neq|\omega_B|$ (i.e., in the presence of asymmetric hoppings). In particular, if $\omega_A \omega_B=0$ (e.g., $\omega_A =0, \omega_B\neq0$), other exotic eta-pairing phenomena can also occur.

For example, as shown in Appendix \ref{subsec:abcde}, some special cases of the general two-sublattice model exhibit non-unique eta-raising or -lowering eigenoperators [i.e., the property $(e)$] and hence have all the other properties $(a)$, $(b)$, $(c)$, and $(d)$ [due to Eq.~\eqref{eq:relations}]. Moreover, for some two-sublattice models in Appendixes \ref{subsec:onlyone-exists} and \ref{subsec:abce-d}, one of the eta-raising and -lowering eigenoperators exists but the other does not exist; thus, the property $(d)$ does not make sense [as mentioned below Eq.~\eqref{eq:relations}], and the properties $(a)$, $(b)$, and $(c)$ occur, with or without the property $(e)$.

We now discuss the localization behavior of eta-pairing eigenstates. Let us first consider the special case when $\omega_A \omega_B=0$ (e.g., $\omega_A =0,\omega_B\neq0$); and the eta-pairing eigenoperators in Eq.~\eqref{eq:twosublattice-3} reduce to
\begin{equation}
\eta^\dag=\sum_{j\in B} c_{j\uparrow}^\dag c_{j\downarrow}^\dag,  \quad  \eta'=\sum_{j\in A} c_{j\downarrow}c_{j\uparrow}.
\label{eq:twosublattice-4}
\end{equation}
Note that the property $(e)$ may or may not hold, depending on the details of the hoppings in Eq.~\eqref{eq:twosublattice-2}; see Appendixes \ref{subsec:abcde} and \ref{subsec:abcd-e} for details. From Eq.~\eqref{eq:twosublattice-4}, we see that the right and left eta-pairing eigenstates $(\eta^\dag)^m \ket{0}$ and $[(\eta')^\dag]^n \ket{0}$ are exactly localized in the regions $B$ and $A$, respectively, where $1\leq m\leq N_B$, $1\leq n\leq N_A$, and $N_A$ ($N_B$) is the number of sites in $A$ ($B$). Since the sublattice $A$ or $B$ can be an arbitrary subregion of $\Lambda$, the eta-pairing eigenstates can be localized in any region (but not necessarily at the boundary) of the system.

Next, we consider the case when $\omega_A \omega_B\neq0$ and $|\omega_A|\neq|\omega_B|$. From Eqs.~\eqref{eq:m-eta-state} and \eqref{eq:twosublattice-3}, we have
\begin{equation}
(\eta^\dag)^m|0\rangle=m!\sum_{j_1,\ldots,j_m\in\Lambda} {\omega_A}^k {\omega_B}^{m-k} \eta_{j_1}^\dag\cdots\eta_{j_m}^\dag|0\rangle,
\label{eq:twosublattice-eta-state}
\end{equation}
where $j_1, \ldots, j_m$ represent $m$ arbitrary and different sites of $\Lambda$, and $k$ denotes the number of sites in the intersection $\{j_1, \ldots, j_m\}\cap A$.
Now, as discussed below Eq.~\eqref{eq:Fock-space}, if we arrange the order of the Fock basis such that the amplitude $|{\omega_A}^k {\omega_B}^{m-k}|$ decreases with respect to the order (for example, when $|\omega_A|<|\omega_B|$, the index $k$ increases with respect to the order), then the many-body eta-pairing eigenstate $(\eta^\dag)^m|0\rangle$ in Eq.~\eqref{eq:twosublattice-eta-state} is localized in the Fock space Eq.~\eqref{eq:Fock-space}.

Besides realizing various non-Hermitian phenomena, our general two-sublattice model can reveal the eta-pairing structure [e.g., the $SO(4)$ symmetry] in systems with Hermitian hoppings, including the original eta-pairing theory for square lattice, the extension to triangular lattice, and some well-known topological systems such as the honeycomb-lattice Haldane model, the Su-Schrieffer-Heeger model, and higher-order topological insulators; see Appendix \ref{subsec:Hermitian-T} for more details.

\section{Conclusion}

In this work, we have established a general eta-pairing theory for non-Hermitian Hubbard models on arbitrary lattices. Our general non-Hermitian eta-pairing theory contains several novel properties that are absent in Hermitian systems [see, e.g., Eq.~\eqref{eq:relations}], which in turn leads to a deeper understanding of the conventional Hermitian eta-pairing theory.
In particular, a direct consequence of these novel properties is that the eta-pairing eigenstates can exhibit the Fock space localization, which becomes the conventional real space localization for special cases.

Another unusual phenomenon is that in non-Hermitian systems, the eta-raising eigenoperator $\eta^\dag$ and the eta-lowering eigenoperator $\eta'$ are in general not Hermitian conjugate of each other (i.e., $\eta'\neq\eta$, up to a trivial factor); see the property $(a)$. We also note that $\eta^\dag$ and $\eta$ [or, $(\eta')^\dag$ and $\eta'$] in general cannot form the angular-momentum operators; see the property $(b)$ and the proposition in Sec.~\ref{sec:general-theory}. Thus, in general one cannot construct the usual \emph{Hermitian} angular-momentum operators from the eta-pairing eigenoperators. However, according to Theorem 11, $\eta^\dag$ and $\eta'$ still satisfy the commutation relations for the usual angular-momentum ladder operators [i.e., Eq.~\eqref{eq:nHcomm}] when all the bonds are strong bonds. This leads us to propose the concept of \emph{non-Hermitian angular-momentum operators}, which may play an essential role in non-Hermitian quantum mechanics.

All of the exotic eta-pairing phenomena can be illustrated with several concrete examples. Note that similar to the two-dimensional Hatano-Nelson-Hubbard (HNH) model in Sec.~\ref{2d-HNH}, it is straightforward to construct a  $d$-dimensional HNH model and we can discuss the associated localization behavior. Moreover, the general two-sublattice model can also be applied to some well-known topological systems (with Hubbard interactions), revealing the eta-pairing structure [e.g., the $SO(4)$ symmetry] in these systems.

Our general eta-pairing theory also has deep implications for the symmetries of Hubbard models: A previously unnoticed fact is that in the context of Hubbard model, the various symmetry properties discussed in Sec.~\ref{sec:symmetries} are equivalent to each other; i.e., they are unified by the algebraic equations \eqref{eq:pseudospin-symmetry-a} and \eqref{eq:pseudospin-symmetry-b}. This observation opens up an interesting question: In addition to the symmetries discussed in Sec.~\ref{sec:symmetries}, does the Hubbard model have some new symmetry whose underlying structure is described by Eqs.~\eqref{eq:pseudospin-symmetry-a} and \eqref{eq:pseudospin-symmetry-b}?

Finally, we would like to emphasize that all the results for interacting non-Hermitian many-body systems in this work are mathematically rigorous, even in arbitrary spatial dimensions and without the bulk translation symmetry.

\begin{acknowledgments}
I am grateful to Zheng-Wei Zuo, Feng Tang, De-Xi Shao, and Iosif Pinelis for valuable discussions on related topics. This work was supported by the National Natural Science Foundation of China under Grant No. 12374195 and the Natural Science Foundation of Henan province of China under Grant No. 262300421875.
\end{acknowledgments}

\bibliography{ref}

\appendix

\onecolumngrid

\section{Proofs of the theorems in the main text}

In this appendix, we prove the various theorems and conclusions that appeared in the main text, which constitute our general non-Hermitian eta-pairing theory.

\subsection{Proofs of Theorems 1 and 2}\label{subsec:proofTheorem1-2}

Here, we present proofs of Theorems 1 and 2, which lay the foundation for other theorems in this paper. To prove Theorem 1, we first need to calculate the commutator $[H,\eta^\dag]=[T,\eta^\dag]+[V,\eta^\dag]$. From Eqs.~\eqref{eq:nH-Hubbard} and \eqref{eq:eta-raising}  and using the fermionic anticommutation relations, we obtain
\begin{equation}
[T,\eta^\dag]=\sum_{i\neq j} (t_{ij}\omega_j+t_{ji}\omega_i) c_{i\uparrow}^\dag c_{j\downarrow}^\dag
\label{eq:T-eta}
\end{equation}
(note that the coefficient $t_{ij}\omega_j+t_{ji}\omega_i$ is invariant under the interchange $i\leftrightarrow j$) and
\begin{equation}
[V,\eta^\dag]=\sum_{j\in\Lambda} (U_j-2\mu_j)\omega_j c_{j\uparrow}^\dag c_{j\downarrow}^\dag.
\label{eq:V-eta}
\end{equation}
Then, from the above results and Eq.~\eqref{eq:eta-raising}, it is clear that the eta-raising eigenoperator-equation $[H,\eta^\dag]=\lambda\eta^\dag$ is now equivalent to
\begin{equation}
\sum_{i\neq j} (t_{ij}\omega_j+t_{ji}\omega_i) c_{i\uparrow}^\dag c_{j\downarrow}^\dag+\sum_{j\in\Lambda} (U_j-2\mu_j-\lambda)\omega_j c_{j\uparrow}^\dag c_{j\downarrow}^\dag=0.
\label{eq:T-V-eta}
\end{equation}
This equation holds \emph{if and only if} all the coefficients in Eq.~\eqref{eq:T-V-eta} vanish, i.e., Eqs.~\eqref{eq:constraints1a} and \eqref{eq:constraints1b}, as can be seen from the following lemma.

\textbf{Lemma 16:}
The pairing operators $c_{i\uparrow}^\dag c_{j\downarrow}^\dag$, $c_{j\uparrow}^\dag c_{j\downarrow}^\dag$ ($i,j\in\Lambda$ and $i\neq j$) are linearly independent; i.e., if $\sum_{i\neq j} x_{ij} c_{i\uparrow}^\dag c_{j\downarrow}^\dag+\sum_{j\in\Lambda} y_j c_{j\uparrow}^\dag c_{j\downarrow}^\dag=0$, then all the coefficients $x_{ij}, y_{j}\in\mathbb{C}$ must be zero.

\textit{Proof.}
Given an arbitrary lattice site $k\in\Lambda$, consider the following commutator
\begin{eqnarray}
[c_{k\uparrow},\sum_{i\neq j} x_{ij} c_{i\uparrow}^\dag c_{j\downarrow}^\dag+\sum_{j\in\Lambda} y_j c_{j\uparrow}^\dag c_{j\downarrow}^\dag]&=&\sum_{i\neq j} x_{ij} [c_{k\uparrow},c_{i\uparrow}^\dag c_{j\downarrow}^\dag]+\sum_{j\in\Lambda} y_j [c_{k\uparrow},c_{j\uparrow}^\dag c_{j\downarrow}^\dag]
\nonumber\\
&=&\sum_{i\neq j} x_{ij} \delta_{ik} c_{j\downarrow}^\dag+\sum_{j\in\Lambda} y_j \delta_{jk} c_{j\downarrow}^\dag
\nonumber\\
&=&\sum_{j\neq k} x_{kj} c_{j\downarrow}^\dag+ y_k c_{k\downarrow}^\dag,
\label{eq:comm-c-pairing}
\end{eqnarray}
where we have used the fermionic anticommutation relations. Clearly, if $\sum_{i\neq j} x_{ij} c_{i\uparrow}^\dag c_{j\downarrow}^\dag+\sum_{j\in\Lambda} y_j c_{j\uparrow}^\dag c_{j\downarrow}^\dag=0$, then the commutator \eqref{eq:comm-c-pairing} is also equal to zero and hence $\sum_{j\neq k} x_{kj} c_{j\downarrow}^\dag+ y_k c_{k\downarrow}^\dag=0$, which implies that $x_{kj}=0$ ($j\in\Lambda$ and $j\neq k$) and $y_k=0$ because the second quantized fermionic (or bosonic) creation and annihilation operators are linearly independent. Finally, note that $k$ is an arbitrary site of $\Lambda$, and hence all the coefficients $x_{ij}, y_{j}$ ($i,j\in\Lambda$ and $i\neq j$) are zero.
\hspace{\fill}$\blacksquare$

Now, from the above discussion, Theorem 1 is obvious: Since the equation $[H,\eta^\dag]=\lambda\eta^\dag$ is equivalent to Eq.~\eqref{eq:T-V-eta} and Eq.~\eqref{eq:T-V-eta} is equivalent to Eqs.~\eqref{eq:constraints1a} and \eqref{eq:constraints1b}, $[H,\eta^\dag]=\lambda\eta^\dag$ is thus equivalent to Eqs.~\eqref{eq:constraints1a} and \eqref{eq:constraints1b}. Furthermore, Lemma 16 implies that \emph{the operators $c_{i\uparrow}^\dag c_{j\downarrow}^\dag$ ($i,j\in\Lambda$ and $i\neq j$) must be linearly independent}, and hence the equation $[T,\eta^\dag]=0$ is equivalent to Eq.~\eqref{eq:constraints1a} [see Eq.~\eqref{eq:T-eta}]. Moreover, \emph{the operators  $c_{j\uparrow}^\dag c_{j\downarrow}^\dag$ ($j\in\Lambda$) are also linearly independent} due to Lemma 16, and hence the equation $[V,\eta^\dag]=\lambda\eta^\dag$ is equivalent to Eq.~\eqref{eq:constraints1b} [see Eq.~\eqref{eq:V-eta}].

Finally, let us prove Theorem 2. There are actually two different proofs. The first proof simply repeats the same procedure as in the above proof of Theorem 1, which is omitted here. The second proof is based on Theorem 1, say, Theorem 2 can be derived from Theorem 1 (and vice versa): Note that the eta-lowering eigenoperator-equation $[H,\eta']=-\lambda'\eta'$ is equivalent to $[H',(\eta')^\dag]=(\lambda')^*(\eta')^\dag$ with $H'=H^\dag$. Thus, using Theorem 1 with the replacements $H\rightarrow H'$ (i.e., $t_{ij}\rightarrow t'_{ij}=t^*_{ji}, U_j\rightarrow U'_j=U^*_j, \mu_j\rightarrow \mu'_j=\mu^*_j$), $\eta\rightarrow\eta'$ (i.e., $\omega_j\rightarrow\omega'_j$), and $\lambda\rightarrow(\lambda')^*$, we obtain Theorem 2.

\subsection{Right and left eta-pairing eigenstates}\label{subsec:RL-eta}

Here, we give a discussion about the right and left eta-pairing eigenstates. Given a non-Hermitian Hamiltonian $H$, the right (left) eigenstates of $H$ are defined as the eigenstates of $H$ ($H^\dag$). Now, let $\eta^\dag$ and $\eta'$ be the eta-raising and -lowering eigenoperators of $H$, respectively; i.e., we have
\begin{equation}
[H,\eta^\dag]=\lambda\eta^\dag, \qquad [H,\eta']=-\lambda'\eta'.
\label{eq:H-eta-sm}
\end{equation}
From the above eigenoperator equations, one can construct many right eta-pairing eigenstates
\begin{equation}
H(\eta^\dag)^m|0\rangle=m\lambda(\eta^\dag)^m|0\rangle,  \qquad H(\eta')^n|2\rangle=[\sum_{j\in\Lambda}(U_j-2\mu_j)-n\lambda'](\eta')^n|2\rangle,
\label{eq:r-state-sm}
\end{equation}
where $ m,n=0, 1, \ldots, N_\Lambda$ and $|2\rangle=\prod_{j\in\Lambda}c_{j\uparrow}^\dag c_{j\downarrow}^\dag|0\rangle$, as defined in the main text. Taking Hermitian conjugate of the eigenoperator equations in Eq.~\eqref{eq:H-eta-sm}, we have
\begin{equation}
[H^\dag,\eta]=-\lambda^*\eta, \qquad [H^\dag,(\eta')^\dag]=(\lambda')^*(\eta')^\dag.
\label{eq:Hdag-eta-sm}
\end{equation}
Namely, $(\eta')^\dag$ and $\eta$ are the eta-raising and -lowering eigenoperators of $H^\dag$, respectively. From Eq.~\eqref{eq:Hdag-eta-sm}, one can construct many left eta-pairing eigenstates
\begin{equation}
H^\dag[(\eta')^\dag]^m|0\rangle=m(\lambda')^*[(\eta')^\dag]^m|0\rangle,  \qquad H^\dag\eta^n|2\rangle=[\sum_{j\in\Lambda}(U^*_j-2\mu^*_j)-n\lambda^*]\eta^n|2\rangle.
\label{eq:l-state-sm}
\end{equation}

Now, we give some remarks about the relations between these eta-pairing eigenstates, and the following general results can be illustrated with the concrete examples studied in this paper.

\textbf{Remark 1:}
In this remark, we assume that all the bonds are strong bonds. From conclusion (b) in Theorem 11 and  conclusion (a) in Lemma 10 (see also the discussion below Theorem 11 in the main text), the two right (left) eta-pairing eigenstates $(\eta^\dag)^m|0\rangle$ and $(\eta')^n|2\rangle$ ($[(\eta')^\dag]^m|0\rangle$ and $\eta^n|2\rangle$) represent the same right (left) eta-pairing eigenstate with eigenvalue $m\lambda$ ($m\lambda^*$), where $m+n=N_\Lambda$. Note that $U_j-2\mu_j$ does not depend on $j$ and we have $\lambda=\lambda'=U_j-2\mu_j$, due to conclusion (c) in Theorem 11 (see also Theorem 15). Thus, the eigenvalues of the right and left eta-pairing eigenstates, namely $m\lambda$ and $m\lambda^*$, are complex conjugate of each other.

Does the right eta-pairing eigenstate $(\eta^\dag)^m|0\rangle=(\eta')^{N_\Lambda-m}|2\rangle$ (up to a trivial factor) coincide with the left eta-pairing eigenstate $[(\eta')^\dag]^m|0\rangle=\eta^{N_\Lambda-m}|2\rangle$ (up to a trivial factor)? The answer is no in the presence of asymmetric hoppings, except when the eta-pairing eigenstates are trivial states $|0\rangle$ (for $m=0$) and $|2\rangle$ (for $m=N_\Lambda$). A short reason is that the eta-raising and -lowering eigenoperators are not Hermitian conjugate of each other (i.e., $\eta\neq\eta'$, up to a trivial factor), as can be seen from Corollary 6 (see also the discussion below it). A more detailed discussion can be found in Remark 2 in Appendix \ref{subsec:proofCorollary14}. Another way to see the distinction is to compare $(\eta^\dag)^m|0\rangle$ with $\eta^{N_\Lambda-m}|2\rangle$, as discussed below Corollary 8 in the main text (note that asymmetric hoppings lead to a site-dependent $|\omega_j|$, due to Corollary 8).

However, when the hopping amplitudes are symmetric for every bond, we have $\eta'=\eta$ (up to a trivial factor), according to conclusion (a) in Theorem 4 (note that the Hamiltonian $H$ is not necessarily Hermitian). Thus, the right and left eta-pairing eigenstates are equal. In other words, the eta-pairing states $(\eta^\dag)^m|0\rangle$ are the common eigenstates of $H$ and $H^\dag$, with eigenvalues $m\lambda$ and $m\lambda^*$, respectively. In particular, we have $(\eta^\dag)^m|0\rangle=\eta^{N_\Lambda-m}|2\rangle$ (up to a trivial factor), which is consistent with the discussion below Lemma 10 [note that, due to conclusion (b) in Theorem 4, $|\omega_j|$ does not depend on $j$].

\textbf{Remark 2:}
If not all the bonds are strong bonds, then, all the eta-pairing eigenstates in Eqs.~\eqref{eq:r-state-sm} and \eqref{eq:l-state-sm} are different from each other, in sharp contrast to the case where all the bonds are strong bonds in Remark 1.

To see this, let us first compare the right and left eta-pairing eigenstates $(\eta^\dag)^m|0\rangle$ and $[(\eta')^\dag]^m|0\rangle$. They have distinct spatial distributions, as discussed in Remark 2 in Appendix \ref{subsec:proofCorollary14}. Next, we compare the right and left eta-pairing eigenstates $(\eta^\dag)^m|0\rangle$ and $\eta^n|2\rangle$. As explained in Remark 1 in Appendix \ref{subsec:proofLemma10}, $(\eta^\dag)^m|0\rangle$ is always different from $\eta^n|2\rangle$ [note that, due to Theorem 12 (see also the discussion below it), we must have $\omega_j=0$ on some site(s) when not all the bonds are strong bonds]. Finally, let us compare the two right eta-pairing eigenstates $(\eta^\dag)^m|0\rangle$ and $(\eta')^n|2\rangle$. Since the two eta-pairing operators $\eta^\dag$ and $\eta'$ creates and annihilates particles in two \emph{disjoint} regions, namely $\Lambda_\omega$ and $\Lambda_{\omega'}$, respectively (see, e.g., Remark 2 in Appendix \ref{subsec:proofCorollary14}), the two eta-pairing states $(\eta^\dag)^m|0\rangle$ and $(\eta')^n|2\rangle$ generally have different spatial distributions of particles. However, there is only one exception: When $\Lambda_\omega\cup\Lambda_{\omega'}=\Lambda$ (see, e.g., the Remark in Appendix \ref{subsec:proofTheorem15}), $(\eta^\dag)^{|\Lambda_\omega|}|0\rangle$ and $(\eta')^{|\Lambda_{\omega'}|}|2\rangle$ represent the same right eigenstate $\prod_{j\in\Lambda_\omega}c_{j\uparrow}^\dag c_{j\downarrow}^\dag|0\rangle$ with eigenvalue $|\Lambda_\omega|\lambda$, where $|\Lambda_\omega|$ ($|\Lambda_{\omega'}|$) denotes the number of sites in $\Lambda_\omega$ ($\Lambda_{\omega'}$) and hence $|\Lambda_\omega|+|\Lambda_{\omega'}|=N_\Lambda$ (note that $\Lambda_\omega\cap\Lambda_{\omega'}=\emptyset$, as mentioned above). Now, the summation $\sum_{j\in\Lambda}(U_j-2\mu_j)$ in Eq.~\eqref{eq:r-state-sm} is equal to $\sum_{j\in\Lambda_\omega}(U_j-2\mu_j)+\sum_{j\in\Lambda_{\omega'}}(U_j-2\mu_j)=|\Lambda_\omega|\lambda+|\Lambda_{\omega'}|\lambda'$, as can be seen from the discussion below Theorem 3 in the main text. The comparisons between the other eta-pairing eigenstates in Eqs.~\eqref{eq:r-state-sm} and \eqref{eq:l-state-sm} are similar to the above discussion and are omitted.

Unlike in the case when all the bonds are strong bonds in Remark 1, $\lambda$ and $\lambda'$ are two independent constants when not all the bonds are strong bonds, due to Theorem 15. Thus, the eigenvalues of the right and left eta-pairing eigenstates are also independent of each other, e.g., they are not complex conjugate of each other when $\lambda\neq\lambda'$ [see Eqs.~\eqref{eq:r-state-sm} and \eqref{eq:l-state-sm}]. As a result, each right eta-pairing eigenstate in Eq.~\eqref{eq:r-state-sm} is orthogonal to each left eta-pairing eigenstate in Eq.~\eqref{eq:l-state-sm}. [Let $\psi_R$ ($\psi_L$) be the right (left) eigenstate of $H$ with eigenvalue $E_R$ ($E_L$), i.e., $H\psi_R=E_R\psi_R$ and $H^\dag\psi_L=E_L\psi_L$, and consider the inner product $(H\psi_R|\psi_L)=(\psi_R|H^\dag\psi_L)$. This equation then becomes $E^*_R(\psi_R|\psi_L)=E_L(\psi_R|\psi_L)$. Thus, if $E^*_R\neq E_L$ (or equivalently $E_R\neq E^*_L$), then $(\psi_R|\psi_L)=0$.] This result is consistent with the distinction between right and left eta-pairing eigenstates discussed above. Note that the eta-pairing eigenstates are insensitive to the values of $U_j$ and $\mu_j$ (or $\lambda$ and $\lambda'$), e.g., the Hubbard interaction (see the discussion above Theorem 3 in the main text).

As another consequence of independent $\lambda$ and $\lambda'$,  from Eqs.~\eqref{eq:r-state-sm} and \eqref{eq:l-state-sm},  $m\lambda'$ ($m\lambda^*$) and $\sum_{j\in\Lambda}(U_j-2\mu_j)-n\lambda$ [$\sum_{j\in\Lambda}(U^*_j-2\mu^*_j)-n(\lambda')^*$] are also eigenvalues of $H$ ($H^\dag$), but the corresponding right (left) eigenstates are generally not eta-pairing states.

\subsection{Proof of Theorem 3}\label{subsec:proofTheorem3}

At first sight, Theorem 3 seems obvious. However, it is nontrivial to give a clear and complete proof of this theorem. Now, let us prove Theorem 3 for the case of eta-raising eigenoperator $\eta^\dag$, and the proof of eta-lowering eigenoperator $\eta'$ would be similar.

Assume that $\eta^\dag=\sum_{j\in\Lambda} \omega_j c_{j\uparrow}^\dag c_{j\downarrow}^\dag$ is an eta-raising eigenoperator of $H$, which means that we have assumed the existence of eta-raising eigenoperator. If there is a lattice site, say $k$, such that $\omega_k=0$, then $\omega_j=0$ for every site $j\in\Lambda$ (see below) and hence $\eta^\dag=0$, which contradicts with the assumption that $H$ has an eta-raising eigenoperator. Thus, conclusion (b) in Theorem 3 must be true.  [The reason for the above implication $\omega_k=0\Rightarrow\omega_j=0$ is as follows: By the assumption of Theorem 3, there exists a connected path of strong bonds between $k$ and $j$; i.e., there is a finite sequence of sites $i_1, \ldots, i_n\in\Lambda$ with $i_1=k$ and $i_n=j$, such that every bond $\{i_l,i_{l+1}\}$ with $l=1, \ldots, n-1$ is a strong bond. From Eq.~\eqref{eq:constraints1a} in Theorem 1, it is obvious that $\omega_{i_l}=0$ implies $\omega_{i_{l+1}}=0$ (and vice versa), since $\{i_l,i_{l+1}\}$ is a strong bond (i.e., both $t_{i_l,i_{l+1}}$ and $t_{i_{l+1},i_l}$ are nonzero). Finally, we have $\omega_{i_1=k}=0\Rightarrow\omega_{i_2}=0\Rightarrow \cdots \Rightarrow\omega_{i_n=j}=0$ and hence $\omega_k=0\Rightarrow\omega_j=0$.]

Next, we prove conclusion (a) in Theorem 3 (e.g., the eta-raising eigenoperator $\eta^\dag$ is unique, up to a trivial factor). Now, consider two arbitrary sites $k$ and $j$ and a connected path of strong bonds between them (e.g., a sequence $i_1, \ldots, i_n\in\Lambda$ with $i_1=k$ and $i_n=j$, as described above). Using Eq.~\eqref{eq:constraints1a} for every strong bond $\{i_l,i_{l+1}\}$ with $l=1, \ldots, n-1$, we have
\begin{equation}
\frac{\omega_j}{\omega_k}=\prod_{l=1}^{n-1} \frac{\omega_{i_{l+1}}}{\omega_{i_l}}=(-1)^{n-1}\prod_{l=1}^{n-1} \frac{t_{i_{l+1},i_l}}{t_{i_l,i_{l+1}}},
\label{eq:theorem3a}
\end{equation}
where the index $n-1$ corresponds to the total number of strong bonds along the path. Note that $\omega_k\neq0$ and $\omega_j\neq0$ due to conclusion (b).

Since the path of strong bonds connecting the two given sites $k$ and $j$ may not be unique, then does the value of the right-hand side (RHS) of Eq.~\eqref{eq:theorem3a} depend on the path we choose? The answer is no: given an eta-raising eigenoperator $\eta^\dag=\sum_{j\in\Lambda} \omega_j c_{j\uparrow}^\dag c_{j\downarrow}^\dag$, the left-hand side of Eq.~\eqref{eq:theorem3a} (i.e., $\omega_j/\omega_k$ with given sites $k$ and $j$) is clearly path independent and hence the RHS is also path independent. (Note that since the operators  $c_{j\uparrow}^\dag c_{j\downarrow}^\dag$ on all sites are linearly independent due to Lemma 16,  an eta-pairing operator is fixed if and only if $\omega_j$ on each site $j$ is fixed.)

Now, the uniqueness of $\eta^\dag$ is obvious: According to Eq.~\eqref{eq:theorem3a}, the ratio $\omega_j/\omega_k$ (with given sites $k$ and $j$) is fixed (i.e., unique) by the hoppings along the path connecting $k$ and $j$, namely the RHS of Eq.~\eqref{eq:theorem3a} which is a path-independent quantity (as explained above). Note that $k$ and $j$ are two arbitrary sites of $\Lambda$, and hence the operator $\eta^\dag$ is unique, up to a site-independent factor.

\textbf{Corollary:}
From the above proof, if the value of the RHS of Eq.~\eqref{eq:theorem3a} depends on the path we choose, then there does not exist any eta-raising eigenoperator; and hence the eta-lowering eigenoperator also does not exist, because: if the eta-lowering eigenoperator $\eta'=\sum_{j\in\Lambda} (\omega'_j)^* c_{j\downarrow}c_{j\uparrow}$ exists, then $\omega'_j\neq0$ ($\forall j\in\Lambda$) due to conclusion (b) in Theorem 3 and one can define the operator $\eta^\dag=\sum_{j\in\Lambda} \frac{1}{(\omega'_j)^*} c_{j\uparrow}^\dag c_{j\downarrow}^\dag$ [e.g., $\omega_j=\frac{1}{(\omega'_j)^*}$, see also Eq.~\eqref{eq:theorem11}] which would be an eta-raising eigenoperator according to Theorems 1 and 2.

For example, consider a simple three-site system with the hoppings $t_{12}=t_{21}\neq0$, $t_{23}=t_{32}\neq0$, and $t_{13}=t_{31}\neq0$. There are two paths connecting the sites 1 and 3: one path is simply the strong bond $\{1,3\}$, and the other consists of the two strong bonds $\{1,2\}$ and $\{2,3\}$. So, for these two paths, the corresponding values of the RHS of Eq.~\eqref{eq:theorem3a} are $(-1)^{1}=-1$ and $(-1)^{2}=1$, respectively, which are clearly not equal (i.e., path dependent). Thus, this three-site system cannot host any eta-pairing eigenoperator. We note that this simple example explains the absence of eta-pairing eigenoperator in the usual Hermitian Hubbard model with real nearest-neighbor hoppings on the triangular lattice.

\textbf{Remark:}
According to Theorem 3 and the discussion below Theorem 12 in the main text, if the eta-raising or -lowering eigenoperator exists, then the following three statements are equivalent to each other: (1) There exists a connected path of strong bonds between two arbitrary sites (i.e., the assumption in Theorem 3); (2) $\forall j\in\Lambda$, $\omega_j\neq0$ ($\omega'_j\neq0$), namely conclusion (b) in Theorem 3; (3) All the bonds are strong bonds.

Therefore, if not all the bonds are strong bonds while there exists a connected path of strong bonds between two arbitrary sites, then there does not exist any eta-pairing eigenoperator. For example, consider a one-dimensional system with the hoppings satisfying $t_{j,j+1}t_{j+1,j}\neq0$ for all $j=1,\ldots, N_\Lambda-1$ and, e.g., $t_{13}\neq0$ (all other hoppings are zero). Clearly, any two lattice sites are connected by a path of strong bonds, but the bond $\{1,3\}$ is not a strong bond. With the above hoppings, Eq.~\eqref{eq:constraints1a} [Eq.~\eqref{eq:constraints2a}] holds if and only if $\omega_j=0$ ($\omega'_j=0$) for every site $j\in\Lambda$, meaning that the eta-pairing eigenoperator does not exist.

\subsection{Proof of Theorem 4}\label{subsec:proofTheorem4}

In the non-Hermitian case, conclusion (a) in Theorem 4 [see also conclusion (a) in Theorem 11] is nontrivial, since it states that the existence of eta-raising (-lowering) eigenoperator implies the existence of eta-lowering (-raising) eigenoperator, which is not always true for non-Hermitian Hamiltonians (see, e.g., the general two-sublattice model we studied in this paper). Now, let us prove Theorem 4 for the case of eta-raising eigenoperator $\eta^\dag$, and the proof of eta-lowering eigenoperator $\eta'$ would be similar.

In Theorem 4, the assumption that the hopping amplitudes are symmetric for every bond implies that all the bonds are strong bonds. Thus, according to conclusion (a) in Theorem 3, the eta-raising eigenoperator $\eta^\dag=\sum_{j\in\Lambda} \omega_j c_{j\uparrow}^\dag c_{j\downarrow}^\dag$ must be unique. Now, we prove that the Hermitian conjugate of $\eta^\dag$, namely the eta-lowering operator $\eta=(\eta^\dag)^\dag$, is also an eigenoperator. According to Theorem 1, $\omega_j$ satisfies Eq.~\eqref{eq:constraints1a}. Taking complex conjugation of Eq.~\eqref{eq:constraints1a}, we have $t_{ij}^*\omega_j^*+t_{ji}^*\omega_i^*=0$ for every bond $\{i,j\}$. Then, multiplying this equation by the product $t_{ij}t_{ji}$ from both sides, we have
\begin{equation}
|t_{ij}|^2 t_{ji}\omega_j^*+|t_{ji}|^2 t_{ij}\omega_i^*=0.
\label{eq:theorem4}
\end{equation}
Under the assumption that $|t_{ij}|=|t_{ji}|$ for every bond $\{i,j\}$, Eq.~\eqref{eq:theorem4} reduces to $ t_{ji}\omega_j^*+ t_{ij}\omega_i^*=0$, meaning that $\omega_j$ also satisfies Eq.~\eqref{eq:constraints2a}. Moreover, Eq.~\eqref{eq:constraints1b} requires that $U_j-2\mu_j$ does not depend on $j\in\Lambda$ and $\lambda=U_j-2\mu_j$, because $\omega_j\neq0$  ($\forall j\in\Lambda$) due to conclusion (b) in Theorem 3. Thus, $\omega_j$ also satisfies Eq.~\eqref{eq:constraints2b} with $\lambda'=\lambda=U_j-2\mu_j$. To summarize, $\omega_j$ satisfies Eqs.~\eqref{eq:constraints2a} and \eqref{eq:constraints2b}. Thus, according to Theorem 2, $\eta=(\eta^\dag)^\dag$ is also an eigenoperator and is the unique eta-lowering eigenoperator due to conclusion (a) in Theorem 3. So, we have proved conclusions (a) and (c) in Theorem 4. Finally, let us prove conclusion (b). Under the assumption that $|t_{ij}|=|t_{ji}|$ for every bond $\{i,j\}$ and using Eq.~\eqref{eq:constraints1a} [or Eq.~\eqref{eq:constraints2a}], we have $|\omega_i|=|\omega_j|$ for every bond $\{i,j\}$. This, together with the lattice connectivity, implies that $|\omega_j|$ on all lattice sites are equal (i.e., $|\omega_j|$ does not depend on $j\in\Lambda$).

\textbf{Remark:}
As mentioned in the main text, Theorem 4 tells us that when the hopping amplitudes are symmetric for every bond, all the basic properties of the Hermitian eta-pairing theory are preserved even in non-Hermitian systems. For example, conclusion (a) in Theorem 4 implies that the right and left eta-pairing eigenstates coincide with each other, i.e., the eta-pairing states $(\eta^\dag)^m \ket{0}$ ($0\leq m\leq N_\Lambda$) are the common eigenstates of $H$ and $H^\dag$, with eigenvalues $m\lambda$ and $m\lambda^*$, respectively; and conclusion (b) in Theorem 4 implies that $(\eta^\dag)^m \ket{0}$ are spatially extended states.

\subsection{Proof of Theorem 5}\label{subsec:proofTheorem5}

This theorem (see also Corollary 6) will touch the first unusual non-Hermitian eta-pairing phenomenon (without any Hermitian analog) studied in this paper: the Hermitian conjugate of an eta-pairing eigenoperator may not be an eigenoperator. Now, let us prove Theorem 5.

By the assumption of the theorem, let the eta-raising and eta-lowering eigenoperators, which are Hermitian conjugate of each other, be $\eta^\dag=\sum_{j\in\Lambda} \omega_j c_{j\uparrow}^\dag c_{j\downarrow}^\dag$ and $\eta=(\eta^\dag)^\dag$, respectively. According to Theorems 1 and 2, $\omega_j$ satisfies Eqs.~\eqref{eq:constraints1a} and \eqref{eq:constraints2a}; i.e., for every bond $\{i,j\}$, we have
\begin{equation}
t_{ij}\omega_j+t_{ji}\omega_i=0
\label{eq:theorem5-1}
\end{equation}
and
\begin{equation}
 t_{ij}\omega_i^*+t_{ji}\omega_j^*=0.
\label{eq:theorem5-2}
\end{equation}
From Eqs.~\eqref{eq:theorem5-1} and \eqref{eq:theorem5-2}, we arrive at
\begin{equation}
|t_{ij}|^2\omega_i\omega_j=|t_{ji}|^2\omega_i\omega_j
\label{eq:theorem5}
\end{equation}
for every bond $\{i,j\}$. Note that we can remove the factor $\omega_i\omega_j$ in Eq.~\eqref{eq:theorem5} and obtain $|t_{ij}|=|t_{ji}|$ only when $\omega_i\omega_j\neq0$. Now we prove that $\omega_j\neq0$ ($\forall j\in\Lambda$). [Note that this does not follow from conclusion (b) in Theorem 3, because in Theorem 5 we did not assume the existence of strong bonds; instead, we are going to prove that the hopping amplitudes are symmetric for every bond and hence all the bonds are strong bonds.] Specifically, consider an arbitrary bond $\{i,j\}$ and, by definition, at least one of the hoppings $t_{ij}$ and $t_{ji}$ is nonzero (e.g., $t_{ij}\neq0$). Thus, from Eq.~\eqref{eq:theorem5-1} [Eq.~\eqref{eq:theorem5-2}], we have the implication $\omega_i=0\Rightarrow\omega_j=0$ ($\omega_j=0\Rightarrow\omega_i=0$). To summarize, for every bond $\{i,j\}$, $\omega_i=0$ if and only if $\omega_j=0$. [Note that this result cannot be derived from only one of the two equations \eqref{eq:theorem5-1} and \eqref{eq:theorem5-2}, because again we did not assume that $\{i,j\}$ is a strong bond.] Therefore, if there is a lattice site, say $k(\in\Lambda)$, such that $\omega_k=0$. Then, according to the above result and  the lattice connectivity, $\omega_j=0$ ($\forall j\in\Lambda$) and hence both $\eta^\dag$ and $\eta$ vanish, which contradicts with the assumption of Theorem 5. Thus, we must have $\omega_j\neq0$ ($\forall j\in\Lambda$). Finally, from Eq.~\eqref{eq:theorem5}, we obtain $|t_{ij}|=|t_{ji}|$ for every bond $\{i,j\}$.

\subsection{Eta-pairing and angular momentum operators}\label{subsec:eta-angular}

In the conventional Hermitian eta-pairing theory, the eta-pairing operator and its Hermitian conjugate can be viewed as the angular momentum ladder operators, since they satisfy the angular-momentum commutation relations.
This property indicates the so-called $SU(2)$  pseudospin symmetry of the Hubbard model. However, in the non-Hermitian case, the eta-pairing operator and its Hermitian conjugate may not be able to form an angular momentum operator when the on-site pairing amplitude $|\omega_j|$ is site dependent.

In the main text, the sufficient and necessary condition for the eta-pairing operators to be an angular momentum operator has been summarized in the proposition.
For completeness, we now repeat the statement of the proposition followed by a proof.

\textbf{Proposition:}
An eta-pairing operator, together with its Hermitian conjugate, can form an angular momentum operator if and only if the nonzero $|\omega_j|$ does not depend on $j\in\Lambda_\omega(\subseteq\Lambda)$.

\textit{Proof.}
The eta-pairing operator in Eq.~\eqref{eq:eta-raising} can be rewritten as
\begin{equation}
\eta^\dag=\sum_{j\in\Lambda} \omega_j c_{j\uparrow}^\dag c_{j\downarrow}^\dag=\sum_{j\in\Lambda}|\omega_j|J_j^+,  \qquad  \omega_j=|\omega_j|e^{{\rm i}\theta_j}
\label{eq:eta-raising-sm}
\end{equation}
with the new operator $J_j^+$ defined as $J_j^+=e^{{\rm i}\theta_j}c_{j\uparrow}^\dag c_{j\downarrow}^\dag$. Thus, its Hermitian conjugate reads $\eta=(\eta^\dag)^\dag=\sum_{j\in\Lambda}|\omega_j|J_j^-$ with $J_j^-=(J_j^+)^\dag$. If we further introduce the operators $J_j^z=\frac{1}{2}(c_{j\uparrow}^\dag c_{j\uparrow}+c_{j\downarrow}^\dag c_{j\downarrow}-1)$, then for each site $j\in\Lambda$, the operators $J_j^+$, $J_j^-$, and $J_j^z$ form an angular momentum operator; i.e., they satisfy
\begin{equation}
[J_i^+,J_j^-]=2\delta_{ij} J_j^z, \qquad [J_i^z,J_j^+]=\delta_{ij} J_j^+, \qquad [J_i^z,J_j^-]=-\delta_{ij} J_j^- ,
\label{eq:local-pseudospin}
\end{equation}
and $[J_i^+,J_j^+]=[J_i^-,J_j^-]=[J_i^z,J_j^z]=0$. Note that these results are independent of the choice of the phase parameter $\theta_j$. In Appendix \ref{subsec:f-fermion}, we discuss the physical meaning of these operators and their relations with the spin operators in more detail.

Because of $|\omega_j|=0$ for $j\notin\Lambda_\omega$ (recall that $\Lambda_\omega$ denotes the set of all the nonzero points of $\omega_j$, as defined in the main text), the summation region in Eq.~\eqref{eq:eta-raising-sm} actually reduces to
\begin{equation}
\eta^\dag=\sum_{j\in\Lambda_\omega}|\omega_j|J_j^+,
\label{eq:eta-raising01-sm}
\end{equation}
and hence $\eta=\sum_{j\in\Lambda_\omega}|\omega_j|J_j^-$. From Eqs.~\eqref{eq:eta-raising01-sm} and \eqref{eq:local-pseudospin}, now we have the commutation relation
\begin{equation}
[\eta^\dag,\eta]=2\sum_{j\in\Lambda_\omega}|\omega_j|^2 J_j^z=2\varepsilon\tilde{J_z},
\label{eq:eta-commutator-sm}
\end{equation}
where we have defined the Hermitian operator
\begin{equation}
\tilde{J_z}=\frac{1}{\varepsilon}\sum_{j\in\Lambda_\omega}|\omega_j|^2 J_j^z
\label{eq:new-Jz-sm}
\end{equation}
with an arbitrary and real constant $\varepsilon\neq0$. Then, using Eqs.~\eqref{eq:eta-raising01-sm}, \eqref{eq:new-Jz-sm}, and \eqref{eq:local-pseudospin}, we get
\begin{equation}
[\tilde{J_z},\eta^\dag]=\frac{1}{\varepsilon}\sum_{j\in\Lambda_\omega}|\omega_j|^3 J_j^+.
\label{eq:z-eta-commutator}
\end{equation}

Since Eq.~\eqref{eq:eta-commutator-sm} actually corresponds to one of the angular-momentum commutation relations, it is clear that the operators $\eta^\dag$, $\eta$, and $\tilde{J_z}$ can form an angular momentum operator if and only if they satisfy
\begin{equation}
[\tilde{J_z},\eta^\dag]=\varepsilon\eta^\dag.
\label{eq:z-eta-commutator01}
\end{equation}
Note that the commutation relation $[\tilde{J_z},\eta]=-\varepsilon\eta$ is equivalent to Eq.~\eqref{eq:z-eta-commutator01}; and the constant $\varepsilon$ plays the role of the Planck's constant $\hbar$ that appears in the conventional angular-momentum commutation relations, and we can make $\varepsilon\rightarrow 1$ by redefining the operators as $\eta^\dag\rightarrow\eta^\dag/\varepsilon$, $\eta\rightarrow\eta/\varepsilon$, and $\tilde{J_z}\rightarrow\tilde{J_z}/\varepsilon$.

Now, using Eqs.~\eqref{eq:z-eta-commutator} and \eqref{eq:eta-raising01-sm}, the commutation relation \eqref{eq:z-eta-commutator01} can be expressed as
\begin{equation}
\frac{1}{\varepsilon}\sum_{j\in\Lambda_\omega}|\omega_j|^3 J_j^+=\varepsilon\sum_{j\in\Lambda_\omega}|\omega_j|J_j^+.
\label{eq:z-eta-eta}
\end{equation}
Because the operators $J_j^+$ on different sites are linearly independent, which follows from the angular-momentum commutation relations \eqref{eq:local-pseudospin} or the fact that the operators $c_{j\uparrow}^\dag c_{j\downarrow}^\dag$ on different sites are linearly independent due to Lemma 16, Eq.~\eqref{eq:z-eta-eta} is thus equivalent to $\frac{1}{\varepsilon}|\omega_j|^3 =\varepsilon|\omega_j|$ ($\forall j\in\Lambda_\omega$), i.e., $|\omega_j|=|\varepsilon|>0$ ($\forall j\in\Lambda_\omega$), which does not depend on $j\in\Lambda_\omega(\subseteq\Lambda)$.
\hspace{\fill}$\blacksquare$

\textbf{Remark:}
When $|\omega_j|=|\varepsilon|>0$ ($\forall j\in\Lambda_\omega$), we have $\eta^\dag=|\varepsilon|\sum_{j\in\Lambda_\omega}J_j^+$, $\eta=|\varepsilon|\sum_{j\in\Lambda_\omega}J_j^-$, and $\tilde{J_z}=\varepsilon\sum_{j\in\Lambda_\omega} J_j^z$.

\subsection{Proof of Lemma 10}\label{subsec:proofLemma10}

Unlike the other theorems and conclusions, Lemma 10 (see also Lemma 16 in Appendix \ref{subsec:proofTheorem1-2} and the proposition in Appendix \ref{subsec:eta-angular}) is simply a statement for the eta-pairing operators and states, which does not rely on any specific Hamiltonian. This lemma plays an essential role in our study of non-Hermitian systems. Now, let us prove Lemma 10, which contains conclusions (a) and (b).

There are actually two different proofs of conclusion (a). We first present one proof, and the other proof is based on conclusion (b). [The proof of conclusion (b) does not rely on conclusion (a), see below.] For our proofs, it is convenient to rewrite the eta-pairing operators [i.e., Eqs.~\eqref{eq:eta-raising} and \eqref{eq:eta-lowering}] as
\begin{equation}
\eta^\dag=\sum_{j\in\Lambda}\omega_j\eta_j^\dag,  \qquad  \eta'=\sum_{j\in\Lambda}(\omega_j')^*\eta_j
\label{eq:eta-operators-sm}
\end{equation}
by introducing the operators $\eta_j^\dag=c_{j\uparrow}^\dag c_{j\downarrow}^\dag$ and $\eta_j=(\eta_j^\dag)^\dag$. Note that
the operator $\eta_j^\dag$  is the special case of the operator $J_j^+=e^{{\rm i}\theta_j}\eta_j^\dag$ [defined below Eq.~\eqref{eq:eta-raising-sm}] with $\theta_j=0$; thus, the operators $\eta_j^\dag$, $\eta_j$, and $J_j^z$ satisfy the usual angular-momentum commutation relations [see Eq.~\eqref{eq:local-pseudospin}].

To prove conclusion (a), consider the eta-pairing state generated by the eta-lowering operator:
\begin{equation}
(\eta')^n|2\rangle=n!\sum_{l_1,\ldots,l_n\in\Lambda} (\omega_{l_1}')^*\cdots (\omega_{l_n}')^*\eta_{l_1}\cdots\eta_{l_n}\prod_{j\in\Lambda}\eta_j^\dag|0\rangle,
\label{eq:eta-pairing-state-sm}
\end{equation}
where $0\leq n\leq N_\Lambda$, $|2\rangle=\prod_{j\in\Lambda}\eta_j^\dag|0\rangle$, and $l_1,\ldots,l_n$ represent $n$ arbitrary and different sites of $\Lambda$. By the assumption that $\omega_j (\omega'_j)^*\equiv C\neq0$, we have $(\omega_{l_1}')^*\cdots (\omega_{l_n}')^*=C^n/(\omega_{l_1}\cdots \omega_{l_n})$. Note that $\prod_{j\in\Lambda}\omega_j=(\omega_{l_1}\cdots \omega_{l_n})\cdot(\omega_{k_1}\cdots \omega_{k_m})$, where $k_1,\ldots,k_m$ denote the other $m=N_\Lambda-n$ different sites of $\Lambda$ besides $l_1,\ldots,l_n$ (i.e., $\Lambda=\{l_1,\ldots,l_n\}\cup\{k_1,\ldots,k_m\}$ and $\{l_1,\ldots,l_n\}\cap\{k_1,\ldots,k_m\}=\emptyset$). Thus, the product $(\omega_{l_1}')^*\cdots (\omega_{l_n}')^*$ can be rewritten as
\begin{equation}
(\omega_{l_1}')^*\cdots (\omega_{l_n}')^*=\frac{C^n}{\prod_{j\in\Lambda}\omega_j}\cdot\omega_{k_1}\cdots \omega_{k_m}.
\label{eq:eta-pairing-state1-sm}
\end{equation}
Note that the factor $\frac{C^n}{\prod_{j\in\Lambda}\omega_j}$ is a constant that does not depend on the sites $l_1,\ldots,l_n$ or $k_1,\ldots,k_m$. Furthermore, using the commutation relations $[\eta_i^\dag,\eta_j]=\delta_{ij}(n_i-1)$ and $[\eta_i^\dag,\eta_j^\dag]=[\eta_i,\eta_j]=0$, we have
\begin{equation}
\eta_{l_1}\cdots\eta_{l_n}\prod_{j\in\Lambda}\eta_j^\dag|0\rangle=\eta_{k_1}^\dag\cdots\eta_{k_m}^\dag|0\rangle.
\label{eq:eta-pairing-state2-sm}
\end{equation}
Therefore, substituting Eqs.~\eqref{eq:eta-pairing-state1-sm} and \eqref{eq:eta-pairing-state2-sm} into Eq.~\eqref{eq:eta-pairing-state-sm}, we obtain
\begin{equation}
(\eta')^n|2\rangle=\frac{n!C^n}{\prod_{j\in\Lambda}\omega_j}\sum_{k_1,\ldots,k_m\in\Lambda} \omega_{k_1}\cdots \omega_{k_m}\eta_{k_1}^\dag\cdots\eta_{k_m}^\dag|0\rangle.
\label{eq:eta-pairing-state3-sm}
\end{equation}
Note that the summation indices $l_1,\ldots,l_n$ have been replaced by $k_1,\ldots,k_m$.

On the other hand, consider the eta-pairing state generated by the eta-raising operator
\begin{equation}
(\eta^\dag)^m|0\rangle=m!\sum_{k_1,\ldots,k_m\in\Lambda} \omega_{k_1}\cdots \omega_{k_m}\eta_{k_1}^\dag\cdots\eta_{k_m}^\dag|0\rangle,
\label{eq:eta-pairing-state4-sm}
\end{equation}
where $k_1,\ldots,k_m$ represent $m$ arbitrary and different sites of $\Lambda$.
Now, comparing Eq.~\eqref{eq:eta-pairing-state4-sm} with Eq.~\eqref{eq:eta-pairing-state3-sm}, we have
\begin{equation}
(\eta')^n|2\rangle=\frac{n!C^n}{m!\prod_{j\in\Lambda}\omega_j}(\eta^\dag)^m|0\rangle,
\label{eq:etastate-nm-sm}
\end{equation}
where $m+n=N_\Lambda$; thus, $(\eta^\dag)^m |0\rangle$ and $(\eta')^n |2\rangle$ represent the same eta-pairing state. So, we have proved conclusion (a) in Lemma 10.

Next, let us prove conclusion (b), i.e., Eq.~\eqref{eq:nHcomm}. From Eq.~\eqref{eq:eta-operators-sm} and using the angular-momentum commutation relations between $\eta_j^\dag$, $\eta_j$, and $J_j^z$ , one can show that $[\eta^\dag,\eta']=\sum_{j\in\Lambda}\omega_j(\omega'_j)^*2J_j^z$. Since the product $\omega_j (\omega'_j)^*\equiv C\neq0$ does not depend on $j$ (by the assumption of Lemma 10), we get $[\eta^\dag,\eta']=2CJ_z$ with $J_z=\sum_{j\in\Lambda}J_j^z$ defined in Lemma 10. [Note the difference between the definitions of the Hermitian operators $\tilde{J_z}$ in Eq.~\eqref{eq:new-Jz-sm} and $J_z$.] Similarly, one can also show that $[J_z, \eta^\dag]=\eta^\dag$ and $[J_z, \eta']=-\eta'$. Note, however, that the commutation relation $[J_z, \eta^\dag]=\eta^\dag$ or $[J_z, \eta']=-\eta'$ does not rely on the constraint $\omega_j (\omega'_j)^*\equiv C\neq0$. Actually, $[J_z, \eta^\dag]=\eta^\dag$ ($[J_z, \eta']=-\eta'$) holds for an arbitrary eta-pairing operator $\eta^\dag$ ($\eta'$), i.e., an arbitrary $\omega_j$ ($\omega'_j$) as a function of $j$.

Finally, we present the second proof of conclusion (a), which is based on conclusion (b). Using Eq.~\eqref{eq:nHcomm}, one can prove the following commutation relation by induction:
\begin{equation}
[(\eta^\dag)^k,\eta']=2Ck(\eta^\dag)^{k-1}J_z+Ck(k-1)(\eta^\dag)^{k-1}, \qquad \text{where $k=1,2, \ldots$ }.
\label{eq:etaoperators-comm-sm}
\end{equation}
From Eq.~\eqref{eq:etaoperators-comm-sm}, we have
\begin{equation}
\eta'(\eta^\dag)^k|0\rangle=Ck(N_\Lambda+1-k)(\eta^\dag)^{k-1}|0\rangle.
\label{eq:etastate-k-sm}
\end{equation}
Now, by repeated use of Eq.~\eqref{eq:etastate-k-sm}, we arrive at
\begin{equation}
(\eta')^n(\eta^\dag)^{N_\Lambda}|0\rangle=\frac{n!N_\Lambda!C^n}{(N_\Lambda-n)!}(\eta^\dag)^{N_\Lambda-n}|0\rangle,
\label{eq:etastate-nN-sm}
\end{equation}
where $0\leq n\leq N_\Lambda$. Note that $(\eta^\dag)^{N_\Lambda}|0\rangle=N_\Lambda!(\prod_{i\in\Lambda}\omega_i)\prod_{j\in\Lambda}\eta_j^\dag|0\rangle=N_\Lambda!(\prod_{i\in\Lambda}\omega_i)|2\rangle$. Substituting this into Eq.~\eqref{eq:etastate-nN-sm}, we obtain
\begin{equation}
(\eta')^n|2\rangle=\frac{n!C^n}{(N_\Lambda-n)!\prod_{i\in\Lambda}\omega_i}(\eta^\dag)^{N_\Lambda-n}|0\rangle,
\label{eq:etastate-n2-sm}
\end{equation}
which means that the two eta-pairing states $(\eta')^n|2\rangle$ and $(\eta^\dag)^{N_\Lambda-n}|0\rangle$ are equal, up to a trivial factor $\frac{n!C^n}{(N_\Lambda-n)!\prod_{i\in\Lambda}\omega_i}$. This completes the second proof of conclusion (a). Note that the above two different proofs of conclusion (a) yield the same equation, namely Eq.~\eqref{eq:etastate-nm-sm} or Eq.~\eqref{eq:etastate-n2-sm}.

\textbf{Remark 1:}
As mentioned below Lemma 10 in the main text, $(\eta^\dag)^m |0\rangle$ and $\eta^{N_\Lambda-m} |2\rangle$ ($0\leq m\leq N_\Lambda$) represent the same eta-pairing state when $|\omega_j|$ does not depend on $j$ (for example, consider the Hermitian systems). On the other hand, if $|\omega_j|$ is site dependent, then the two states $(\eta^\dag)^m |0\rangle$ and $\eta^{N_\Lambda-m} |2\rangle$ (or more generally, $\eta^n |2\rangle$ with $n$ independent of $m$) are different from each other. 

To see this, let us first consider the case of $\omega_j\neq0$ ($\forall j\in\Lambda$). By defining $\omega'_j=1/\omega_j^*$ ($ j\in\Lambda$), we introduce the eta-pairing operator $(\eta')^\dag=\sum_{j\in\Lambda}\omega'_j\eta_j^\dag$. So, by definition, the on-site product $\omega'_j \omega_j^*$ ($ j\in\Lambda$) is a nonzero constant equal to, e.g., 1 (note that this constant can be an arbitrary nonzero complex number). Thus, according to conclusion (a) in Lemma 10, we have $[(\eta')^\dag]^m |0\rangle=\eta^{N_\Lambda-m} |2\rangle$ (up to a trivial factor). Since $|\omega'_j|=1/|\omega_j|$ and $|\omega_j|$ is site dependent, $|\omega'_j|$ and $|\omega_j|$ (as a function of $j$) must have distinct spatial distributions. Thus, $(\eta^\dag)^m |0\rangle$ and $[(\eta')^\dag]^m |0\rangle=\eta^{N_\Lambda-m} |2\rangle$ are different from each other. Moreover, $(\eta^\dag)^m |0\rangle$ and $\eta^n |2\rangle$ (with $n\neq N_\Lambda-m$) are different states because they have different number of particles.

Second, if $\omega_j=0$ on some sites, then these sites are empty (full) of particles for the state $(\eta^\dag)^m |0\rangle$ ($\eta^n |2\rangle$). Thus, $(\eta^\dag)^m |0\rangle$ and $\eta^n |2\rangle$ are always different from each other. Note that in this case $|\omega_j|$ is allowed to be site independent for $ j\in\Lambda_\omega$, and hence the eta-pairing operators $\eta^\dag$ and $\eta$ can form an angular momentum operator.

To summarize, when $|\omega_j|$ depends on $j\in\Lambda$, the two eta-pairing states $(\eta^\dag)^m |0\rangle$ and $\eta^n |2\rangle$ are different from each other.

\textbf{Remark 2:}
Eq.~\eqref{eq:nHcomm} in the main text  may be viewed as a non-Hermitian generalization of the conventional angular-momentum commutation relations, which we call \emph{non-Hermitian angular-momentum commutation relations}. Here, non-Hermitian generalization means that unlike the usual angular-momentum ladder operators, the eta-raising operator $\eta^\dag$ and the eta-lowering operator $\eta'$ in Eq.~\eqref{eq:nHcomm} are not Hermitian conjugate of each other (i.e., $\eta'\neq\eta$, up to a trivial factor) in general; in other words, the on-site pairing amplitude $|\omega_j|$ (or equivalently, $|\omega'_j|$) is site dependent in general, as discussed below Eq.~\eqref{eq:nHcomm} in the main text.

Therefore, by using the eta-pairing operators $\eta^\dag$, $\eta$, $(\eta')^\dag$, and $\eta'$ in Lemma 10, in general one cannot construct the Hermitian angular-momentum operators that satisfy the usual angular-momentum commutation relations (see also the proposition in Appendix \ref{subsec:eta-angular}). For this reason, the operators $\eta^\dag$ and $\eta'$ obeying Eq.~\eqref{eq:nHcomm} are called \emph{non-Hermitian angular-momentum operators}.


\subsection{Proof of Theorem 11}\label{subsec:proofTheorem11}

This theorem can be viewed as a generalization of Theorem 4 to more general non-Hermitian Hamiltonians. Namely, in Theorem 11, the hopping amplitudes are not required to be symmetric for any bond, while the only assumption is that all the bonds are strong bonds. Now, let us prove Theorem 11 for the case of eta-raising eigenoperator $\eta^\dag$, and the proof of eta-lowering eigenoperator $\eta'$ would be similar.

Since all the bonds are strong bonds (by the assumption of Theorem 11), the eta-raising eigenoperator $\eta^\dag=\sum_{j\in\Lambda} \omega_j c_{j\uparrow}^\dag c_{j\downarrow}^\dag$ must be unique according to conclusion (a) in Theorem 3. Furthermore, since $\omega_j\neq0$  ($\forall j\in\Lambda$) due to conclusion (b) in Theorem 3, we can define
\begin{equation}
\omega'_j=\frac{1}{\omega_j^*}, \qquad  j\in\Lambda .
\label{eq:theorem11}
\end{equation}
Now according to Theorem 1, $\omega_j$ satisfies Eq.~\eqref{eq:constraints1a}; substituting Eq.~\eqref{eq:theorem11} into Eq.~\eqref{eq:constraints1a}, one can thus show that $\omega'_j$ satisfies Eq.~\eqref{eq:constraints2a}. Moreover, Eq.~\eqref{eq:constraints1b} requires that $U_j-2\mu_j$ does not depend on $j\in\Lambda$ and $\lambda=U_j-2\mu_j$, because $\omega_j\neq0$  ($\forall j\in\Lambda$) due to conclusion (b) in Theorem 3. Thus,  $\omega'_j$ also satisfies Eq.~\eqref{eq:constraints2b} with $\lambda'=\lambda=U_j-2\mu_j$. To summarize, $\omega'_j$ satisfies Eqs.~\eqref{eq:constraints2a} and \eqref{eq:constraints2b}. Thus, according to Theorem 2, the eta-lowering operator $\eta'$ defined as $\eta'=\sum_{j\in\Lambda} (\omega'_j)^*  c_{j\downarrow}c_{j\uparrow}$ must be an eigenoperator and is the unique eta-lowering eigenoperator due to conclusion (a) in Theorem 3. [Note that from Eq.~\eqref{eq:theorem11}, it is clear that $\omega'_j\neq0$  ($\forall j\in\Lambda$), which is consistent with conclusion (b) in Theorem 3.] So, we have proved conclusions (a) and (c) in Theorem 11.

Finally, from Eq.~\eqref{eq:theorem11}, the on-site product $\omega_j (\omega'_j)^*$ ($ j\in\Lambda$) is clearly a nonzero constant equal to, e.g., 1 (note that this constant can be an arbitrary nonzero complex number); thus, according to Lemma 10, $\eta^\dag$ and $\eta'$ obey Eq.~\eqref{eq:nHcomm}. So, conclusion (b) in Theorem 11 is true.

\textbf{Remark 1:}
From the above proof of Theorem 11, one can show that conclusions (a) and (b) in Theorem 4 are equivalent to each other. Specifically, if conclusion (a) in Theorem 4 is true (i.e., $\eta'=\eta$ or $\forall j\in\Lambda, \omega'_j=\omega_j$, up to a constant factor), then Eq.~\eqref{eq:theorem11} implies that $|\omega_j|^2=\omega_j\omega_j^*=\omega'_j\omega_j^*=1$ ($\forall j\in\Lambda$), i.e., conclusion (b) in Theorem 4 is true. Conversely,
conclusion (b) in Theorem 4 implies that we can set $|\omega_j|=1$ ($\forall j\in\Lambda$) without loss of generality. So, we have $\omega_j\omega_j^*=|\omega_j|^2=1$ and hence $\omega_j=1/\omega_j^*$. Now, from Eq.~\eqref{eq:theorem11}, we have $\omega'_j=\omega_j$ ($j\in\Lambda$), i.e., $\eta'=\eta=(\eta^\dag)^\dag$. Namely, conclusion (a) in Theorem 4 is true. Therefore,  Theorem 4 can be viewed as a special case of Theorem 11.

\textbf{Remark 2:}
As mentioned in Appendix \ref{subsec:proofTheorem4}, conclusion (a) in Theorem 4 or 11 is nontrivial, since it states that the existence of eta-raising (-lowering) eigenoperator implies the existence of eta-lowering (-raising) eigenoperator. For example, as can be seen from Eq.~\eqref{eq:theorem11}, the above implication is based on the fact that $\omega_j\neq0$  ($\forall j\in\Lambda$), which follows from the assumption that all the bonds are strong bonds. By contrast, when not all the bonds are strong bonds, $\omega_j$ must be zero on some lattice site(s) (see Theorem 12), and hence we cannot define, e.g., the eta-lowering eigenoperator from the eta-raising eigenoperator via Eq.~\eqref{eq:theorem11}. In fact, in the presence of bonds that are not strong bonds, there are non-Hermitian Hamiltonians in which one of the eta-raising and -lowering eigenoperators exists but the other does not exist (see the related discussion in the main text), such as some concrete two-sublattice models we studied later.

\subsection{Proof of Theorem 13}\label{subsec:proofTheorem13}

This theorem can be viewed as a generalization of Theorem 5 to more general cases. Namely, in Theorem 13, the eta-raising and -lowering eigenoperators are not necessarily Hermitian conjugate of each other. Now, let us prove Theorem 13.

By the assumption of the theorem, for any bond $\{i,j\}$, we have $t_{ij}\omega_j=-t_{ji}\omega_i$ [i.e., Eq.~\eqref{eq:constraints1a}] and $ t_{ij}(\omega'_i)^*=-t_{ji}(\omega'_j)^*$ [i.e., Eq.~\eqref{eq:constraints2a}], according to Theorems 1 and 2, respectively.
Thus, if both $\omega_i$ and $\omega'_i$ are nonzero (i.e., $\omega_i\omega'_i\neq0$), then $\{i,j\}$ must be a strong bond (i.e., $t_{ij}t_{ji}\neq0$) and hence $\omega_j\omega'_j\neq0$. This is because Eq.~\eqref{eq:constraints1a} [Eq.~\eqref{eq:constraints2a}] does not hold if $t_{ji}\neq0$ but $t_{ij}=0$ ($t_{ij}\neq0$ but $t_{ji}=0$). (Note that $t_{ij}$ and $t_{ji}$ cannot be both zero, since $\{i,j\}$ is a bond.) Similarly, $\omega_j\omega'_j\neq0$ implies $\omega_i\omega'_i\neq0$.

Therefore, due to the lattice connectivity, if there is a site $k$ such that $\omega_k\omega'_k\neq0$, then $\omega_j\omega'_j\neq0$ ($\forall j\in\Lambda$), or equivalently, all the bonds must be strong bonds (see the discussion below Theorem 12 in the main text).

\textbf{Remark:}
As mentioned above, Theorem 5 is actually a special case of Theorem 13. Specifically, when the eta-raising and -lowering eigenoperators are Hermitian conjugate of each other, namely $\eta'=\eta$ (i.e., $\forall j\in\Lambda, \omega'_j=\omega_j$), the site-independent product $\omega_j (\omega'_j)^*\neq0$ (see the discussion below Theorem 13) means that the on-site pairing amplitude $|\omega_j|$ is site independent. Now, as can be seen from Theorem 7, the hopping amplitudes must be symmetric for every bond (and hence all the bonds are strong bonds).

\subsection{Proof of Corollary 14}\label{subsec:proofCorollary14}

This corollary is a very general result, since it tells us that for any system [described by Eq.~\eqref{eq:nH-Hubbard}], the eta-raising and -lowering eigenoperators (provided that they both exist) are related to each other via the site-independent product $\omega_j (\omega'_j)^*$. Now, we provide a proof of Corollary 14.

As can be seen from conclusion (b) of Theorem 11, the product $\omega_j (\omega'_j)^*$ is a nonzero constant when all the bonds are strong bonds. By contrast, when not all the bonds are strong bonds, the product $\omega_j (\omega'_j)^*$ must be zero for each site $j\in\Lambda$ (i.e., a zero constant). This is because if there exists a lattice site, say $k$, such that $\omega_k (\omega'_k)^*\neq0$ (or equivalently, $\omega_k\omega'_k\neq0$), then, according to Theorem 13, all the bonds must be strong bonds, which contradicts with the assumption that not all the bonds are strong bonds.
This completes the proof.

Note that when not all the bonds are strong bonds, the eta-raising (or -lowering) eigenoperator may or may not be unique (see, e.g., the concrete two-sublattice models in Appendix \ref{app:two-sublattice-model}).

\textbf{Remark 1:}
For the Hermitian case, as discussed above Theorem 4 in the main text, the on-site pairing amplitude $|\omega_j|$ is site independent, i.e., the product $\omega_j \omega_j^*=|\omega_j|^2$ is a nonzero constant. Thus, the nonzero constant $\omega_j (\omega'_j)^*$ due to Corollary 14 now leads to the constraint $\omega_j=\omega'_j$ (up to a site-independent factor), i.e., $\eta^\dag=(\eta')^\dag$ (up to a trivial factor). In other words, the eta-raising and -lowering eigenoperators are related to each other by Hermitian conjugate, which is simply one of the basic properties of the Hermitian eta-pairing theory.

\textbf{Remark 2:}
In the presence of asymmetric hoppings, both $|\omega_j|$ and $|\omega'_j|$ are site dependent, as can be seen from Theorem 7 or Corollary 8. Therefore, when all the bonds are strong bonds, the nonzero constant $\omega_j (\omega'_j)^*$ implies that $|\omega_j|$ and $|\omega'_j|$ are inversely proportional to each other and hence they must have distinct spatial distributions. Thus, the right and left eta-pairing eigenstates, say $(\eta^\dag)^m |0\rangle$ and $[(\eta')^\dag]^m |0\rangle$, have distinct spatial distributions.

On the other hand, when not all the bonds are strong bonds, the zero constant $\omega_j (\omega'_j)^*$ implies that: If $\omega_k\neq0$ ($\omega'_k\neq0$) for some site $k$, then we must have $\omega'_k=0$ ($\omega_k=0$) for the same site $k$. In other words, we have $\Lambda_\omega\cap\Lambda_{\omega'}=\emptyset$ (see also Appendix \ref{subsec:proofTheorem15}). Thus, the right and left eta-pairing eigenstates, e.g., $(\eta^\dag)^m |0\rangle$ and $[(\eta')^\dag]^m |0\rangle$, must have distinct spatial distributions: because the two eta-pairing operators $\eta^\dag$ and $(\eta')^\dag$ create particles in two disjoint regions, namely $\Lambda_\omega$ and $\Lambda_{\omega'}$, respectively.

\subsection{Proof of Theorem 15}\label{subsec:proofTheorem15}

As we saw in the main text, an essential property of the Hermitian eta-pairing theory is that the two constants in the eta-raising and eta-lowering eigenoperator equations are always equal. By contrast, this is not always true for non-Hermitian systems, as revealed by Theorem 15. Now, we give a proof of this theorem.

From Eqs.~\eqref{eq:constraints1b} and \eqref{eq:constraints2b} or the discussion below Theorem 3 in the main text, we have $\lambda=U_i-2\mu_i$ ($\forall i\in\Lambda_\omega$) and $\lambda'=U_j-2\mu_j$ ($\forall j\in\Lambda_{\omega'}$). Recall that $\Lambda_\omega$ ($\Lambda_{\omega'}$) denotes the set of all the nonzero points of $\omega_j$ ($\omega'_j$), as defined in the main text. Thus, if $\Lambda_\omega\cap\Lambda_{\omega'}\neq\emptyset$, then we must have $\lambda=\lambda'$; if $\Lambda_\omega\cap\Lambda_{\omega'}=\emptyset$, then $\lambda$ and $\lambda'$ are two independent constants (i.e., they are not necessarily equal). To summarize, $\lambda$ and $\lambda'$ can be independent of each other if and only if $\Lambda_\omega\cap\Lambda_{\omega'}=\emptyset$. Now, we see that proving Theorem 15 is equivalent to proving that: $\Lambda_\omega\cap\Lambda_{\omega'}=\emptyset$ if and only if not all the bonds are strong bonds.

In other words, Theorem 15 is now equivalent to the following statement: $\Lambda_\omega\cap\Lambda_{\omega'}\neq\emptyset$ if and only if all the bonds are strong bonds. Now, let us prove this statement. If $\Lambda_\omega\cap\Lambda_{\omega'}\neq\emptyset$, then, due to Theorem 13, all the bonds must be strong bonds. Conversely, if all the bonds are strong bonds, then, due to conclusion (b) in Theorem 11, we must have $\Lambda_\omega=\Lambda_{\omega'}=\Lambda$ and hence $\Lambda_\omega\cap\Lambda_{\omega'}=\Lambda\neq\emptyset$.
This completes the proof.

\textbf{Remark:}
As can be seen from the above proof, when $\Lambda_\omega\cap\Lambda_{\omega'}\neq\emptyset$, we must have $\Lambda_\omega=\Lambda_{\omega'}=\Lambda$. On the other hand, when $\Lambda_\omega\cap\Lambda_{\omega'}=\emptyset$, we have $\Lambda_\omega\cup\Lambda_{\omega'}\subseteq\Lambda$, which means that $\Lambda_\omega\cup\Lambda_{\omega'}$ is either a subset (i.e., sublattice) of $\Lambda$ (see, e.g., the one-dimensional Hatano-Nelson-Hubbard model with $\gamma=1$ in the main text) or equal to $\Lambda$ (see, e.g., the general two-sublattice model with $\omega_A\omega_B=0$ in the main text).

\subsection{Proof of Eq. \eqref{eq:relations} in the main text}\label{subsec:proof-edabc}

Our general non-Hermitian eta-pairing theory revealed many exotic non-Hermitian eta-pairing phenomena [i.e., the five properties labeled by $(a)$, $(b)$, $(c)$, $(d)$, and $(e)$, as summarized above Eq. \eqref{eq:relations} in the main text] that are impossible in Hermitian systems. In fact, these novel non-Hermitian phenomena are closely related to each other by the theorems of our general non-Hermitian eta-pairing theory, as summarized in Eq. \eqref{eq:relations}:
\begin{equation}
(e)\Rightarrow (d) \Rightarrow(a)\Leftrightarrow(b) \Leftrightarrow (c).
\label{eq:relations-sm}
\end{equation}
Now, we give a proof of this equation.

The three properties $(a)$, $(b)$, and $(c)$ are equivalent to each other, as can be seen from Corollaries 6, 8, and 9. Due to Theorem 15, the property $(d)$ implies the property $(a)$ [or equivalently, $(b)$ or $(c)$], i.e., $(d) \Rightarrow(a)$, but the converse is not true (since a bond with asymmetric hoppings does not necessarily imply that it is not a strong bond). Furthermore, according to conclusion (a) in Theorem 11, the property $(e)$ would imply that not all the bonds are strong bonds and hence, due to Theorem 15, we have $(e)\Rightarrow (d)$ [note that the converse is not true: the property $(d)$ is equivalent to that not all the bonds are strong bonds (again, due to Theorem 15). As stated below Theorem 3 in the main text, the assumption in Theorem 3 is a sufficient, but not necessary, condition for the uniqueness of eta-raising or -lowering eigenoperator; in other words, when the assumption in Theorem 3 is violated (i.e., not all the bonds are strong bonds; see the Remark in Appendix \ref{subsec:proofTheorem3}), it is possible that both the eta-raising and -lowering eigenoperators are unique. Thus, the property $(d)$ does not necessarily imply the property $(e)$].

\textbf{Remark:}
Eq.~\eqref{eq:relations} can  be illustrated with the concrete examples studied in this paper, such as the Hatano-Nelson-Hubbard models and the general two-sublattice model (together with its variants). Specifically, some of these models host the three properties $(a)$, $(b)$, and $(c)$ but without the properties $(d)$ and $(e)$; some can realize all the five properties $(a)$, $(b)$, $(c)$, $(d)$, and $(e)$; some have the properties $(a)$, $(b)$, $(c)$, and $(d)$ but without the property $(e)$; and some possess the properties $(a)$, $(b)$, $(c)$, and $(e)$ but without the property $(d)$.

\section{Symmetries of the Hubbard model}\label{app:symmetries}

In this appendix, we study symmetries of the general non-Hermitian Hubbard model Eq.~\eqref{eq:nH-Hubbard} on an arbitrary lattice. Specifically, we will focus on the $SO(4)=\frac{SU(2)\times SU(2)}{\mathbb{Z}_2}$ symmetry (i.e., the combination of the $SU(2)$ spin and $SU(2)$  pseudospin symmetries) and its subgroups.  In particular, we find many different kinds of particle-hole (PH) symmetries, i.e., a general $\mathbb{Z}_{2k}$ PH symmetry for any positive integer $k$, including the usual $\mathbb{Z}_2$ PH symmetry of the Hubbard model. Note that all of these discrete PH symmetries belong to the $SO(4)$ symmetry.

However, there is another kind of discrete PH transformation, the so-called $\mathbb{Z}_2$ Shiba (or partial PH) transformation, which does not belong to the $SO(4)$ symmetry; i.e., it is not the combination of the $SU(2)$ spin and $SU(2)$  pseudospin rotations. It is even not a symmetry of the Hubbard model. In fact, the $\mathbb{Z}_2$ Shiba symmetry can be respected by the off-site part $T$ of $H$ in Eq.~\eqref{eq:nH-Hubbard}, but is explicitly broken by the on-site part $V$ of $H$. More precisely, $T$ has a global $O(4)=\mathbb{Z}_2\ltimes SO(4)$ symmetry with $\mathbb{Z}_2$ representing the Shiba symmetry, and $V$ has a local $SO(4)$ symmetry. Thus, the total Hamiltonian $H=T+V$ has a global $SO(4)$ symmetry. These symmetries become explicit in the Majorana representation.

Interestingly, our general non-Hermitian eta-pairing theory enables us to find a previously unnoticed unification of the above various symmetry properties, such as the $\mathbb{Z}_2$ Shiba symmetry, the $SU(2)$  pseudospin or $SO(4)$ symmetry, and all the PH symmetries belonging to the $SO(4)$ symmetry. More specifically, for the Hubbard model Eq.~\eqref{eq:nH-Hubbard}, all these symmetry properties are equivalent to each other: if the Hamiltonian $H$ possesses any one of these symmetry properties, then $H$ would possess all the other symmetry properties (see Theorems 17 and 21 and the related discussion in this appendix). This observation is nontrivial because, for example, if $H$ has the usual $\mathbb{Z}_2$ PH symmetry, then $H$ would have the bigger $SO(4)$ symmetry that contains the $\mathbb{Z}_2$ PH symmetry.

\subsection{$\mathbb{Z}_2$ Shiba transformation}\label{subsec:Shiba}

The $\mathbb{Z}_2$ Shiba transformation plays an important role in understanding the symmetry properties of Hubbard model. It is a partial PH transformation (for spin-down electrons):
\begin{equation}
c_{j\uparrow}\rightarrow c_{j\uparrow}, \qquad  c_{j\downarrow}\rightarrow e^{{\rm i}\theta_j}c_{j\downarrow}^\dag, \qquad     j\in\Lambda,
\label{eq:shiba1}
\end{equation}
where the phase parameter $\theta_j\in\mathbb{R}$ is allowed to depend on $j$. Importantly, we find that the above transformation can be realized by the following unitary operator
\begin{equation}
  U_{\downarrow} = \left\{
  \begin{array}{ll}
    \prod_{j\in\Lambda}(c_{j\downarrow}^\dag+c_{j\downarrow})e^{{\rm i}(\pi-\theta_j)c_{j\downarrow}^\dag c_{j\downarrow}} & \text{for even  $N_\Lambda$} \\
   \prod_{j\in\Lambda}(c_{j\downarrow}^\dag+c_{j\downarrow})e^{-{\rm i}\theta_jc_{j\downarrow}^\dag c_{j\downarrow}}e^{{\rm i}\pi c_{j\uparrow}^\dag c_{j\uparrow}} & \text{for odd  $N_\Lambda$}
  \end{array}
  \right.,
  \label{eq:shiba2}
\end{equation}
which we call the \emph{Shiba operator}. So, we have $U_\downarrow c_{j\uparrow}U^{-1}_\downarrow=c_{j\uparrow}$ and $U_\downarrow c_{j\downarrow}U^{-1}_\downarrow= e^{{\rm i}\theta_j}c_{j\downarrow}^\dag$, namely Eq.~\eqref{eq:shiba1}.

As can be seen from Eq.~\eqref{eq:shiba1}, the Shiba transformation is a $\mathbb{Z}_2$ transformation:
\begin{equation}
U^2_\downarrow c_{j\sigma}U^{-2}_\downarrow=c_{j\sigma}.
\label{eq:shiba03}
\end{equation}
This implies that the Shiba operator $U_\downarrow$ should square to the identity, up to a phase factor (e.g., a gauge transformation). Indeed, from Eq.~\eqref{eq:shiba2}, one can verify that
\begin{equation}
U^2_\downarrow= e^{{\rm i}[\frac{\pi N_\Lambda(N_\Lambda-1)}{2}-\sum_j\theta_j]} \quad \text{for both even $N_\Lambda$ and odd $N_\Lambda$}.
\label{eq:shiba3}
\end{equation}

Now, let us examine how the (non-Hermitian) Hamiltonian $H=T+V$ in Eq.~\eqref{eq:nH-Hubbard} transforms under the Shiba transformation. Under the Shiba transformation \eqref{eq:shiba1}, the spin-down hopping terms in $T$ transform as
\begin{equation}
U_\downarrow (t_{ij}c_{i\downarrow}^\dag c_{j\downarrow}) U^{-1}_\downarrow=-e^{{-\rm i}\theta_i}e^{{\rm i}\theta_j}t_{ij}c_{j\downarrow}^\dag c_{i\downarrow}, \quad   \text{$\forall i,j\in\Lambda$ and $i\neq j$};
\label{eq:shiba3-1}
\end{equation}
whereas the spin-up terms are unchanged. Thus, the off-site part $T$ of Eq.~\eqref{eq:nH-Hubbard} has the $\mathbb{Z}_2$ Shiba symmetry, say
\begin{equation}
U_\downarrow T U^{-1}_\downarrow=T
\label{eq:shiba04}
\end{equation}
if and only if $t_{ji}=-e^{{-\rm i}\theta_i}e^{{\rm i}\theta_j}t_{ij}$, or equivalently
\begin{equation}
t_{ij}e^{{\rm i}\theta_j}+t_{ji}e^{{\rm i}\theta_i}=0     \qquad     \text{for each bond $\{i,j\}$}.
\label{eq:shiba4}
\end{equation}
It is worth emphasizing that Eq.~\eqref{eq:shiba4} implies that the hopping amplitudes must be symmetric for every bond. By contrast, the on-site part $V$ of Eq.~\eqref{eq:nH-Hubbard} breaks the $\mathbb{Z}_2$ Shiba symmetry explicitly:
\begin{equation}
U_\downarrow(U_jn_{j\uparrow}n_{j\downarrow}-\mu_jn_j)U^{-1}_\downarrow=-U_jn_{j\uparrow}n_{j\downarrow}+\mu_jn_j+(U_j-2\mu_j)n_{j\uparrow}-\mu_j,     \quad     \forall j\in\Lambda,
\label{eq:shiba5}
\end{equation}
where $n_j=n_{j\uparrow}+n_{j\downarrow}$. From Eq.~\eqref{eq:shiba5}, when $\mu_j=U_j/2$ ($\forall j\in\Lambda$), the $\mathbb{Z}_2$ Shiba transformation changes $V$ to $-V$, up to an additive constant $\sum_j(-\mu_j)=\sum_j(-U_j/2)$. Thus, if we rewrite the on-site part $V$ with $\mu_j=U_j/2$ as
\begin{equation}
V'=\sum_{j\in\Lambda}U_j(n_{j\uparrow}-\frac{1}{2})(n_{j\downarrow}-\frac{1}{2}),
\label{eq:shiba6}
\end{equation}
then we have
\begin{equation}
U_\downarrow V'U^{-1}_\downarrow=-V'.
\label{eq:shiba7}
\end{equation}
Using $V'$ in Eq.~\eqref{eq:shiba6}, we can rewrite the Hamiltonian $H=T+V$ in Eq.~\eqref{eq:nH-Hubbard} with $\mu_j=U_j/2$ ($j\in\Lambda$) as
\begin{equation}
H'=T+V'.
\label{eq:shiba8}
\end{equation}

According to the above discussion and writing $H'(U_j)$ for $H'$, we have
\begin{equation}
U_\downarrow H'(U_j)U^{-1}_\downarrow=H'(-U_j),   \qquad     j\in\Lambda
\label{eq:shiba9}
\end{equation}
if and only if $T$ has the $\mathbb{Z}_2$ Shiba symmetry, i.e., Eq.~\eqref{eq:shiba04}, or equivalently Eq.~\eqref{eq:shiba4}.

On the other hand, from the statement in Sec.~\ref{symmetry-pseudospin}, we see that $T$ has the $SU(2)$ pseudospin symmetry if and only if Eq.~\eqref{eq:pseudospin-symmetry-a} holds (since $T$ can be regarded as the Hamiltonian $H$ with $\mu_j=U_j=0$), and $V'$ in Eq.~\eqref{eq:shiba6} always has the $SU(2)$ pseudospin symmetry (since $V'$ can be regarded as $H$ with $t_{ij}=0$ and $\mu_j=U_j/2$). Importantly, \emph{by identifying $\theta_j$ in the Shiba transformation \eqref{eq:shiba1} with the phase of $\omega_j$ in Eq.~\eqref{eq:pseudospin-symmetry-a}, Eqs.~\eqref{eq:shiba4} and \eqref{eq:pseudospin-symmetry-a} are actually the same equation}.

The above results can be summarized in the following theorem.

\textbf{Theorem 17:}
Note that the (non-Hermitian) Hubbard model Eq.~\eqref{eq:nH-Hubbard} with $\mu_j=U_j/2$ ($j\in\Lambda$) can be represented by $H'=T+V'$ in Eq. \eqref{eq:shiba8}. The following four statements are equivalent to each other: (1) $H'$ has the $\mathbb{Z}_2$ Shiba symmetry property, i.e., Eq.~\eqref{eq:shiba9}; (2) $T$ has the $\mathbb{Z}_2$ Shiba symmetry; (3) $H'$ has the $SU(2)$ pseudospin symmetry; (4) $T$ has the $SU(2)$ pseudospin symmetry.

\textbf{Remark:}
From Theorem 17 and the discussion in this subsection, it is interesting to see that although the $\mathbb{Z}_2$ Shiba symmetry and the $SU(2)$ pseudospin symmetry are clearly two distinct symmetries by definition, they are unified by Eqs.~\eqref{eq:pseudospin-symmetry-a} and \eqref{eq:pseudospin-symmetry-b} in the context of Hubbard model.

\subsection{$f$ fermions, spin, and pseudospin}\label{subsec:f-fermion}

As will become clear later, it is helpful to introduce a new fermion (dubbed the $f$-fermion) via the Shiba transformation \eqref{eq:shiba1}, which provides a physical understanding of the pseudospin (symmetry),
\begin{equation}
f_{j\uparrow}=U_\downarrow c_{j\uparrow}U^{-1}_\downarrow=c_{j\uparrow}, \qquad  f_{j\downarrow}=U_\downarrow c_{j\downarrow}U^{-1}_\downarrow= e^{{\rm i}\theta_j}c_{j\downarrow}^\dag.
\label{eq:f1}
\end{equation}
Since the Shiba transformation is a $\mathbb{Z}_2$ transformation [see Eq.~\eqref{eq:shiba03}, or Eq.~\eqref{eq:shiba3}], the above equation implies
\begin{equation}
c_{j\uparrow}=U_\downarrow f_{j\uparrow}U^{-1}_\downarrow, \qquad  c_{j\downarrow}=U_\downarrow f_{j\downarrow}U^{-1}_\downarrow.
\label{eq:f2}
\end{equation}
Thus, we have the following correspondence between the occupation-number basis (for each given site) of the electrons and that of the $f$ fermions:
\begin{equation}
|\uparrow\rangle_e= |\uparrow\downarrow\rangle_f, \qquad  |\downarrow\rangle_e= |0\rangle_f, \qquad |\uparrow\downarrow\rangle_e=|\uparrow\rangle_f, \qquad |0\rangle_e=|\downarrow\rangle_f,
\label{eq:f3}
\end{equation}
where the notations, for example, $|\uparrow\rangle_e$ and $|0\rangle_f$ represent the state of one spin-up electron and the vacuum state of $f$ fermions, respectively. We note that the two states $|\uparrow\rangle_e$ and $|\uparrow\downarrow\rangle_e$ (or, $|\downarrow\rangle_e$ and $|0\rangle_e$) are mapped to each other by the Shiba operator $U_S$.

Now, let us consider the (total) spin of $f$ fermions (namely the pseudospin)
\begin{equation}
J_\alpha=\sum_{j\in\Lambda}J^\alpha_j, \qquad J^\alpha_j=\frac{1}{2}f_j^\dag\tau_\alpha f_j    ,
\label{eq:f4}
\end{equation}
where $\tau_\alpha$ ($\alpha=x, y, z$) are the Pauli matrices and $f_j^\dag=(f_{j\uparrow}^\dag, f_{j\downarrow}^\dag)$; and the (total) spin of electrons
\begin{equation}
S_\alpha=\sum_{j\in\Lambda}S^\alpha_j, \qquad S^\alpha_j=\frac{1}{2}c_j^\dag\tau_\alpha c_j   ,
\label{eq:f6}
\end{equation}
where $c_j^\dag=(c_{j\uparrow}^\dag, c_{j\downarrow}^\dag)$. From Eqs.~\eqref{eq:f1} and \eqref{eq:f2}, we see that the Shiba operator $U_\downarrow$ interchanges the spin and pseudospin operators, say
\begin{equation}
J_\alpha=U_\downarrow S_\alpha U^{-1}_\downarrow, \qquad S_\alpha=U_\downarrow J_\alpha U^{-1}_\downarrow.
\label{eq:f5}
\end{equation}
More specifically, we have $J^\alpha_j=U_\downarrow S^\alpha_j U^{-1}_\downarrow$ and $S^\alpha_j=U_\downarrow J^\alpha_j U^{-1}_\downarrow$ at each site $j$. Furthermore, the spin operators commute with the pseudospin operators:
\begin{equation}
[S_\alpha,J_\beta]=0 \quad \text{with $\alpha,\beta=x, y, z$},
\label{eq:f7}
\end{equation}
meaning that the pseudospin (spin) operators have the $SU(2)$ spin (pseudospin) symmetry.  This can be seen by using Eq.~\eqref{eq:f1} and rewriting the pseudospin operators Eq.~\eqref{eq:f4} as (in terms of the electron operators)
\begin{equation}
J_+=J_x+{\rm i}J_y=\sum_{j\in\Lambda} e^{{\rm i}\theta_j}c_{j\uparrow}^\dag c_{j\downarrow}^\dag, \qquad J_-=J_+^\dag,  \qquad J_z=\frac{1}{2}\sum_{j\in\Lambda}(c_{j\uparrow}^\dag c_{j\uparrow}+c_{j\downarrow}^\dag c_{j\downarrow}-1),
\label{eq:f8}
\end{equation}
where the operator $J_z$ has already appeared in Lemma 10. Now, since each on-site term in Eq.~\eqref{eq:f8} is (locally) $SU(2)$ spin symmetric, the $SU(2)$ spin symmetry of pseudospin operators becomes explicit in the electron representation. Similarly, the $SU(2)$ pseudospin symmetry of spin operators would become explicit in the $f$-fermion representation. In fact, these $SU(2)$ symmetries hold locally (e.g., at each lattice site); i.e., we have a more stronger result than Eq.~\eqref{eq:f7}:
\begin{equation}
[S^\alpha_i,J^\beta_j]=0, \quad \text{where $i,j\in\Lambda$}.
\label{eq:f9}
\end{equation}
In particular, we have $S^\alpha_j J^\beta_j=J^\beta_j S^\alpha_j=0$ at each site $j$. It is also worth mentioning that the $SU(2)$ pseudospin symmetry of spin operators corresponds to the local $SU(2)$ gauge symmetry in the context of Schwinger-fermion representation of spin-1/2 systems.

Importantly, the pseudospin operator $J_+$ in Eq.~\eqref{eq:f8} is nothing but the eta-pairing operator [see Eq.~\eqref{eq:eta-raising} or Eq.~\eqref{eq:hermitian-eta}]:
\begin{equation}
J_+=\eta^\dag=\sum_{j\in\Lambda} \omega_jc_{j\uparrow}^\dag c_{j\downarrow}^\dag \quad (\text{with $\omega_j=e^{{\rm i}\theta_j}$}),
\label{eq:f10}
\end{equation}
which becomes an eigenoperator of the Hubbard model \eqref{eq:nH-Hubbard} if and only if Eq.~\eqref{eq:shiba4}  [or Eq.~\eqref{eq:pseudospin-symmetry-a}] holds and $U_j-2\mu_j$ does not depend on $j\in\Lambda$.

As a direct consequence of Eq.~\eqref{eq:f5}, the Shiba operator $U_\downarrow$ also interchanges the spin and pseudospin symmetries:

\textbf{Theorem 18:}
Let $O$ and $O'$ be two generic (non-Hermitian) operators, which are related to each other by the Shiba operator $U_\downarrow$: $O'=U_\downarrow O U_\downarrow^{-1}$, or equivalently $O=U_\downarrow O' U_\downarrow^{-1}$ [due to the $\mathbb{Z}_2$ property of $U_\downarrow$, i.e., Eq.~\eqref{eq:shiba3}]. Now, if $O$ has the $SU(2)$ spin (pseudospin) symmetry, i.e., $[O, S_\alpha]=0$ ($[O, J_\alpha]=0$) for all $\alpha=x, y, z$, then $O'$ must have the $SU(2)$ pseudospin (spin) symmetry, i.e., $[O', J_\alpha]=0$ ($[O', S_\alpha]=0$) for all $\alpha=x, y, z$.

\textbf{Remark:}
Consider either the Hamiltonian $H'$ or the off-site part $T$ in Eq.~\eqref{eq:shiba8}. Because of the presence of $SU(2)$ spin symmetry, the existence of $\mathbb{Z}_2$ Shiba symmetry [or symmetry property, e.g., Eq. \eqref{eq:shiba9}] implies the existence of $SU(2)$ pseudospin symmetry according to Theorem 18. In fact, this conclusion is already contained in Theorem 17.

Now, suppose that Eq.~\eqref{eq:shiba4} holds. Then, using Eq.~\eqref{eq:f1},  $T$  can be rewritten as (in terms of the $f$-fermion operators)
\begin{equation}
T=\sum_{i\neq j,\sigma} t_{ij}c_{i\sigma}^\dag c_{j\sigma}=\sum_{i\neq j,\sigma} t_{ij}f_{i\sigma}^\dag f_{j\sigma}.
\label{eq:f11}
\end{equation}
This equation is nothing but Eq.~\eqref{eq:shiba04}, meaning that $T$ has the $\mathbb{Z}_2$ Shiba symmetry; and hence the $SU(2)$ pseudospin symmetry is respected by $T$ (see the above Remark below Theorem 18). In fact, the $SU(2)$ pseudospin symmetry becomes explicit in the $f$-fermion representation of $T$ in Eq.~\eqref{eq:f11}.

For the on-site part $V'$, i.e., Eq.~\eqref{eq:shiba6}, using Eq.~\eqref{eq:f1},  we can rewrite $V'$ as (in terms of the $f$-fermion operators)
\begin{equation}
V'=\sum_{j\in\Lambda}U_j(n_{j\uparrow}-\frac{1}{2})(n_{j\downarrow}-\frac{1}{2})=-\sum_{j\in\Lambda}U_j(n^f_{j\uparrow}-\frac{1}{2})(n^f_{j\downarrow}-\frac{1}{2}),
\label{eq:f12}
\end{equation}
where $n^f_{j\sigma}=f_{j\sigma}^\dag f_{j\sigma}$. This equation is nothing but Eq.~\eqref{eq:shiba7}; thus, according to Theorem 18, the $SU(2)$ pseudospin symmetry is respected by $V'$. In fact, the $SU(2)$ pseudospin symmetry becomes explicit in the $f$-fermion representation of $V'$ in Eq.~\eqref{eq:f12}. [Actually, from Eq.~\eqref{eq:f12}, it is clear that $V'$ has \emph{local} $SU(2)$ spin and $SU(2)$ pseudospin symmetries; i.e., $V'$ is invariant under the $SU(2)$ (pseudo)spin transformations that at different lattice sites are independent of each other.]

Thus, for the total Hamiltonian $H'=T+V'$ [i.e., Eq.~\eqref{eq:shiba8}], we have
\begin{equation}
H'=H'_c(t_{ij},U_j)=H'_f(t_{ij},-U_j),
\label{eq:f13}
\end{equation}
provided that Eq.~\eqref{eq:shiba4} holds. Here, $H'_c$ and $H'_f$ denote the electron and the $f$-fermion representations of $H'$, respectively. Note that Eq.~\eqref{eq:f13} is nothing but Eq.~\eqref{eq:shiba9}.

Finally, it is instructive to represent the interacting part $V$ of Eq.~\eqref{eq:nH-Hubbard} in terms of the spin and pseudospin operators; i.e., using the identity $n_{j\uparrow}n_{j\downarrow}=-\frac{2}{3}\bm{S}_j^2+\frac{1}{2}(n_{j\uparrow}+n_{j\downarrow})$ with $\bm{S}_j^2=\sum_\alpha (S_j^\alpha)^2$, we can rewrite $V$ as
\begin{equation}
V=-\frac{2}{3}\sum_{j\in\Lambda}U_j\bm{S}_j^2+\sum_{j\in\Lambda}(U_j-2\mu_j)J^z_j+\frac{1}{2}\sum_{j\in\Lambda}(U_j-2\mu_j).
\label{eq:f14}
\end{equation}
It is convenient to use this expression of $V$ to calculate its commutator with the eta-pairing operator, e.g., Eq.~\eqref{eq:V-eta}.
Now, it is clear that the first term containing $\bm{S}_j^2$ in Eq. \eqref{eq:f14} respects both the $SU(2)$ pseudospin symmetry [see Eq.~\eqref{eq:f9}] and the $SU(2)$ spin symmetry; and the second term containing $J^z_j$ respects the $SU(2)$ spin symmetry [see Eq.~\eqref{eq:f9}] but explicitly breaks the $SU(2)$ pseudospin symmetry. Thus, when $U_j-2\mu_j=0$, i.e., $\mu_j=U_j/2$ $(\forall j\in\Lambda)$, $V$ (or equivalently $V'$) would have the $SU(2)$ pseudospin symmetry.

\textbf{$SO(4)$ symmetric interactions:}
From the above discussion, we see that the (spin) Hubbard interaction $\bm{S}_j^2$ possesses both the $SU(2)$ spin and $SU(2)$ pseudospin symmetries, or equivalently, the $SO(4)$ symmetry. Similarly, one can write down other interacting terms with the $SO(4)$ symmetry, such as the pseudospin Hubbard interaction $\bm{J}_j^2=\sum_\alpha (J_j^\alpha)^2$, the (spin) Heisenberg interaction $\bm{S}_i\cdot\bm{S}_j=\sum_\alpha S_i^\alpha S_j^\alpha$, and the pseudospin Heisenberg interaction $\bm{J}_i\cdot\bm{J}_j=\sum_\alpha J_i^\alpha J_j^\alpha$.

Interestingly, the on-site spin and pseudospin interactions $\bm{S}_j^2$ and $\bm{J}_j^2$ satisfy the following constraint:
\begin{equation}
\bm{S}_j^2+\bm{J}_j^2=\frac{3}{4}, \quad \forall j\in\Lambda.
\label{eq:f15}
\end{equation}
In other words, we have $\bm{J}_j^2=\frac{3}{4}-\bm{S}_j^2$, where the minus sign is nothing but the one appearing in Eqs.~\eqref{eq:shiba7}, \eqref{eq:shiba9}, and \eqref{eq:f13}. Recall that the spin and pseudospin operators can be transformed into each other by the Shiba operator $U_\downarrow$ [see Eq. \eqref{eq:f5} and the equations below it].

\subsection{The physical $SO(4)$ symmetry}\label{subsec:physicalSO4}

For quantum systems, a symmetry transformation can be described by a unitary (or an antiunitary, e.g., time reversal) operator. For example, the $SU(2)$ spin symmetry of the Hubbard model can be represented by the unitary spin-rotation operator
\begin{equation}
R_S=e^{{\rm i}\phi_1S_z}e^{{\rm i}\phi_2S_y}e^{{\rm i}\phi_3S_z}
\label{eq:so4-1}
\end{equation}
with the three Euler angles $\phi_{1,2,3}$ and the spin operators $S_{y,z}$ given by Eq.~\eqref{eq:f6}.

In the following discussion, it is convenient to use the matrix representation of $R_S$ in the occupation-number basis Eq.~\eqref{eq:f3}, say
\begin{equation}
R_S=\left(\begin{array}{cc} U_S & 0 \\ 0 & \mathbf{I} \\\end{array} \right)^{\otimes N_\Lambda}, \qquad U_S=\left(\begin{array}{cc} e^{{\rm i}\frac{\phi_1+\phi_3}{2}}\cos\frac{\phi_2}{2} & e^{{\rm i}\frac{\phi_1-\phi_3}{2}}\sin\frac{\phi_2}{2} \\ -e^{{\rm i}\frac{\phi_3-\phi_1}{2}}\sin\frac{\phi_2}{2} & e^{-{\rm i}\frac{\phi_1+\phi_3}{2}}\cos\frac{\phi_2}{2} \\\end{array} \right)\in SU(2),
\label{eq:so4-2}
\end{equation}
where $U_S$ is an arbitrary $SU(2)$ matrix and $\mathbf{I}$ is the $2\times2$ identity matrix. Now, the action of $R_S$ on electron operators reads
\begin{equation}
R_Sc_j^\dag R_S^{-1}=c_j^\dag U_S, \quad \text{where $c_j^\dag=(c_{j\uparrow}^\dag, c_{j\downarrow}^\dag)$}.
\label{eq:so4-3}
\end{equation}
This is the meaning of $SU(2)$ spin symmetry. Alternatively, we can understand it by looking at the set of the unitary operators $R_S$
\begin{equation}
G_S=\{R_S=\left(\begin{array}{cc} U_S & 0 \\ 0 & \mathbf{I} \\\end{array} \right)^{\otimes N_\Lambda}\mid  U_S\in SU(2)\}.
\label{eq:so4-4}
\end{equation}
Because of Eq.~\eqref{eq:so4-3}, the unitary operators $R_S$ are in one-to-one correspondence with the $SU(2)$ matrices $U_S$.  More precisely, $G_S$ is isomorphic to the group $SU(2)$
\begin{equation}
G_S\cong SU(2).
\label{eq:so4-4-1}
\end{equation}
Importantly, $G_S$ does not contain gauge transformations (e.g., nontrivial phase factors): If there is a gauge transformation $R_S\in G_S$, which leaves the electron operators unchanged. Then, according to Eq.~\eqref{eq:so4-3}, the corresponding $U_S$ must be the $2\times2$ identity matrix $\mathbf{I}$, and hence $R_S$ must be the identity operator (i.e., the trivial phase factor $1$), which contradicts with the gauge transformation assumption. Thus, the group $G_S$ indeed represents the $SU(2)$ spin symmetry.

Similarly, we can describe the $SU(2)$ pseudospin symmetry by introducing the unitary pseudospin-rotation operator
\begin{equation}
R_J=e^{{\rm i}\varphi_1J_z}e^{{\rm i}\varphi_2J_y}e^{{\rm i}\varphi_3J_z}
\label{eq:so4-5}
\end{equation}
with the three Euler angles $\varphi_{1,2,3}$ and the pseudospin operators $J_{y,z}$ given by Eq.~\eqref{eq:f4}. Now, the action of $R_J$ on the $f$-fermion operators reads
\begin{equation}
R_Jf_j^\dag R_J^{-1}=f_j^\dag U_J, \quad \text{where $f_j^\dag=(f_{j\uparrow}^\dag, f_{j\downarrow}^\dag)$ and $U_J=\left(\begin{array}{cc} e^{{\rm i}\frac{\varphi_1+\varphi_3}{2}}\cos\frac{\varphi_2}{2} & e^{{\rm i}\frac{\varphi_1-\varphi_3}{2}}\sin\frac{\varphi_2}{2} \\ -e^{{\rm i}\frac{\varphi_3-\varphi_1}{2}}\sin\frac{\varphi_2}{2} & e^{-{\rm i}\frac{\varphi_1+\varphi_3}{2}}\cos\frac{\varphi_2}{2} \\\end{array} \right)\in SU(2)$}.
\label{eq:so4-6}
\end{equation}
Alternatively, we can understand the $SU(2)$ pseudospin symmetry by looking at the $SU(2)$ group formed by the unitary operators $R_J$
\begin{equation}
G_J=\{R_J=\left(\begin{array}{cc} \mathbf{I} & 0 \\ 0 & U_J \\\end{array} \right)^{\otimes N_\Lambda}\mid  U_J\in SU(2)\}\cong SU(2),
\label{eq:so4-7}
\end{equation}
which does not contain gauge transformations and hence represents the $SU(2)$ pseudospin symmetry. Note that the above matrix representation of $R_J$ is written down in the same basis Eq.~\eqref{eq:f3} as $R_S$.
It is also worth noting that both the $SU(2)$ matrices $U_S$ and $U_J$ do not depend on $j\in\Lambda$, i.e., the $SU(2)$ spin and $SU(2)$ pseudospin symmetries are global symmetries.

Now, let us study the combination of the above two $SU(2)$ symmetries, which can be described by the product of $R_S$ and $R_J$, namely the unitary operator
\begin{equation}
R=R_SR_J=R_JR_S,
\label{eq:so4-8}
\end{equation}
where $R_S$ and $R_J$ commute with each other because of Eq.~\eqref{eq:f7}. Using Eqs.~\eqref{eq:so4-3}, \eqref{eq:so4-6}, and \eqref{eq:f1}, we can obtain the action of $R$ on electron operators
\begin{eqnarray}
Rc_{j\uparrow}R^{-1}&=&g_Sg_Jc_{j\uparrow}+h_Sg_Jc_{j\downarrow}-h_Sh_Je^{{\rm i}\theta_j}c_{j\uparrow}^\dag+g_Sh_Je^{{\rm i}\theta_j}c_{j\downarrow}^\dag,
\nonumber\\
Rc_{j\downarrow}R^{-1}&=&-h^*_Sg_Jc_{j\uparrow}+g^*_Sg_Jc_{j\downarrow}-g^*_Sh_Je^{{\rm i}\theta_j}c_{j\uparrow}^\dag-h^*_Sh_Je^{{\rm i}\theta_j}c_{j\downarrow}^\dag,
\label{eq:so4-9}
\end{eqnarray}
where the complex numbers $g_{S,J}$ and $h_{S,J}$ satisfying $|g_S|^2+|h_S|^2=|g_J|^2+|h_J|^2=1$ represent the $SU(2)$ matrices $U_S$ and $U_J$, say
\begin{equation}
U_S=\left(\begin{array}{cc} g^*_S & -h_S \\ h_S^* & g_S \\\end{array} \right), \qquad U_J=\left(\begin{array}{cc} g^*_J & -h_J \\ h_J^* & g_J \\\end{array} \right).
\label{eq:so4-10}
\end{equation}
As an aside, clearly, Eq.~\eqref{eq:so4-9} does not involve the $\mathbb{Z}_2$ Shiba transformation \eqref{eq:shiba1}, i.e., the $\mathbb{Z}_2$ Shiba symmetry does not belong to the $SO(4)$ symmetry [see also Eq.~\eqref{eq:mr-22}].

Unlike in the case of $SU(2)$ spin (or pseudospin) symmetry, from the transformations in Eq.~\eqref{eq:so4-9}, it is difficult to identify the group structure of the combined symmetry $R$ of spin $R_S$ and pseudospin $R_J$. [However, as shown later, the $SO(4)$ group structure becomes explicit from the action of $R$ on Majorana operators.] Instead, let us consider the set of the unitary operators $R$
\begin{equation}
G(N_\Lambda)=\{R=R_SR_J=R_JR_S=\left(\begin{array}{cc} U_S & 0 \\ 0 & U_J \\\end{array} \right)^{\otimes N_\Lambda}\mid   U_{S,J} \in SU(2)\}.
\label{eq:so4-11}
\end{equation}
So, it is clear that $G(N_\Lambda)$  [i.e., Eq.~\eqref{eq:so4-11}] can be written as the product of the groups $G_S$ [i.e., Eq.~\eqref{eq:so4-4}] and $G_J$ [i.e., Eq.~\eqref{eq:so4-7}]:
\begin{equation}
G(N_\Lambda)=G_SG_J=G_JG_S=\{R_SR_J=R_JR_S \mid R_S\in G_S,  R_J\in G_J \} .
\label{eq:so4-011}
\end{equation}
Does the above product $G_SG_J$ coincide with the \emph{direct product} of $G_S$ and $G_J$ (i.e., $G_S\times G_J$)? The answer depends on $N_\Lambda$ (namely the total number of lattice sites). To see this, we first recall that the direct product means that every $R$ in $G(N_\Lambda)=G_SG_J$ can be expressed \emph{uniquely} as the product of $R_S\in G_S$ and $R_J\in G_J$ (i.e., if $R=R_SR_J=R'_SR'_J$, then $R'_S=R_S$ and $R'_J=R_J$), or equivalently, the intersection $G_S\cap G_J$ is trivial (i.e., $G_S\cap G_J=\{1\}$, where $1$ is the identity operator).
In fact, there is a general relation between $G_SG_J$ and $G_S\times G_J$. Roughly, we have
\begin{equation}
G_SG_J=\frac{G_S\times G_J}{G_S\cap G_J}.
\label{eq:so4-0011}
\end{equation}
The precise mathematical meaning of Eq.~\eqref{eq:so4-0011} is as follows. Let $A$ and $B$ be two subgroups of some bigger group. If every element of $A$ commutes with every element of $B$, then we have the following group isomorphism
\begin{equation}
AB\cong\frac{A\times B}{C},\qquad C\cong A\cap B,
\label{eq:so4-00011}
\end{equation}
where $AB=BA=\{ab=ba \mid a\in A,  b\in B \}$ is the product of $A$ and $B$, $A\times B$ denotes the \emph{external} direct product of $A$ and $B$, and $C=\{(c,c^{-1})\in A\times B\mid  c\in A\cap B\}$ is a normal subgroup of $A\times B$. [Eq.~\eqref{eq:so4-00011} is a known formula that can be proved in several different ways.] In particular, when the intersection $A\cap B$ is trivial, the product $AB$ becomes the internal direct product of $A$ and $B$, and Eq.~\eqref{eq:so4-00011} reduces to $AB\cong A\times B$, meaning that the internal and external direct products are isomorphic (and hence both of which are collectively called direct product).
Before proceeding further, we give a useful conclusion:

\textbf{Theorem 19:}
Suppose $R,R'\in G(N_\Lambda)$ and let
\begin{equation*}
R=\left(\begin{array}{cc} U_S & 0 \\ 0 & U_J \\\end{array} \right)^{\otimes N_\Lambda}, \qquad R'=\left(\begin{array}{cc} U'_S & 0 \\ 0 & U'_J \\\end{array} \right)^{\otimes N_\Lambda}.
\end{equation*}
Then, the following three statements are equivalent to each other: (1) $Rc_{j\sigma}R^{-1}=R'c_{j\sigma}R'^{-1}$; (2) $U'_S=U_S$ and $U'_J=U_J$, or $U'_S=-U_S$ and $U'_J=-U_J$; (3) $R'=R$ or $R'=(-1)^{N_\Lambda}R$.

This theorem can be obtained by using Eqs.~\eqref{eq:so4-9} and \eqref{eq:so4-10}. Now, let us find the spin-rotation operator $R_S\in G_S$ and the pseudospin-rotation operator $R_J\in G_J$ that have the same action on electron operators: $R_Sc_{j\sigma}R_S^{-1}=R_Jc_{j\sigma}R_J^{-1}$. According to Theorem 19, these nontrivial operators (i.e., not the identity operator) turn out to be
\begin{equation}
e^{{\rm i}2\pi S_z}=\left(\begin{array}{cc} -\mathbf{I} & 0 \\ 0 & \mathbf{I} \\\end{array} \right)^{\otimes N_\Lambda}\in G_S, \qquad e^{{\rm i}2\pi J_z}=\left(\begin{array}{cc} \mathbf{I} & 0 \\ 0 & -\mathbf{I} \\\end{array} \right)^{\otimes N_\Lambda}\in G_J.
\label{eq:so4-12}
\end{equation}
Note that $e^{{\rm i}2\pi S_z}=e^{{\rm i}2\pi S_x}=e^{{\rm i}2\pi S_y}$ and $e^{{\rm i}2\pi J_z}=e^{{\rm i}2\pi J_x}=e^{{\rm i}2\pi J_y}$. Thus, under the unitary operator $e^{{\rm i}2\pi S_z}$ or $e^{{\rm i}2\pi J_z}$, the electron operators transform as $c_{j\sigma}\rightarrow -c_{j\sigma}$. From the matrix representation in Eq.~\eqref{eq:so4-12}, it is clear that
\begin{equation}
e^{{\rm i}2\pi (S_z+J_z)}=e^{{\rm i}2\pi S_z}e^{{\rm i}2\pi J_z}= (-1)^{N_\Lambda} ,
\label{eq:so4-13}
\end{equation}
or equivalently
\begin{equation}
e^{{\rm i}2\pi S_z}= (-1)^{N_\Lambda} e^{{\rm i}2\pi J_z},
\label{eq:so4-14}
\end{equation}
which is consistent with the statement (3) of Theorem 19.

Based on the above results, the intersection $G_S\cap G_J$ can be readily obtained as
\begin{equation}
  G_S\cap G_J = \left\{
  \begin{array}{ll}
    \{1\} \quad (\text{i.e., trivial}) & \text{for odd  $N_\Lambda$} \\
   \{1, e^{{\rm i}2\pi S_z}= e^{{\rm i}2\pi J_z}\}=\mathbb{Z}_2 & \text{for even  $N_\Lambda$}
  \end{array}
  \right..
  \label{eq:so4-15}
\end{equation}
Now, from Eqs.~\eqref{eq:so4-011}, \eqref{eq:so4-0011}, and \eqref{eq:so4-15}, we arrive at
\begin{equation}
  G(N_\Lambda) = \left\{
  \begin{array}{ll}
    SU(2)\times SU(2) & \text{for odd  $N_\Lambda$} \\
   (SU(2)\times SU(2)) / \mathbb{Z}_2=SO(4) & \text{for even  $N_\Lambda$}
  \end{array}
  \right..
  \label{eq:so4-16}
\end{equation}
This result can also be obtained from the \emph{fundamental homomorphism theorem of groups}: The set $G(N_\Lambda)$ [see Eq.~\eqref{eq:so4-11}] can be viewed as the image of a group homomorphism $f: \{ (U_S,U_J) \mid U_{S,J} \in SU(2)\}=SU(2)\times SU(2)\rightarrow G(N_\Lambda)$, which is defined by $f(U_S,U_J)=\left(\begin{array}{cc} U_S & 0 \\ 0 & U_J \\\end{array} \right)^{\otimes N_\Lambda}$; i.e., we have $G(N_\Lambda)=\text{im} (f) $. Due to Theorem 19, the kernel of $f$ is $\ker(f)=\{ (\mathbf{I},\mathbf{I})\}$ (i.e., trivial) for odd  $N_\Lambda$ and is $\ker(f)=\{ (\mathbf{I},\mathbf{I}), (-\mathbf{I},-\mathbf{I})\}=\mathbb{Z}_2$ for even  $N_\Lambda$. Now, the fundamental homomorphism theorem states that $\text{im} (f) \cong (SU(2)\times SU(2)) /\ker(f)$, which gives rise to Eq.~\eqref{eq:so4-16}.

Does Eq.~\eqref{eq:so4-16} represent the true physical symmetry? If the answer is yes, then it would be somewhat elusive: $G(N_\Lambda)$ describes the combination of the $SU(2)$ spin and $SU(2)$  pseudospin symmetries, which is a global and internal symmetry and should thus not rely on spatial degrees of freedom, e.g., $N_\Lambda$.

Remarkably, the gauge freedom hidden in $G(N_\Lambda)$ is crucial for clarifying this issue. According to statements (1) and (3) in Theorem 19, when $N_\Lambda$ is even, different unitary operators of $G(N_\Lambda)$ always describe different physical transformations of electron operators, and hence there is no gauge freedom in $G(N_\Lambda)$ with even $N_\Lambda$; however, when $N_\Lambda$ is odd, two different unitary operators of $G(N_\Lambda)$, say $R$ and $R'$, describe the same physical transformation if and only if $R'=-R$, meaning that $G(N_\Lambda)$ with odd $N_\Lambda$ contains a $\mathbb{Z}_2$ gauge transformation (i.e., $-1$).
[It is worth noting that for even $N_\Lambda$, if $R\in G(N_\Lambda)$, then $-R\notin G(N_\Lambda)$; i.e., $R$ and $-R$ cannot simultaneously belong to $G(N_\Lambda)$ with even $N_\Lambda$.]

Therefore, the true symmetry is simply $G(N_\Lambda)=SO(4)$ for even $N_\Lambda$ and is $G(N_\Lambda)/\mathbb{Z}_2=\frac{SU(2)\times SU(2)}{\mathbb{Z}_2}=SO(4)$ for odd $N_\Lambda$ (with $\mathbb{Z}_2$ corresponding to the gauge transformation). So, the true symmetry is always $SO(4)$, which does not depend on $N_\Lambda$.

Nevertheless, some consequences of the $SO(4)$ symmetry may depend on $N_\Lambda$:

\textbf{Theorem 20:}
If a generic non-Hermitian (or Hermitian) operator $H$ has the $SO(4)$ symmetry (i.e., $[H, J_\alpha]=[H, S_\alpha]=0$ for all $\alpha=x, y, z$) and the total number of lattice sites $N_\Lambda$ is odd. Then, every eigenvalue of $H$ is degenerate.

\textit{Proof.}
If there is a nondegenerate eigenstate of $H$, namely $\psi$, then $\psi$ is also a common eigenstate of $J_\alpha$ and $S_\alpha$ since $[H, J_\alpha]=[H, S_\alpha]=0$. Moreover, one can show that $J_z \psi=S_z \psi=0$ by using the angular-momentum commutation relations: Since $\psi$ is a common eigenstate of $J_x$ and $J_y$, we must have $J_x J_y \psi=J_y J_x \psi$ and hence ${\rm i} J_z \psi=[J_x, J_y] \psi=0$, i.e., $J_z \psi=0$; similarly, we also have $S_z \psi=0$.

Recall that
\begin{equation}
J_z=\frac{1}{2}(N_\uparrow+N_\downarrow-N_\Lambda), \quad  S_z=\frac{1}{2}(N_\uparrow-N_\downarrow), \quad \text{where $N_\uparrow=\sum_{j\in\Lambda} c_{j\uparrow}^\dag c_{j\uparrow}$ and $N_\downarrow=\sum_{j\in\Lambda} c_{j\downarrow}^\dag c_{j\downarrow}$}.
\label{eq:so4-17}
\end{equation}
Now, the equations $J_z \psi=S_z \psi=0$ imply that $N_\uparrow \psi=N_\downarrow \psi=\frac{N_\Lambda}{2} \psi$; i.e., $\psi$ is an eigenstate of $N_{\uparrow,\downarrow}$ with eigenvalue $\frac{N_\Lambda}{2}$. It is clear that all the eigenvalues of $N_{\uparrow,\downarrow}$ must be integers. Thus, when $N_\Lambda$ is odd, $\frac{N_\Lambda}{2}$ is not an integer and cannot be an eigenvalue of $N_{\uparrow,\downarrow}$. So, our assumption that there is a nondegenerate eigenstate of $H$ (with $SO(4)$ symmetry) is impossible for odd $N_\Lambda$.
\hspace{\fill}$\blacksquare$

\subsection{Particle-hole symmetries belonging to the $SO(4)$ symmetry}\label{subsec:PHsymmetries}

In this subsection, we study a class of discrete symmetries, namely the particle-hole (PH) symmetries, which are the combination of an arbitrary spin rotation and a particular pseudospin rotation, and hence belong to the $SO(4)$ symmetry.

The particular pseudospin rotation mentioned above can be chosen as the pseudospin rotation by $\pi$ around the $J_y$ axis \cite{Kai2020}, i.e., the unitary operator $R_J$ in Eq.~\eqref{eq:so4-5} with $\varphi_1=\varphi_3=0$ and $\varphi_2=\pi$
\begin{equation}
e^{{\rm i}\pi J_y}=\left(\begin{array}{cc} \mathbf{I} & 0 \\ 0 & {\rm i}\tau_y \\\end{array} \right)^{\otimes N_\Lambda}\in G_J,
\label{eq:ph-1}
\end{equation}
where $\tau_y$ is the second Pauli matrix; see Eq.~\eqref{eq:so4-7} for the matrix representation of $R_J$. From Eqs.~\eqref{eq:so4-6} and \eqref{eq:f1}, we see that Eq.~\eqref{eq:ph-1} actually represents a $\mathbb{Z}_4$ PH transformation:
\begin{equation}
(e^{{\rm i}\pi J_y})^4=1, \qquad e^{{\rm i}\pi J_y}c_{j\uparrow}e^{-{\rm i}\pi J_y}=-e^{{\rm i}\theta_j}c_{j\downarrow}^\dag, \qquad e^{{\rm i}\pi J_y}c_{j\downarrow}e^{-{\rm i}\pi J_y}=e^{{\rm i}\theta_j}c_{j\uparrow}^\dag.
\label{eq:ph-2}
\end{equation}
Note that $e^{{\rm i}\pi J_y}$ squares to the charge parity: $(e^{{\rm i}\pi J_y})^2=(-1)^Q$, under which the electron operators transform as $c_{j\sigma}\rightarrow -c_{j\sigma}$, where $Q$ denotes the charge operator $Q=N_\uparrow+N_\downarrow-N_\Lambda=2J_z$ [see Eq.~\eqref{eq:so4-17}]. Moreover, $e^{{\rm i}\pi J_y}$ is also called charge conjugation, since it flips the sign of $Q$: $e^{{\rm i}\pi J_y}Qe^{-{\rm i}\pi J_y}=-Q$.

Now, from Eqs.~\eqref{eq:ph-2} and \eqref{eq:so4-3}, it follows that the product of $e^{{\rm i}\pi J_y}$ and an arbitrary spin rotation $R_S\in G_S$, namely
\begin{equation}
\mathcal{C}_S=e^{{\rm i}\pi J_y}R_S\in G(N_\Lambda),
\label{eq:ph-3}
\end{equation}
can realize a very general PH transformation [see also Eq.~\eqref{eq:so4-9}]:
\begin{equation}
\mathcal{C}_Sc_{j\uparrow}\mathcal{C}_S^{-1}=h_Se^{{\rm i}\theta_j}c_{j\uparrow}^\dag-g_Se^{{\rm i}\theta_j}c_{j\downarrow}^\dag, \qquad
\mathcal{C}_Sc_{j\downarrow}\mathcal{C}_S^{-1}=g^*_Se^{{\rm i}\theta_j}c_{j\uparrow}^\dag+h^*_Se^{{\rm i}\theta_j}c_{j\downarrow}^\dag
\label{eq:ph-4}
\end{equation}
with the complex numbers $g_{S}$ and $h_{S}$ satisfying $|g_S|^2+|h_S|^2=1$. Note that Eq.~\eqref{eq:ph-3} also represents a charge conjugation: $\mathcal{C}_SQ\mathcal{C}_S^{-1}=-Q$. In fact, the unitary operator $\mathcal{C}_S$ [or more generally, $R$ in Eq.~\eqref{eq:so4-8}] describes an internal (or, on-site) symmetry and we have
\begin{equation}
\mathcal{C}_SQ_j\mathcal{C}_S^{-1}=-Q_j, \quad \text{where $Q_j= c_{j\uparrow}^\dag c_{j\uparrow}+ c_{j\downarrow}^\dag c_{j\downarrow}-1$}.
\label{eq:ph-4-1}
\end{equation}
Furthermore, the general PH transformation \eqref{eq:ph-4} includes the usual $\mathbb{Z}_2$ PH transformation
\begin{equation}
c_{j\uparrow}\rightarrow e^{{\rm i}\theta_j}c_{j\uparrow}^\dag, \qquad  c_{j\downarrow}\rightarrow e^{{\rm i}\theta_j}c_{j\downarrow}^\dag
\label{eq:ph-5}
\end{equation}
as a special case (with $g_{S}=0$ and $h_{S}=1$), which is realized by the unitary operator $\mathcal{C}_S=e^{{\rm i}\pi J_y}e^{-{\rm i}\pi S_y}$ [i.e., Eq.~\eqref{eq:ph-3} with $R_S=e^{-{\rm i}\pi S_y}$] that squares to $(-1)^{N_\Lambda}$,  consistent with the $\mathbb{Z}_2$ nature of the transformation \eqref{eq:ph-5}. Note that when $g_{S}=1$ and $h_{S}=0$ (i.e., $R_S=1$), Eqs.~\eqref{eq:ph-3} and \eqref{eq:ph-4} reduce to Eqs.~\eqref{eq:ph-1} and \eqref{eq:ph-2}, respectively.

Therefore, different spin rotations $R_S$ in Eq.~\eqref{eq:ph-3}, or equivalently, different choices of the parameters $g_{S}$ and $h_{S}$ in Eq.~\eqref{eq:ph-4}, may give rise to different kinds of PH transformations [i.e., PH transformations with different symmetry structures, such as the above $\mathbb{Z}_4$ and $\mathbb{Z}_2$ PH transformations described by Eqs.~\eqref{eq:ph-2} and \eqref{eq:ph-5}, respectively]. Note, however, that different spin rotations may also give rise to the same kind of PH transformations: For example, $R_S=e^{{\rm i}\phi S_z}e^{-{\rm i}\pi S_y}$ (i.e., $g_{S}=0$ and $h_{S}=e^{{\rm i}\phi/2}$) with an arbitrary angle $\phi$ leads to a $\mathbb{Z}_2$ PH transformation, e.g., the operator $\mathcal{C}_S=e^{{\rm i}\pi J_y}e^{{\rm i}\phi S_z}e^{-{\rm i}\pi S_y}$ squares to $(-1)^{N_\Lambda}$; in particular, the case of $\phi=0$ corresponds to Eq.~\eqref{eq:ph-5}.

Given these results, we then ask whether Eq.~\eqref{eq:ph-3}, or  Eq.~\eqref{eq:ph-4}, contains a $\mathbb{Z}_{2k}$ PH transformation for any positive integer $k$. [It is clear that there does not exist a $\mathbb{Z}_{2k-1}$ PH transformation.] That is, how we choose $R_S$ in Eq.~\eqref{eq:ph-3}, or equivalently, $g_{S}$ and $h_{S}$ in Eq.~\eqref{eq:ph-4}, to realize a general $\mathbb{Z}_{2k}$ PH transformation. This can be done by a $U(1)$ spin rotation, namely $R_S=e^{{\rm i}\phi S_z}$ (i.e., $g_{S}=e^{-{\rm i}\phi/2}$ and $h_{S}=0$) with $0\leq \phi<4\pi$. Thus, using Eq.~\eqref{eq:ph-4} with $g_{S}=e^{-{\rm i}\phi/2}$ and $h_{S}=0$, one obtains
\begin{equation}
(\mathcal{C}_\phi)^{2p} c_{j\uparrow}(\mathcal{C}_\phi)^{-2p}=e^{-{\rm i}p(\phi+\pi)}c_{j\uparrow}, \qquad (\mathcal{C}_\phi)^{2p} c_{j\downarrow}(\mathcal{C}_\phi)^{-2p}=e^{{\rm i}p(\phi+\pi)}c_{j\downarrow},
\label{eq:ph-6}
\end{equation}
where $p$ is an arbitrary integer (i.e., $p=0,\pm 1,\pm2,\ldots$) and
\begin{equation}
\mathcal{C}_\phi=e^{{\rm i}\pi J_y}e^{{\rm i}\phi S_z}.
\label{eq:ph-6-1}
\end{equation}
Now,  according to Eq.~\eqref{eq:ph-6}, a $\mathbb{Z}_{2k}$ PH transformation realized by $\mathcal{C}_\phi$ means that the smallest positive integer $p$ (if one exists) that satisfies $e^{{\rm i}p(\phi+\pi)}=1$ [i.e., the electron operators $c_{j\sigma}$ are invariant under $(\mathcal{C}_\phi)^{2p}$] is $p=k$.

The condition $e^{{\rm i}p(\phi+\pi)}=1$ with a nonzero integer $p$ requires that $\phi+\pi$ is a rational multiple of $2\pi$. Specifically, when $(\phi+\pi)/(2\pi)$ is a rational number expressed as an irreducible fraction, say
\begin{equation}
\frac{\phi+\pi}{2\pi}=\frac{l}{k} \quad (k>0),
\label{eq:ph-7}
\end{equation}
where $l$ and $k$ are coprime integers, the equation $e^{{\rm i}p(\phi+\pi)}=1$ holds if and only if $p/k$ is an integer. [Note that $e^{{\rm i}p(\phi+\pi)}=1\Leftrightarrow (pl/k)\in\mathbb{Z}$, and $(pl/k)\in\mathbb{Z}\Leftrightarrow (p/k)\in\mathbb{Z}$. Mathematically, the fact that $(pl/k)\in\mathbb{Z}\Rightarrow (p/k)\in\mathbb{Z}$ is due to the \emph{fundamental theorem of arithmetic}, which states that every integer greater than 1 can be represented uniquely as a product of prime numbers (up to the order of the factors).] So, the smallest positive $p$ for $e^{{\rm i}p(\phi+\pi)}=1$ is $p=k$, i.e., $\mathcal{C}_\phi$ with $\phi$ satisfying Eq.~\eqref{eq:ph-7} realizes a $\mathbb{Z}_{2k}$ PH transformation. This implies that $(\mathcal{C}_\phi)^{2k}$ should be equal to the identity, up to a phase factor (e.g., a gauge transformation). To check this, using the matrix representation of Eq.~\eqref{eq:ph-6-1} [see, e.g., Eq.~\eqref{eq:so4-11}], one can show that $(\mathcal{C}_\phi)^{2p}=[\text{diag}(e^{{\rm i}p\phi},e^{-{\rm i}p\phi},(-1)^p,(-1)^p)]^{\otimes N_\Lambda}$. Thus, when $\phi$ satisfies Eq.~\eqref{eq:ph-7} and $p=k$, we have
\begin{equation}
(\mathcal{C}_\phi)^{2k}=(-1)^{kN_\Lambda}=e^{{\rm i}\pi kN_\Lambda}.
\label{eq:ph-8}
\end{equation}
[Without using the matrix representation, one can also obtain Eq.~\eqref{eq:ph-8} by using the relation $e^{{\rm i}2\pi S_z}e^{{\rm i}2\pi J_z}= (-1)^{N_\Lambda}$, i.e., Eq.~\eqref{eq:so4-13}; note that $e^{{\rm i}2\pi J_y}=e^{{\rm i}2\pi J_z}$.] It is clear that $k$ can be any positive integer because $\frac{1}{2}\leq\frac{\phi+\pi}{2\pi}=\frac{l}{k}<\frac{5}{2}$, due to $0\leq \phi<4\pi$. Moreover, for any given $k$, the corresponding rational number $\frac{\phi+\pi}{2\pi}=\frac{l}{k}$ (or simply $\phi$) is not unique; in other words, different PH transformations, described by $\mathcal{C}_\phi$ with different $\phi$, may share the same symmetry group $\mathbb{Z}_{2k}$.
For example, for $k=1$, the corresponding rational number $\frac{\phi+\pi}{2\pi}=\frac{l}{k}\in[\frac{1}{2},\frac{5}{2})$ is an integer and hence $\frac{\phi+\pi}{2\pi}=1$ or 2, i.e., $\phi=\pi$ or $3\pi$. Thus, both $\mathcal{C}_{\phi=\pi}: c_{j\uparrow}\rightarrow {\rm i}e^{{\rm i}\theta_j}c_{j\downarrow}^\dag,  c_{j\downarrow}\rightarrow {\rm i}e^{{\rm i}\theta_j}c_{j\uparrow}^\dag$ and $\mathcal{C}_{\phi=3\pi}: c_{j\uparrow}\rightarrow -{\rm i}e^{{\rm i}\theta_j}c_{j\downarrow}^\dag,  c_{j\downarrow}\rightarrow -{\rm i}e^{{\rm i}\theta_j}c_{j\uparrow}^\dag$ realize a $\mathbb{Z}_{2}$ PH transformation [compare with Eq.~\eqref{eq:ph-5}]. As another example, consider the case of $k=2$. The corresponding rational number $\frac{\phi+\pi}{2\pi}=\frac{l}{k}\in[\frac{1}{2},\frac{5}{2})$ is thus $\frac{\phi+\pi}{2\pi}=\frac{1}{2}$ or $\frac{3}{2}$, i.e., $\phi=0$ or $2\pi$. So, both $\mathcal{C}_{\phi=0}$ and $\mathcal{C}_{\phi=2\pi}$ realize a $\mathbb{Z}_{4}$ PH transformation. Note that $\mathcal{C}_{\phi=0}$ is nothing but the particular pseudospin rotation \eqref{eq:ph-1}.

In contrast to Eq.~\eqref{eq:ph-7}, when $(\phi+\pi)/(2\pi)$ is an irrational number, $e^{{\rm i}p(\phi+\pi)}$ cannot be equal to 1 for any nonzero integer $p$. This means that for any even integer $2p\neq0$, the unitary operator $(\mathcal{C}_\phi)^{2p}$ transforms the electron operators $c_{j\sigma}$ nontrivially, see Eq.~\eqref{eq:ph-6}. Clearly, for every odd integer $2m+1$  ($m=0,\pm 1,\pm2,\ldots$), $(\mathcal{C}_\phi)^{2m+1}$ transforms the electron operators $c_{j\sigma}$ nontrivially. Thus, the unitary operator $\mathcal{C}_\phi$ generates a $\mathbb{Z}$ symmetry, i.e., a $\mathbb{Z}$ PH transformation, where $\mathbb{Z}$ denotes the additive group of the integers. Note that the infinite cyclic group $\mathbb{Z}$ may be viewed as the above group $\mathbb{Z}_{2k}$ with $k=\infty$; and all these discrete cyclic groups $\mathbb{Z}_{2k}$ (including $\mathbb{Z}$) are subgroups of $SO(4)$.

At the end of this subsection, let us study the relations between the various PH symmetries and the $SU(2)$ pseudospin symmetry of the Hubbard model, which are summarized in the following theorem.

\textbf{Theorem 21:}
For the (non-Hermitian) Hubbard model with the Hamiltonian $H=T+V$ in Eq.~\eqref{eq:nH-Hubbard}, all the PH symmetries described by Eq.~\eqref{eq:ph-3} are equivalent to each other. Moreover, the Hamiltonian $H$ has these PH symmetries if and only if $H$ has the $SU(2)$ pseudospin [or $SO(4)$] symmetry.

\textit{Proof.}
The first conclusion in Theorem 21 is actually a consequence of the $SU(2)$ spin symmetry of the Hamiltonian $H$: Consider two arbitrary PH symmetries described by Eq.~\eqref{eq:ph-3}, say, $\mathcal{C}_1=e^{{\rm i}\pi J_y}R_1$ and $\mathcal{C}_2=e^{{\rm i}\pi J_y}R_2$, where $R_{1,2}\in G_S$. Now, $H$ has the PH symmetry $\mathcal{C}_1$ if and only if $H$ has the PH symmetry $\mathcal{C}_2$, i.e., $[H,\mathcal{C}_1]=0 \Leftrightarrow [H,\mathcal{C}_2]=0$. This is because $\mathcal{C}_1$ and $\mathcal{C}_2$ differ by a spin rotation, namely $\mathcal{C}_2=\mathcal{C}_1 R_3$ with $R_3=R_1^{-1}R_2\in G_S$, and $H$ has the $SU(2)$ spin symmetry, i.e., $[H,R_S]=0$ ($\forall R_S\in G_S$).

To prove the second conclusion in Theorem 21, we need to examine how the Hamiltonian $H=T+V$ transforms under the PH transformation. We first note that the action on $H$ by $\mathcal{C}_S$ in Eq.~\eqref{eq:ph-3} does not depend on the choice of $R_{S}$, or equivalently, $g_{S}$ and $h_{S}$ in Eq.~\eqref{eq:ph-4}, due to the $SU(2)$ spin symmetry of $H$; i.e., for two arbitrary PH operators $\mathcal{C}_1$ and $\mathcal{C}_2$ (as described above), we have
\begin{equation}
\mathcal{C}_1 H \mathcal{C}_1^{-1}=\mathcal{C}_2 H \mathcal{C}_2^{-1}.
\label{eq:ph-9}
\end{equation}
It is also worth mentioning that the fact that $[H,\mathcal{C}_1]=0 \Leftrightarrow [H,\mathcal{C}_2]=0$ (i.e., the first conclusion in Theorem 21 as discussed above) is a direct consequence of Eq.~\eqref{eq:ph-9}.

In fact, since both $T$ and $V$ in $H$ respect the $SU(2)$ spin symmetry, the action on $T$ or $V$ by $\mathcal{C}_S$ does not depend on the choice of $R_{S}$, or equivalently, $g_{S}$ and $h_{S}$. Specifically, under the PH transformation Eq.~\eqref{eq:ph-4}, the off-site hopping terms in $T$ transform as
\begin{equation}
\mathcal{C}_S: t_{ij}(c_{i\uparrow}^\dag c_{j\uparrow}+c_{i\downarrow}^\dag c_{j\downarrow})\rightarrow -e^{{-\rm i}\theta_i}e^{{\rm i}\theta_j}t_{ij}(c_{j\uparrow}^\dag c_{i\uparrow}+c_{j\downarrow}^\dag c_{i\downarrow}),
\quad     \text{$\forall i,j\in\Lambda$ and $i\neq j$};
\label{eq:ph-10}
\end{equation}
and the on-site terms in $V$ transform as
\begin{equation}
\mathcal{C}_S: U_jn_{j\uparrow}n_{j\downarrow}-\mu_jn_j \rightarrow U_jn_{j\uparrow}n_{j\downarrow}-(U_j-\mu_j)n_j+U_j-2\mu_j,     \quad     \forall j\in\Lambda,
\label{eq:ph-11}
\end{equation}
where $n_j=n_{j\uparrow}+n_{j\downarrow}$, and we have used the equations $\mathcal{C}_S n_j \mathcal{C}_S^{-1}=2-n_j$, which is equivalent to Eq.~\eqref{eq:ph-4-1}, and $n_{j\uparrow}n_{j\downarrow}=n_j(n_j-1)/2$.

Thus, from Eqs.~\eqref{eq:ph-10} and \eqref{eq:ph-11}, the total Hamiltonian $H=T+V$ has the PH symmetries (i.e., $\mathcal{C}_S H \mathcal{C}_S^{-1}=H$) if and only if $t_{ij}e^{{\rm i}\theta_j}+t_{ji}e^{{\rm i}\theta_i}=0$ (for each bond $\{i,j\}$) and $\mu_j=U_j/2$ ($\forall j\in\Lambda$), i.e., Eqs.~\eqref{eq:pseudospin-symmetry-a} and \eqref{eq:pseudospin-symmetry-b}, which is also the sufficient and necessary condition for $H$ to have the $SU(2)$ pseudospin symmetry, as stated in Sec.~\ref{symmetry-pseudospin}.
\hspace{\fill}$\blacksquare$

\textbf{Remarks:}
First, since the Hubbard model described by Eq.~\eqref{eq:nH-Hubbard} always has the $SU(2)$ spin symmetry (by assumption), the $SU(2)$ pseudospin symmetry and the $SO(4)$ symmetry are equivalent to each other.
Second, it is not surprising that the $SO(4)$ symmetry implies the PH symmetry, because the latter belongs to the former. While, a nontrivial result from Theorem 21 is that the PH symmetry also implies the bigger $SO(4)$ symmetry.
Third, from Theorems 17 and 21 and the related discussion, we see that the $\mathbb{Z}_2$ Shiba symmetry (property), all the PH symmetries, and the $SU(2)$ pseudospin [or $SO(4)$] symmetry are unified by Eqs.~\eqref{eq:pseudospin-symmetry-a} and \eqref{eq:pseudospin-symmetry-b} in the context of Hubbard model.

\subsection{The Majorana representation}\label{subsec:Majorana-rep}

As we mentioned before, the $SO(4)$ symmetry would become explicit in the Majorana representation. To see this, let us introduce the Majorana operators
\begin{eqnarray}
\gamma_{1j}= {\rm i}e^{{\rm i}\theta_j/2}c_{j\uparrow}^\dag-{\rm i}e^{{-\rm i}\theta_j/2}c_{j\uparrow},  \qquad
\gamma_{2j}= e^{{\rm i}\theta_j/2}c_{j\uparrow}^\dag+e^{{-\rm i}\theta_j/2}c_{j\uparrow},
\nonumber\\
\gamma_{3j}= {\rm i}e^{{\rm i}\theta_j/2}c_{j\downarrow}^\dag-{\rm i}e^{{-\rm i}\theta_j/2}c_{j\downarrow},  \qquad
\gamma_{4j}= e^{{\rm i}\theta_j/2}c_{j\downarrow}^\dag+e^{{-\rm i}\theta_j/2}c_{j\downarrow},
\label{eq:mr-1}
\end{eqnarray}
which are Hermitian operators and satisfy the anticommutation relations $\{\gamma_{\mu i}, \gamma_{\nu j}\}=2\delta_{\mu\nu}\delta_{ij}$ with $\mu, \nu=1,2,3,4$ and $i,j\in\Lambda$. Recall that the phase parameter $\theta_j$ has been defined in the Shiba transformation \eqref{eq:shiba1}, or equivalently, as the phase of $\omega_j=|\omega_j|e^{{\rm i}\theta_j}$ (as stated above Theorem 17 in Appendix \ref{subsec:Shiba}). Note that $\gamma_{1j}$ and $\gamma_{2j}$ ($\gamma_{3j}$ and $\gamma_{4j}$) may be viewed as the spin-up (spin-down) real fermions.

Now, from Eq.~\eqref{eq:so4-9}, one can obtain the action of $R\in G(N_\Lambda)$ [see Eqs.~\eqref{eq:so4-8} and \eqref{eq:so4-11}] on Majorana operators
\begin{equation}
R\gamma_{j}R^{-1}=\gamma_j A, \quad     \text{where $\gamma_j=(\gamma_{1j}, \gamma_{2j}, \gamma_{3j}, \gamma_{4j})$}.
\label{eq:mr-2}
\end{equation}
The above $4\times4$ matrix $A$ reads
\begin{equation}
A=\left(\begin{array}{cccc}
 \text{Re}(g_Sg_J+h_Sh_J) & -\text{Im}(g_Sg_J+h_Sh_J) & \text{Re}(g_Sh^*_J-h_Sg^*_J) & -\text{Im}(g^*_Sh_J+h_Sg^*_J) \\
 \text{Im}(g_Sg_J-h_Sh_J) & \text{Re}(g_Sg_J-h_Sh_J) & -\text{Im}(h^*_Sg_J+g^*_Sh_J) & -\text{Re}(g_Sh^*_J+h_Sg^*_J) \\
 \text{Re}(h_Sg_J-g_Sh_J) & \text{Im}(g_Sh_J-h_Sg_J) & \text{Re}(g_Sg^*_J+h_Sh^*_J) & -\text{Im}(g^*_Sg_J+h^*_Sh_J) \\
 \text{Im}(g_Sh_J+h_Sg_J) & \text{Re}(g_Sh_J+h_Sg_J) & -\text{Im}(g_Sg^*_J+h^*_Sh_J) & \text{Re}(g^*_Sg_J-h^*_Sh_J)
\end{array} \right),
\label{eq:mr-3}
\end{equation}
where $\text{Re}(\omega)$ and $\text{Im}(\omega)$ respectively denote the real and imaginary parts of a complex number $\omega$. [In Eq.~\eqref{eq:mr-2}, the operators $R$ are (not) in one-to-one correspondence with the matrices $A$ for even (odd) $N_\Lambda$, according to Theorem 19; see also the discussion of gauge freedom above Theorem 20 in Appendix \ref{subsec:physicalSO4}.] Note that the matrix $A$ does not depend on $j\in\Lambda$, i.e., the global transformation \eqref{eq:mr-2}; and recall that the complex parameters $g_{S,J}$ and $h_{S,J}$ satisfy $|g_S|^2+|h_S|^2=|g_J|^2+|h_J|^2=1$. Before proceeding, let us clarify the relations between Majorana operators and real orthogonal matrices.

\textbf{Lemma 22:}
Let $\gamma=(\gamma_{1}, \gamma_{2}, \ldots , \gamma_{n})$ be Majorana operators, i.e., $\gamma_p^\dag=\gamma_p$ and $\{\gamma_p, \gamma_q\}=2\delta_{pq}$ with $p,q=1,2,\ldots,n$. And we define the operators $\gamma'=(\gamma'_{1}, \gamma'_{2},\ldots, \gamma'_{n})=\gamma A$, where $A$ is an $n\times n$ matrix. Then (1) $\gamma'$ are Majorana operators if and only if $A$ is a real orthogonal matrix, i.e., $A\in O(n)$; (2) when $\gamma'$ are Majorana operators [or equivalently, $A\in O(n)$, due to conclusion (1)], the product of Majorana operators obeys $\gamma'_{1}\gamma'_{2}\cdots\gamma'_{n}=(\det A)\gamma_{1}\gamma_{2}\cdots\gamma_{n}$.

\textbf{Remark:}
The conclusions in Lemma 22 can be obtained by using the Hermiticity and anticommutation relations of Majorana operators. Note that in conclusion (2), since $A$ is a real orthogonal matrix, its determinant $\det A$ is either $+1$ or $-1$, and hence $\gamma'_{1}\gamma'_{2}\cdots\gamma'_{n}=\pm\gamma_{1}\gamma_{2}\cdots\gamma_{n}$. In particular, $\gamma'_{1}\gamma'_{2}\cdots\gamma'_{n}=\gamma_{1}\gamma_{2}\cdots\gamma_{n}$ if and only if $A\in SO(n)$.

In the following discussion, we explain why the $4\times4$ matrix $A$ in Eq.~\eqref{eq:mr-2} is a real orthogonal matrix with determinant $+1$, i.e., $A\in SO(4)$, although one may check this by merely looking at the expression of $A$ in Eq.~\eqref{eq:mr-3} (which seems to be an awkward approach). Let us introduce the operators $\gamma'_j=(\gamma'_{1j}, \gamma'_{2j}, \gamma'_{3j}, \gamma'_{4j})$ defined as $\gamma'_j=R\gamma_{j}R^{-1}$ [see Eq.~\eqref{eq:mr-2}]. Since $R$ is a unitary operator (i.e., $R^{-1}=R^\dag$), which preserves the Hermiticity and anticommutation relations of the Majorana operators $\gamma_j$, then $\gamma'_j$ are also Majorana operators. On the other hand, from Eq.~\eqref{eq:mr-2}, we have $\gamma'_j=\gamma_j A$. Thus, according to conclusion (1) in Lemma 22, $A$ is a real orthogonal matrix, i.e., $A\in O(4)$.

Next, based on conclusion (2) in Lemma 22, we are going to prove that $\det A=1$, namely $A\in SO(4)$. From Eq.~\eqref{eq:mr-1}, we have
\begin{equation}
\gamma_{1j}\gamma_{2j}=2{\rm i}(c_{j\uparrow}^\dag c_{j\uparrow}-\frac{1}{2}),     \qquad     \gamma_{3j}\gamma_{4j}=2{\rm i}(c_{j\downarrow}^\dag c_{j\downarrow}-\frac{1}{2}),
\label{eq:mr-4}
\end{equation}
and thus
\begin{equation}
\gamma_{1j}\gamma_{2j}\gamma_{3j}\gamma_{4j}=-4(c_{j\uparrow}^\dag c_{j\uparrow}-\frac{1}{2})(c_{j\downarrow}^\dag c_{j\downarrow}-\frac{1}{2}),
\label{eq:mr-5}
\end{equation}
which is proportional to the on-site term $V'_j=U_j(c_{j\uparrow}^\dag c_{j\uparrow}-\frac{1}{2})(c_{j\downarrow}^\dag c_{j\downarrow}-\frac{1}{2})$ in $V'=\sum_{j\in\Lambda}V'_j$ [see Eq.~\eqref{eq:shiba6}]. We already know that $V'_j$ (for each site $j\in\Lambda$) respects both the $SU(2)$ spin symmetry and the $SU(2)$ pseudospin symmetry, i.e., $\forall R_S\in G_S$ and $\forall R_J\in G_J$, $[V'_j, R_S]=[V'_j, R_J]=0$. Thus, we have $[V'_j, R]=0$, where $R=R_S R_J$. In other words, Eq.~\eqref{eq:mr-5} also commutes with $R$, i.e.,
\begin{equation}
R\gamma_{1j}\gamma_{2j}\gamma_{3j}\gamma_{4j}R^{-1}=\gamma_{1j}\gamma_{2j}\gamma_{3j}\gamma_{4j}.
\label{eq:mr-6}
\end{equation}
Note that according to the definition $\gamma'_j=R\gamma_{j}R^{-1}$, we have $\gamma'_{1j}\gamma'_{2j}\gamma'_{3j}\gamma'_{4j}=R\gamma_{1j}\gamma_{2j}\gamma_{3j}\gamma_{4j}R^{-1}$. Thus, Eq.~\eqref{eq:mr-6} means that $\gamma'_{1j}\gamma'_{2j}\gamma'_{3j}\gamma'_{4j}=\gamma_{1j}\gamma_{2j}\gamma_{3j}\gamma_{4j}$. Now, from conclusion (2) in Lemma 22 (see also the Remark below Lemma 22), we conclude that $\det A=1$, i.e., $A\in SO(4)$ (recall that $\gamma'_j=\gamma_j A$).

As subgroups of $SO(4)$, both the $SU(2)$ spin transformation [e.g., Eq.~\eqref{eq:so4-3}] and the $SU(2)$ pseudospin transformation [e.g., Eq.~\eqref{eq:so4-6}] can be described by Eq.~\eqref{eq:mr-2}. Specifically, let us first consider the spin transformation
\begin{equation}
R_S\gamma_{j}R_S^{-1}=\gamma_j A_S, \quad     \text{where $R_S\in G_S$}.
\label{eq:mr-7}
\end{equation}
The matrix $A_S$ in Eq.~\eqref{eq:mr-7} can be obtained from the matrix $A$ in Eq.~\eqref{eq:mr-3} by setting $g_J=1$ and $h_J=0$ (i.e., $R_J=1$), namely
\begin{equation}
A_S=A(g_J=1,h_J=0)=\text{Re}(g_S) I+\text{Re}(h_S)W_1+\text{Im}(g_S)W_2+\text{Im}(h_S)W_3,
\label{eq:mr-8}
\end{equation}
where $I$ is the $4\times4$ identity matrix and
\begin{equation}
W_1=\left(\begin{array}{cccc}
 0 & 0 & -1 & 0 \\
 0 & 0 & 0 & -1 \\
 1 & 0 & 0 & 0 \\
 0 & 1 & 0 & 0
\end{array} \right), \qquad W_2=\left(\begin{array}{cccc}
 0 & -1 & 0 & 0 \\
 1 & 0 & 0 & 0 \\
 0 & 0 & 0 & 1 \\
 0 & 0 & -1 & 0
\end{array} \right), \qquad W_3=\left(\begin{array}{cccc}
 0 & 0 & 0 & -1 \\
 0 & 0 & 1 & 0 \\
 0 & -1 & 0 & 0 \\
 1 & 0 & 0 & 0
\end{array} \right).
\label{eq:mr-9}
\end{equation}
Importantly, the matrices $W_1$, $W_2$ and $W_3$ obey the quaternion algebra
\begin{equation}
W^2_1=W^2_2=W^2_3=W_1W_2W_3=-I,
\label{eq:mr-10}
\end{equation}
which leads to the relations $W_1W_2=-W_2W_1=W_3$, and so on. Recall that the four real parameters $\text{Re}(g_S)$, $\text{Re}(h_S)$, $\text{Im}(g_S)$ and $\text{Im}(h_S)$ satisfy the constraint
\begin{equation}
(\text{Re}g_S)^2+(\text{Re}h_S)^2+(\text{Im}g_S)^2+(\text{Im}h_S)^2=|g_S|^2+|h_S|^2=1.
\label{eq:mr-11}
\end{equation}
Thus, Eq.~\eqref{eq:mr-8} is nothing but the quaternion representation of the group $SU(2)$. In other words, the set of all the matrices $A_S$, denoted by $\Gamma_S$, is an $SU(2)$ subgroup of $SO(4)$. In fact, without the explicit form of $A_S$ [i.e., Eq.~\eqref{eq:mr-8}], one can infer that $\Gamma_S\cong G_S$, because in Eq.~\eqref{eq:mr-7} the operators $R_S\in G_S$ are in one-to-one correspondence with the matrices $A_S\in\Gamma_S$ (see the discussion in the preceding subsections). Thus, we have $\Gamma_S\cong SU(2)$ because $G_S\cong SU(2)$ [see Eq.~\eqref{eq:so4-4-1}].

Another $SU(2)$ subgroup can correspond to the pseudospin transformation
\begin{equation}
R_J\gamma_{j}R_J^{-1}=\gamma_j A_J, \quad     \text{where $R_J\in G_J$}.
\label{eq:mr-12}
\end{equation}
The above matrix $A_J$ can be obtained by setting $g_S=1$ and $h_S=0$ (i.e., $R_S=1$) in Eq.~\eqref{eq:mr-3}, namely
\begin{equation}
A_J=A(g_S=1,h_S=0)=\text{Re}(g_J) I+\text{Re}(h_J)K_1+\text{Im}(g_J)K_2+\text{Im}(h_J)K_3,
\label{eq:mr-13}
\end{equation}
where
\begin{equation}
K_1=\left(\begin{array}{cccc}
 0 & 0 & 1 & 0 \\
 0 & 0 & 0 & -1 \\
 -1 & 0 & 0 & 0 \\
 0 & 1 & 0 & 0
\end{array} \right), \qquad K_2=\left(\begin{array}{cccc}
 0 & -1 & 0 & 0 \\
 1 & 0 & 0 & 0 \\
 0 & 0 & 0 & -1 \\
 0 & 0 & 1 & 0
\end{array} \right), \qquad K_3=\left(\begin{array}{cccc}
 0 & 0 & 0 & -1 \\
 0 & 0 & -1 & 0 \\
 0 & 1 & 0 & 0 \\
 1 & 0 & 0 & 0
\end{array} \right).
\label{eq:mr-14}
\end{equation}
Importantly, the matrices $K_1$, $K_2$ and $K_3$ obey the quaternion algebra
\begin{equation}
K^2_1=K^2_2=K^2_3=K_1K_2K_3=-I,
\label{eq:mr-15}
\end{equation}
which leads to the relations $K_1K_2=-K_2K_1=K_3$, and so on. Recall that the four real parameters $\text{Re}(g_J)$, $\text{Re}(h_J)$, $\text{Im}(g_J)$ and $\text{Im}(h_J)$ satisfy the constraint
\begin{equation}
(\text{Re}g_J)^2+(\text{Re}h_J)^2+(\text{Im}g_J)^2+(\text{Im}h_J)^2=|g_J|^2+|h_J|^2=1.
\label{eq:mr-16}
\end{equation}
Thus, similar to  Eq.~\eqref{eq:mr-8}, Eq.~\eqref{eq:mr-13} also forms the quaternion representation of the group $SU(2)$; i.e., the set of all the matrices $A_J$, denoted by $\Gamma_J$, is also an $SU(2)$ subgroup of $SO(4)$.
Similar to the above discussion of the spin transformation, without the explicit form of $A_J$ [i.e., Eq.~\eqref{eq:mr-13}], one can infer that $\Gamma_J\cong G_J$ and hence $\Gamma_J\cong SU(2)$.

Since the spin and pseudospin operators commute with each other (e.g., $R_S R_J=R_JR_S$), we have $A_SA_J=A_JA_S$, or equivalently
\begin{equation}
W_aK_b=K_b W_a, \quad     \text{where $a,b=1,2,3$},
\label{eq:mr-17}
\end{equation}
as can be seen from Eqs.~\eqref{eq:mr-8} and \eqref{eq:mr-13} (compare with the relations $W_1W_2=-W_2W_1$, $K_1K_2=-K_2K_1$, and so on). According to the relation $R=R_S R_J$, Eq.~\eqref{eq:mr-3} can be decomposed into the matrix multiplication of Eqs.~\eqref{eq:mr-8} and \eqref{eq:mr-13}, i.e., $A=A_SA_J$. Furthermore, we expect that every $SO(4)$ matrix can be written in the form \eqref{eq:mr-3}; in other words, the set of all the matrices $A$ coincides with the group $SO(4)$. Thus, we conclude that
\begin{equation}
\Gamma_S \Gamma_J=SO(4),
\label{eq:mr-18}
\end{equation}
where the product $\Gamma_S \Gamma_J=\Gamma_J \Gamma_S=\{A_SA_J=A_JA_S \mid A_S\in\Gamma_S , A_J\in\Gamma_J\}$. Moreover, the two $SU(2)$ groups $\Gamma_S$ and $\Gamma_J$ have a nontrivial intersection
\begin{equation}
\Gamma_S\cap\Gamma_J=\{I, -I\}\cong\mathbb{Z}_2,
\label{eq:mr-19}
\end{equation}
as can be seen from Eqs.~\eqref{eq:mr-8} and \eqref{eq:mr-13}. Now, from Eqs.~\eqref{eq:mr-18} and \eqref{eq:mr-19} and using Eq.~\eqref{eq:so4-00011}, we arrive at the well-known formula
\begin{equation}
SO(4)\cong\frac{SU(2)\times SU(2)}{\mathbb{Z}_2}
\label{eq:mr-20}
\end{equation}
that has been used in this paper.

Eq.~\eqref{eq:mr-12} contains the charge $U(1)$ transformation $e^{-{\rm i}2\alpha J_z}\gamma_j e^{{\rm i}2\alpha J_z}=\gamma_j A_c$, where $0\leq\alpha<2\pi$ and the matrix $A_c$ can be obtained by setting $g_J=e^{{\rm i}\alpha}$ and $h_J=0$ (i.e., $R_J=e^{-{\rm i}2\alpha J_z}$) in Eq.~\eqref{eq:mr-13}
\begin{equation}
A_c=\left(\begin{array}{cc}
 A_\alpha &  0 \\
 0 & A_\alpha
\end{array} \right)=(\cos\alpha) I+(\sin\alpha) K_2=e^{\alpha K_2}, \quad
\text{where $A_\alpha=\left(\begin{array}{cc}
 \cos\alpha & -\sin\alpha  \\
 \sin\alpha & \cos\alpha
\end{array} \right)\in SO(2)$}.
\label{eq:mr-21}
\end{equation}
We thus see that the $U(1)$ transformation is an $SO(2)$ rotation in the Majorana representation, e.g., $U(1)\cong SO(2)$. Note that the equation $(\cos\alpha) I+(\sin\alpha) K_2=e^{\alpha K_2}$ coincides with the famous Euler formula $\cos\alpha+{\rm i}\cdot\sin\alpha=e^{{\rm i}\alpha}$, where the matrix $K_2$ plays the role of the imaginary unit ${\rm i}=\sqrt{-1}$ since $K_2^2=-I$ [see Eq.~\eqref{eq:mr-15}].

Moreover, as discussed in Appendix \ref{subsec:PHsymmetries}, the pseudospin transformation \eqref{eq:mr-12} also contains the $\mathbb{Z}_4$ PH transformation \eqref{eq:ph-2}: $e^{{\rm i}\pi J_y}\gamma_j e^{-{\rm i}\pi J_y}=\gamma_j (-K_1)$, where the matrix $-K_1$ can be obtained by setting $g_J=0$ and $h_J=-1$ (i.e., $R_J=e^{{\rm i}\pi J_y}$) in Eq.~\eqref{eq:mr-13}. From Eq.~\eqref{eq:mr-15}, we have $(-K_1)^4=I$, namely a $\mathbb{Z}_4$ transformation.

In the Majorana representation, the usual $\mathbb{Z}_2$ PH transformation \eqref{eq:ph-5} can be described by Eq.~\eqref{eq:mr-2} with $R=e^{{\rm i}\pi J_y}e^{-{\rm i}\pi S_y}$. The corresponding matrix $A$ can be obtained by setting $g_J=0, h_J=-1$ and $g_S=0, h_S=1$ in Eq.~\eqref{eq:mr-3}: $A=\text{diag}(-1,1,-1,1)$, which satisfies $A^2=I$, indicating a $\mathbb{Z}_2$ transformation.

As the last example of symmetry transformations, let us consider the $\mathbb{Z}_2$ Shiba transformation \eqref{eq:shiba1}. Unlike the symmetries discussed above, the $\mathbb{Z}_2$ Shiba symmetry does not belong to the $SO(4)$ symmetry and hence cannot be described by Eq.~\eqref{eq:mr-2}. Using Eqs.~\eqref{eq:mr-1} and \eqref{eq:shiba1}, one can obtain the Shiba transformation of Majorana operators:
\begin{equation}
\gamma_j\rightarrow U_\downarrow\gamma_j U^{-1}_\downarrow=\gamma_j A_\downarrow, \quad     \text{where $A_\downarrow=\text{diag}(1,1,-1,1)$}.
\label{eq:mr-22}
\end{equation}
It is clear that the above matrix $A_\downarrow$ satisfies $A_\downarrow^2=I$ (namely a $\mathbb{Z}_2$ transformation) and $A_\downarrow\in O(4)$ but $A_\downarrow\notin SO(4)$ because $\det (A_\downarrow)=-1$.

After discussing the above various symmetry transformations of Majorana operators, we now switch to the study of the Majorana representation of the Hamiltonian, which makes the symmetries of the Hubbard model more explicit. In the following discussion, we assume the relations $t_{ij}e^{{\rm i}\theta_j}+t_{ji}e^{{\rm i}\theta_i}=0$ and $\mu_j=U_j/2$ [i.e., Eqs.~\eqref{eq:pseudospin-symmetry-a} and \eqref{eq:pseudospin-symmetry-b}], which unify the different symmetries of the Hubbard model (see Sec.~\ref{symmetry-unification}).

In terms of the Majorana operators \eqref{eq:mr-1}, we can rewrite the spin-up hopping terms as
\begin{equation}
t_{ij}c_{i\uparrow}^\dag c_{j\uparrow}+t_{ji}c_{j\uparrow}^\dag c_{i\uparrow}=\frac{1}{2}e^{{\rm i}(\theta_j-\theta_i)/2}t_{ij}(\gamma_{1i}\gamma_{1j}+\gamma_{2i}\gamma_{2j}),
\label{eq:mr-23}
\end{equation}
where we have used the relation $t_{ji}=-e^{{\rm i}(\theta_j-\theta_i)}t_{ij}$ due to Eq.~\eqref{eq:pseudospin-symmetry-a}. Similarly, the spin-down hoppings can be rewritten as
\begin{equation}
t_{ij}c_{i\downarrow}^\dag c_{j\downarrow}+t_{ji}c_{j\downarrow}^\dag c_{i\downarrow}=\frac{1}{2}e^{{\rm i}(\theta_j-\theta_i)/2}t_{ij}(\gamma_{3i}\gamma_{3j}+\gamma_{4i}\gamma_{4j}).
\label{eq:mr-24}
\end{equation}
Combining Eqs.~\eqref{eq:mr-23} and \eqref{eq:mr-24}, the off-site part $T$ of Eq.~\eqref{eq:nH-Hubbard} then becomes
\begin{equation}
T=\frac{1}{2}\sum_{\{i,j\}}e^{{\rm i}(\theta_j-\theta_i)/2}t_{ij}(\gamma_{1i}\gamma_{1j}+\gamma_{2i}\gamma_{2j}+\gamma_{3i}\gamma_{3j}+\gamma_{4i}\gamma_{4j})=\frac{1}{2}\sum_{\{i,j\}}e^{{\rm i}(\theta_j-\theta_i)/2}t_{ij}\gamma_i\gamma_j^t,
\label{eq:mr-25}
\end{equation}
where $\gamma_j^t$ denotes the transposition of the row vector $\gamma_j=(\gamma_{1j}, \gamma_{2j}, \gamma_{3j}, \gamma_{4j})$, i.e., a column vector. Note that the summation $\sum_{\{i,j\}}$ in Eq.~\eqref{eq:mr-25} runs over every bond $\{i,j\}=\{j,i\}$. Now, $T$ in Eq.~\eqref{eq:mr-25} is manifestly invariant under global $O(4)$ transformations: $\gamma_j\rightarrow \gamma_j A$ with $A\in O(4)$. The rotational subgroup $SO(4)$, which is actually a normal subgroup of $O(4)$, simply corresponds to the $SO(4)$ symmetry discussed in this paper [see, e.g., Eq.~\eqref{eq:mr-2}]; and any matrix $A\in O(4)$ but $A\notin SO(4)$ can be written as $A=A_\downarrow (A_\downarrow A)$, where $A_\downarrow A\in SO(4)$ [see Eq.~\eqref{eq:mr-22}]. Clearly, the Shiba transformation $A_\downarrow$ does not commute with a general $SO(4)$ transformation. [Recall that $A_\downarrow\in O(4)$ satisfies $A_\downarrow^2=I$ and $A_\downarrow\notin SO(4)$.] These results can be summarized as the following formula
\begin{equation}
O(4)=\mathbb{Z}_2 \ltimes SO(4), \quad     \text{where $\mathbb{Z}_2=\{I, A_\downarrow\}$},
\label{eq:mr-26}
\end{equation}
i.e., the group $O(4)$ can be decomposed into the semidirect product of its subgroups $SO(4)$ and $\mathbb{Z}_2$. Here, the group $\mathbb{Z}_2=\{I, A_\downarrow\}$ is chosen to represent the Shiba transformation. But, mathematically, there are many other choices of $\mathbb{Z}_2$: any matrix $A_-\in O(4)$ satisfying $A_-^2=I$ and $\det (A_-)=-1$ can be chosen to generate the group $\mathbb{Z}_2$ in Eq.~\eqref{eq:mr-26}.

\textbf{Remark:}
From the preceding subsections (without using the Majorana representation), we know that the hopping term $T$ possesses both the $\mathbb{Z}_2$ Shiba symmetry and the $SO(4)$ symmetry if Eq.~\eqref{eq:pseudospin-symmetry-a} holds (see, e.g., Theorems 17 and 21), which is consistent with the above discussion based on the Majorana representation. But the global $O(4)$ symmetry of $T$ is made explicit only by the Majorana representation.

Now, let us consider the symmetry of the on-site part $V$ of Eq.~\eqref{eq:nH-Hubbard}. When $\mu_j=U_j/2$ [i.e., Eq.~\eqref{eq:pseudospin-symmetry-b}], $V$ can be represented by $V'=\sum_{j\in\Lambda}U_j(c_{j\uparrow}^\dag c_{j\uparrow}-\frac{1}{2})(c_{j\downarrow}^\dag c_{j\downarrow}-\frac{1}{2})$ [see Eq.~\eqref{eq:shiba6}]; as stated below Eq.~\eqref{eq:f12}, $V'$ possesses the local $SU(2)$ spin and $SU(2)$ pseudospin symmetries, and hence a \emph{local} $SO(4)$ symmetry. Another way to see the local symmetries is to rewrite $V'$ as $V'=-\frac{2}{3}\sum_{j\in\Lambda}U_j(\bm{S}_j^2-\frac{3}{8})$ [see Eq.~\eqref{eq:f14}] that now manifestly commutes with the symmetry generators $S^\alpha_i$ and $J^\alpha_i$ for every site $i\in\Lambda$ and for all $\alpha=x,y,z$ [see Eq.~\eqref{eq:f9}].

The local $SO(4)$ symmetry of $V'$ becomes more explicit in the Majorana representation. Using Eq.~\eqref{eq:mr-5}, we can rewrite $V'$ as
\begin{equation}
V'=-\frac{1}{4}\sum_{j\in\Lambda}U_j\gamma_{1j}\gamma_{2j}\gamma_{3j}\gamma_{4j},
\label{eq:mr-27}
\end{equation}
e.g., interacting Majorana fermions. Now, Eq.~\eqref{eq:mr-27} is manifestly invariant under the local $SO(4)$ transformations: $\gamma_j\rightarrow \gamma_j A_j$, where the matrices $A_j\in SO(4)$ at different sites are independent of each other (see Lemma 22 and the Remark below it). The group structure of the local $SO(4)$ symmetry is described by the direct product of $N_\Lambda$ $SO(4)$ groups for every site, i.e., $SO(4)\times SO(4)\times\cdots\times SO(4)$.

On the other hand, unlike for the hopping $T$, it is clear that $O(4)$ transformations with determinant $-1$ would change the sign of $V'$ in Eq.~\eqref{eq:mr-27}, as can be seen from Lemma 22 (see also the Remark below it): for example, the Shiba transformation \eqref{eq:mr-22} changes $V'$ to $-V'$, or equivalently, Eq.~\eqref{eq:shiba7}.

To summarize, when Eqs.~\eqref{eq:pseudospin-symmetry-a} and \eqref{eq:pseudospin-symmetry-b} hold, the off-site hopping $T$ has a global $O(4)$ symmetry and the on-site interaction $V$ has a local $SO(4)$ symmetry. Thus, the total Hamiltonian $H=T+V$ possesses a global $SO(4)$ symmetry.

\section{Two-sublattice models}\label{app:two-sublattice-model}

In this appendix, we consider a general two-sublattice model and its variants to realize all of the exotic non-Hermitian eta-pairing phenomena with no counterparts in Hermitian systems, including the ones that do not exist in the Hatano-Nelson-Hubbard models studied in the main text: (1) one of the eta-raising and -lowering eigenoperators exists but the other does not exist, which belongs to the property $(a)$; and (2) the eta-raising or -lowering eigenoperator is not unique, namely the property $(e)$. Besides realizing various non-Hermitian phenomena, our general two-sublattice model can also be applied to a class of Hubbard models with Hermitian hoppings, revealing the eta-pairing structure [e.g., the $SO(4)$ symmetry and various symmetries belonging to it] in these systems.

For completeness, here, we restate the general two-sublattice model, which is defined on an arbitrary lattice $\Lambda$ that is divided into two disjoint sublattices $A$ and $B$.  The model is described by Eq.~\eqref{eq:nH-Hubbard} with the hoppings between different sublattices
\begin{equation}
  t_{ij} = \left\{
  \begin{array}{ll}
    \omega_A \lambda_{ij} & \text{if  $i\in A$ and $j\in B$} \\
   -\omega_B \lambda_{ij} & \text{if  $i\in B$ and $j\in A$}
  \end{array}
  \right.,
  \label{eq:tsm-1}
\end{equation}
where $\lambda_{ij}=\lambda_{ji}$ and $\omega_A, \omega_B, \lambda_{ij}\in\mathbb{C}$; and the hoppings within the same sublattice
\begin{equation}
  t_{ij} =-t_{ji} \quad \text{if  $i,j\in A$ or $B$},
  \label{eq:tsm-2}
\end{equation}
where $i\neq j$ and $t_{ij}\in\mathbb{C}$. Note that when $\omega_A=\omega_B(\neq0)$, the above two-sublattice model reduces to a very simple model characterized by the hoppings $t_{ij} =-t_{ji}$ ($\forall i,j\in\Lambda$ but $i\neq j$); see also the discussion below Theorem 2 in Sec.~\ref{sec:general-theory}. Thus, in this special case, the two-sublattice partition $\Lambda=A\cup B$ is trivial and does not make sense.

Before we proceed, let us clarify the concept of bipartite lattice. It is really a property of a \emph{graph}, i.e., a collection of a lattice $\Lambda$ and edges (or bonds) connecting paired sites of $\Lambda$. Here, a lattice $\Lambda$ is said to be bipartite if it \emph{can} be divided into two disjoint sublattices such that no bond (e.g., hopping) connects two lattice sites from the same sublattice; otherwise, $\Lambda$ is called nonbipartite. [Note a slightly different meaning of bipartite lattice in Ref. \cite{Lieb1989}, where the bipartite condition does not allow the on-site hoppings (i.e., the chemical potential terms). However, in this paper, the presence or absence of the on-site chemical potentials is irrelevant to the definition of (non)bipartite lattice.] So, given an arbitrary lattice $\Lambda$, it can be bipartite or nonbipartite, depending on the details of the bonds or hoppings. (When we say that, for example, the square or honeycomb lattice is a bipartite lattice whereas the triangular or kagome lattice is a nonbipartite lattice, these regular lattices are implicitly thought of as graphs, e.g., the collection of the lattice sites and the edges connecting nearest-neighbor sites.)

Therefore, our two-sublattice model is not necessarily a bipartite model (see the concrete examples below): the model is bipartite if all the hoppings in Eq.~\eqref{eq:tsm-2} vanish; otherwise, it may be nonbipartite. This means that in the latter case when there exists a bond (i.e., hopping) within the same sublattice $A$ or $B$, the model may also be bipartite: For example, consider a simple three-site system with the hoppings $t_{12}\neq0$ and $t_{23}\neq0$ (all other hoppings are zero). If we choose the two sublattices to be $A=\{1, 2\}$ and $B=\{3\}$, then there is a nonzero hopping (i.e., $t_{12}\neq0$) within the same sublattice $A$. However, we can also divide the lattice into $A=\{1, 3\}$ and $B=\{2\}$ and hence there is clearly no hopping within the same sublattice. Thus, by definition, this three-site model is actually bipartite.

In the following, we present some concrete examples (or variants) of the general two-sublattice model to illustrate the various exotic non-Hermitian eta-pairing phenomena (see, e.g., the Remark in Appendix \ref{subsec:proof-edabc}), especially the two novel phenomena that do not exist in the Hatano-Nelson-Hubbard models (as mentioned at the beginning of this appendix). Then, we show the applications of our general two-sublattice model to systems with Hermitian hoppings.

\subsection{Models with all the five properties $(a)$, $(b)$, $(c)$, $(d)$, and $(e)$}\label{subsec:abcde}

\emph{Model 1}.---
The model with all the five properties $(a)$, $(b)$, $(c)$, $(d)$, and $(e)$ can be obtained by setting, e.g., $\omega_A=0$ ($\omega_B\neq0$) in Eq.~\eqref{eq:tsm-1}. Specifically, consider the following model
\begin{eqnarray}
t_{ij}&=&0 \quad \text{and} \quad t_{ji}\neq0 \quad \text{  $\forall i\in A$ and $\forall j\in B$},
\nonumber\\
 t_{ij}&=&-t_{ji} \quad \text{if  $i,j\in A$ },  \qquad    t_{ij}=0 \quad \text{if  $i,j\in B$ };
\label{eq:tsm-3}
\end{eqnarray}
and the sublattice $A$ (as an isolated lattice) is assumed to be connected by itself. [Thus, this two-sublattice model is nonbipartite since it contains triangles: each triangle consists of three lattice sites (e.g., two $A$ sites and one $B$ site) and three bonds connecting the three sites.] Moreover, $U_j$ and $\mu_j$ satisfy
\begin{equation}
  U_j-2\mu_j = \left\{
  \begin{array}{ll}
    C_A  & \text{if  $j\in A$ } \\
   C_B  & \text{if  $j\in B$}
  \end{array}
  \right.,
  \label{eq:tsm-4}
\end{equation}
where $C_A$ and $C_B$ are two arbitrary complex numbers.

Now, solving Eq.~\eqref{eq:constraints1a} with Eq.~\eqref{eq:tsm-3}, one can obtain the eta-raising eigenoperator
\begin{equation}
\eta^\dag=\sum_{j\in B} \omega_jc_{j\uparrow}^\dag c_{j\downarrow}^\dag ,
\label{eq:tsm-5}
\end{equation}
where $\omega_j$ ($\forall j\in B$) is an arbitrary complex number and $\omega_j$ at different sites of sublattice $B$ are independent of each other. Thus, the eta-raising eigenoperator \eqref{eq:tsm-5} of the above two-sublattice model is not unique; i.e., the property $(e)$ occurs. On the other hand, using Eqs.~\eqref{eq:constraints2a} and \eqref{eq:tsm-3} (together with the assumption that the sublattice $A$ is connected by itself), we obtain the unique eta-lowering eigenoperator
\begin{equation}
\eta'=\sum_{j\in A} c_{j\downarrow}c_{j\uparrow}.
\label{eq:tsm-6}
\end{equation}

In conclusion, the above two-sublattice model possesses non-unique eta-raising eigenoperators but a unique eta-lowering eigenoperator, and hence the property $(e)$ holds.
Finally, from Eqs.~\eqref{eq:constraints1b} and \eqref{eq:constraints2b}, together with Eqs.~\eqref{eq:tsm-4}, \eqref{eq:tsm-5}, and \eqref{eq:tsm-6}, we have
\begin{equation}
\lambda=C_B, \qquad \lambda'=C_A.
\label{eq:tsm-7}
\end{equation}
So, $\lambda$ and $\lambda'$ are two independent constants; i.e., the property $(d)$ holds. Clearly, the three properties $(a)$, $(b)$, and $(c)$ also hold. [Note that the property $(e)$ actually implies the other properties $(a)$, $(b)$, $(c)$, and $(d)$, according to Eq.~\eqref{eq:relations} in the main text.]

\emph{Model 2}.---
There are also models where both the eta-raising and -lowering eigenoperators are not unique. For example, consider the following model
\begin{eqnarray}
t_{ij}&=&0 \quad \text{and} \quad t_{ji}\neq0 \quad \text{  $\forall i\in A$ and $\forall j\in B$},
\nonumber\\
 t_{ij}&=&0 \quad \text{if  $i,j\in A$ or $B$ }.
\label{eq:tsm-8}
\end{eqnarray}
Thus, different from the model \eqref{eq:tsm-3}, the two-sublattice model \eqref{eq:tsm-8} is clearly bipartite. Moreover, $U_j$ and $\mu_j$ also satisfy Eq.~\eqref{eq:tsm-4}.

Now, solving Eqs.~\eqref{eq:constraints1a}, \eqref{eq:constraints1b}, \eqref{eq:constraints2a}, and \eqref{eq:constraints2b}, together with Eq.~\eqref{eq:tsm-8}, one can obtain the eta-pairing eigenoperators
\begin{equation}
\eta^\dag=\sum_{j\in B} \omega_jc_{j\uparrow}^\dag c_{j\downarrow}^\dag , \qquad \eta'=\sum_{j\in A} (\omega'_j)^* c_{j\downarrow}c_{j\uparrow}
\label{eq:tsm-9}
\end{equation}
with $\lambda=C_B$ and $\lambda'=C_A$, where the arbitrary complex numbers $\omega_j$ ($\omega'_j$) at different sites of sublattice $B$ ($A$) are independent of each other. Thus, both the eta-raising eigenoperator $\eta^\dag$ and the eta-lowering eigenoperator $\eta'$ in Eq.~\eqref{eq:tsm-9} are not unique; i.e., the property $(e)$ occurs. Like the first two-sublattice model \eqref{eq:tsm-3}, our second model \eqref{eq:tsm-8} also exhibits the other properties $(a)$, $(b)$, $(c)$, and $(d)$ [see, e.g., the discussion below Eq.~\eqref{eq:tsm-7}].

\textbf{Remarks:}
(1) Both the above two models are special examples of the general two-sublattice model described by Eqs.~\eqref{eq:tsm-1} and \eqref{eq:tsm-2}, e.g., the case where $\omega_A=0$ and $\omega_B\neq0$. So, the non-unique eta-pairing eigenoperators in these two models contain $\eta^\dag=\sum_{j\in B} c_{j\uparrow}^\dag c_{j\downarrow}^\dag$ and $\eta'=\sum_{j\in A}  c_{j\downarrow}c_{j\uparrow}$ in Eq.~\eqref{eq:twosublattice-4} as a special case. It is also worth emphasizing that $\eta\neq\eta'$ [see the property $(a)$].

(2) In the above two models, the bonds connecting $A$ and $B$ sites are not strong bonds; and from Eqs.~\eqref{eq:tsm-5}, \eqref{eq:tsm-6}, and \eqref{eq:tsm-9}, the on-site product $\omega_j (\omega'_j)^*\equiv 0$ is clearly a zero constant, which is consistent with Corollary 14. Moreover, as discussed below Eq.~\eqref{eq:twosublattice-4} in the main text, the right and left eta-pairing eigenstates $(\eta^\dag)^m \ket{0}$ and $[(\eta')^\dag]^n \ket{0}$ are exactly localized in the regions $B$ and $A$, respectively.

(3) Although the above two models have all the five properties $(a)$, $(b)$, $(c)$, $(d)$, and $(e)$, one exotic non-Hermitian eta-pairing phenomenon, which belongs to the property $(a)$, is not involved: one of the eta-raising and -lowering eigenoperators exists but the other does not exist. [In fact, if this exotic eta-pairing phenomenon occurs, then the property $(d)$ does not make sense, as mentioned below Eq.~\eqref{eq:relations} in the main text.]

\subsection{An example with the properties $(a)$, $(b)$, $(c)$, and $(d)$ but without the property $(e)$}\label{subsec:abcd-e}

As another special example of the general two-sublattice model (e.g., $\omega_A=0, \omega_B\neq0$), let us consider the following model
\begin{eqnarray}
t_{ij}&=&0 \quad \text{and} \quad t_{ji}\neq0 \quad \text{  $\forall i\in A$ and $\forall j\in B$},
\nonumber\\
 t_{ij}&=&-t_{ji} \quad \text{if  $i,j\in A$ or $B$ };
\label{eq:tsm-10}
\end{eqnarray}
and both the sublattices $A$ and $B$ are assumed to be connected by themselves. Similar to the model \eqref{eq:tsm-3}, the two-sublattice model \eqref{eq:tsm-10} is also nonbipartite. Moreover, $U_j$ and $\mu_j$ satisfy Eq.~\eqref{eq:tsm-4}.

Now, solving Eqs.~\eqref{eq:constraints1a}, \eqref{eq:constraints1b}, \eqref{eq:constraints2a}, and \eqref{eq:constraints2b} for the above model, one can obtain the unique eta-raising and -lowering eigenoperators
\begin{equation}
\eta^\dag=\sum_{j\in B} c_{j\uparrow}^\dag c_{j\downarrow}^\dag , \qquad \eta'=\sum_{j\in A} c_{j\downarrow}c_{j\uparrow}
\label{eq:tsm-11}
\end{equation}
with $\lambda=C_B$ and $\lambda'=C_A$, respectively. Thus, the property $(e)$ is absent, while the properties $(a)$, $(b)$, $(c)$, and $(d)$ clearly hold [see also Eq.~\eqref{eq:relations} in the main text].

\textbf{Remarks:}
As stated below Theorem 3 in the main text, the assumption in Theorem 3 is a sufficient, but not necessary, condition for the uniqueness of eta-raising or -lowering eigenoperator. This can be illustrated with the two-sublattice model studied in this subsection (see also the one-dimensional Hatano-Nelson-Hubbard model with $\gamma=1$ in the main text):  Eq.~\eqref{eq:tsm-10} clearly violates the assumption in Theorem 3, e.g., not all the bonds are strong bonds (see the Remark in Appendix \ref{subsec:proofTheorem3}), but both the eta-raising and -lowering eigenoperators in Eq.~\eqref{eq:tsm-11} are unique.

Moreover, a similar discussion about the eta-pairing eigenoperators and eta-pairing eigenstates of the model in this subsection can be found in the Remarks in Appendix \ref{subsec:abcde}.

\subsection{Models where one of the eta-raising and -lowering eigenoperators exists but the other does not exist}\label{subsec:onlyone-exists}

Among the various non-Hermitian eta-pairing phenomena that have no counterparts in Hermitian systems, perhaps the most exotic phenomenon is that the eta-raising (-lowering) eigenoperator exists but the eta-lowering (-raising) eigenoperator does not exist, which belongs to the property $(a)$. To see this, let us consider two concrete examples.

\emph{Model 1}.---
The first example is actually a variant of the general two-sublattice model described by Eqs.~\eqref{eq:tsm-1} and \eqref{eq:tsm-2}:
\begin{eqnarray}
t_{ij}&=&0 \quad \text{and} \quad t_{ji}\neq0 \quad \text{  $\forall i\in A$ and $\forall j\in B$},
\nonumber\\
 t_{ij}&=&t_{ji}\neq0 \quad \text{for nearest-neighbor sites $i,j\in A$ },
 \nonumber\\
 t_{ij}&=&-t_{ji} \quad \text{if  $i,j\in B$ },
\label{eq:tsm-12}
\end{eqnarray}
where the sublattice $A$ is a triangular lattice and the sublattice $B$ is assumed to be connected by itself. No extra assumption about the hoppings and sublattices is made. (Clearly, this two-sublattice model is nonbipartite.) Moreover,  $U_j-2\mu_j=C_B$ ($j\in B$) is a constant in sublattice $B$, but there is no constraint on $U_j$ and $\mu_j$ for $j\in A$.

Now, solving Eqs.~\eqref{eq:constraints1a} and \eqref{eq:constraints1b} for the above model, one can obtain the unique eta-raising eigenoperator
\begin{equation}
\eta^\dag=\sum_{j\in B} c_{j\uparrow}^\dag c_{j\downarrow}^\dag  \quad \text{with $\lambda=C_B$ }.
\label{eq:tsm-13}
\end{equation}
In the following, we demonstrate the absence of eta-lowering eigenoperator $\eta'=\sum_{j\in \Lambda} (\omega'_j)^* c_{j\downarrow}c_{j\uparrow}$, which also explains the absence of eta-pairing eigenoperator in the usual Hermitian Hubbard model with real nearest-neighbor hoppings on the triangular lattice (see also the Corollary in Appendix \ref{subsec:proofTheorem3}).

First, from Eq.~\eqref{eq:constraints2a} and the ansatz $t_{ij}=0 , t_{ji}\neq0$  ($\forall i\in A, \forall j\in B$) in Eq.~\eqref{eq:tsm-12}, we have $\omega'_j=0$ ($\forall j\in B$); and hence Eq.~\eqref{eq:constraints2a} holds for $i,j\in B$. Next, due to Eq.~\eqref{eq:constraints2a} and the ansatz $t_{ij}=t_{ji}\neq0$  (for nearest-neighbor sites $i,j\in A$) in Eq.~\eqref{eq:tsm-12}, we also have $\omega'_j=0$ ($\forall j\in A$): Consider three arbitrary sites of the triangular sublattice $A$ denoted as 1, 2, and 3 (without loss of generality) such that they are nearest neighbored to each other. Now, from the assumption $t_{12}=t_{21}\neq0$ and Eq.~\eqref{eq:constraints2a}, e.g., $t_{12}(\omega'_1)^*+t_{21}(\omega'_2)^*=0$, we have $\omega'_2=-\omega'_1$. Similarly, $\omega'_3=-\omega'_1$ and $\omega'_3=-\omega'_2$, and hence $\omega'_1=\omega'_2=\omega'_3=0$. Finally, recall that the sites 1, 2, and 3 represent three arbitrary nearest-neighbor sites of $A$, and thus we get $\omega'_j=0$ ($\forall j\in A$).

To summarize, $\omega'_j=0$ for every lattice site $j\in \Lambda(=A\cup B)$, meaning that the eta-lowering eigenoperator $\eta'=\sum_{j\in \Lambda} (\omega'_j)^* c_{j\downarrow}c_{j\uparrow}$ does not exist.

\emph{Model 2}.---
In fact, there are simpler models in which one of the eta-raising and -lowering eigenoperators exists but the other does not exist. For example, consider the following two-sublattice model: The hoppings satisfy Eq.~\eqref{eq:tsm-10} and both the sublattices $A$ and $B$ are assumed to be connected by themselves. [Thus, this two-sublattice model is nonbipartite and is a special example of the general two-sublattice model (e.g., $\omega_A=0, \omega_B\neq0$).] However, $U_j$ and $\mu_j$ do not fully satisfy Eq.~\eqref{eq:tsm-4}.

Specifically, if $U_j-2\mu_j$ is a constant in sublattice $B$ whereas it is not a constant in sublattice $A$, then there exists a unique eta-raising eigenoperator (i.e., $\eta^\dag=\sum_{j\in B} c_{j\uparrow}^\dag c_{j\downarrow}^\dag$) but the eta-lowering eigenoperator $\eta'=\sum_{j\in \Lambda} (\omega'_j)^* c_{j\downarrow}c_{j\uparrow}$ does not exist. The absence of eta-lowering eigenoperator is due to the absence of (nonzero) solutions $\omega'_j$ of the coupled equations \eqref{eq:constraints2a} and \eqref{eq:constraints2b}; i.e., if Eq.~\eqref{eq:constraints2a} holds, which gives the unique solution (up to a trivial factor): $\omega'_j=1$ for $j\in A$ and $\omega'_j=0$ for $j\in B$, then Eq.~\eqref{eq:constraints2b} cannot hold because $\omega'_j\neq0$ for every site $j\in A$ but $U_j-2\mu_j$ is not a constant in sublattice $A$.

On the other hand, if $U_j-2\mu_j$ is a constant in sublattice $A$ whereas it is not a constant in sublattice $B$, then there exists a unique eta-lowering eigenoperator (i.e., $\eta'=\sum_{j\in A} c_{j\downarrow}c_{j\uparrow}$) but the eta-raising eigenoperator does not exist.

\textbf{Remark:}
Both the above two models violate the assumption in Theorem 3, e.g., not all the bonds are strong bonds (see the Remark in Appendix \ref{subsec:proofTheorem3}), but they have unique eta-raising or -lowering eigenoperator (see also the Remarks in Appendix \ref{subsec:abcd-e}). Clearly, these two models possess the properties $(a)$, $(b)$, and $(c)$ but without the properties $(e)$ and $(d)$.

\subsection{An example with the properties $(a)$, $(b)$, $(c)$, and $(e)$ but without the property $(d)$}\label{subsec:abce-d}

The most salient non-Hermitian model may be the one with the properties $(a)$, $(b)$, $(c)$, and $(e)$ but without the property $(d)$, which exhibits both the two novel non-Hermitian eta-pairing phenomena that have been mentioned at the beginning of this appendix.

To see this, consider the following two-sublattice model: The hoppings satisfy Eq.~\eqref{eq:tsm-3} and the sublattice $A$ is assumed to be connected by itself. [Thus, this two-sublattice model is nonbipartite and is a special example of the general two-sublattice model (e.g., $\omega_A=0, \omega_B\neq0$).] However, $U_j$ and $\mu_j$ do not fully satisfy Eq.~\eqref{eq:tsm-4}. Specifically, if $U_j-2\mu_j$ is a constant in sublattice $B$ whereas it is not a constant in sublattice $A$, then the model has non-unique eta-raising eigenoperators $\eta^\dag=\sum_{j\in B} \omega_j c_{j\uparrow}^\dag c_{j\downarrow}^\dag$ [see Eq.~\eqref{eq:tsm-5}], but the eta-lowering eigenoperator does not exist. The absence of eta-lowering eigenoperator is due to the same reason as discussed in the second model in Appendix \ref{subsec:onlyone-exists}.

\textbf{Remarks:}
In fact, no matter how $U_j-2\mu_j$ depends on sites $j\in \Lambda$ (e.g., for arbitrary $U_j$ and $\mu_j$ as a function of $j$), the eta-raising eigenoperator always exists and is always not unique. For example, if $U_j-2\mu_j$ at different sites of sublattice $B$ are different from each other (no constraint on $U_j$ and $\mu_j$ for $j\in A$), i.e., $U_i-2\mu_i\neq U_j-2\mu_j$  ($\forall i,j\in B$ and $i\neq j$), then the non-unique eta-raising eigenoperators are
\begin{equation}
\eta^\dag=\eta_j^\dag= c_{j\uparrow}^\dag c_{j\downarrow}^\dag \quad \text{with $\lambda=U_j-2\mu_j$ } \quad (j\in B),
\label{eq:tsm-14}
\end{equation}
which are actually special cases of Eq.~\eqref{eq:tsm-5}; and there are no other eta-raising eigenoperators [e.g., any linear combination of the eigenoperators in Eq.~\eqref{eq:tsm-14} is not an eigenoperator, because $U_j-2\mu_j$ at different $B$ sites take different values].

On the other hand, the (unique) eta-lowering eigenoperator $\eta'=\sum_{j\in A} c_{j\downarrow}c_{j\uparrow}$ [see Eq.~\eqref{eq:tsm-6}] exists if and only if $U_j-2\mu_j$ is a constant in sublattice $A$ (no constraint on $U_j$ and $\mu_j$ for $j\in B$).

\subsection{Applications to Hubbard models with Hermitian hopping $T$}\label{subsec:Hermitian-T}

In the previous subsections, we studied several concrete two-sublattice models that realize various exotic non-Hermitian eta-pairing phenomena. In addition to these models with non-Hermitian hopping $T$, our general two-sublattice model described by Eqs.~\eqref{eq:tsm-1} and \eqref{eq:tsm-2} can also be applied to a class of Hubbard models with Hermitian $T$ (i.e., $T^\dag=T$), revealing the eta-pairing structure [e.g., the $SO(4)$ symmetry and various symmetries belonging to it] in these systems.

In the following models we are going to study, since $T$ is always Hermitian and hence all the bonds are strong bonds, we assume that $U_j-2\mu_j=C$ is a site-independent constant (see Theorem 4 or 11). Note that the total Hamiltonian can be non-Hermitian because $U_j$ is allowed to be complex valued. Furthermore, Hermitian $T$ requires that the hopping amplitudes are symmetric (i.e., $|t_{ij}|=|t_{ji}|$) for every bond $\{i,j\}$, e.g., $|\omega_A|=|\omega_B|$ in Eq.~\eqref{eq:tsm-1}. Specifically, let us consider the case when
\begin{equation}
\omega_A=-\omega_B=1.
\label{eq:herhop-1}
\end{equation}
Substituting this into Eq.~\eqref{eq:twosublattice-3}, we obtain the eta-raising and -lowering eigenoperators
\begin{equation}
\eta^\dag=\sum_{j\in A} c_{j\uparrow}^\dag c_{j\downarrow}^\dag- \sum_{j\in B} c_{j\uparrow}^\dag c_{j\downarrow}^\dag , \qquad \eta'=\eta=(\eta^\dag)^\dag,
\label{eq:herhop-2}
\end{equation}
respectively, both of which are unique due to Theorem 4 or 11. Because of Eq.~\eqref{eq:herhop-1}, Eq.~\eqref{eq:tsm-1} reduces to
\begin{equation}
  t_{ij} =\lambda_{ij}=\lambda_{ji}=t_{ji} \quad (\text{for  $i\in A$ and $j\in B$}),
  \label{eq:herhop-3}
\end{equation}
where we have used the assumption $\lambda_{ij}=\lambda_{ji}$.
Thus, based on Eqs.~\eqref{eq:herhop-3} and \eqref{eq:tsm-2}, the Hermiticity of $T$ (i.e., $t_{ij}=t_{ji}^*$ for every bond $\{i,j\}$) implies that \emph{the hoppings between different sublattices are real and the hoppings within the same sublattice are purely imaginary}.

Now, let us apply the above results to some concrete models defined on regular lattices.

\textbf{Square lattice:}
The original eta-pairing theory established by C. N. Yang \cite{Yang1989} is defined on the square lattice. For square lattice, the two disjoint sublattices $A$ and $B$ are chosen such that any two nearest-neighbor sites belong to different sublattices (so both $A$ and $B$ are also square lattices by themselves).

Now, the hoppings between different sublattices [see Eq.~\eqref{eq:herhop-3}] are
\begin{equation}
  \lambda_{ij} = \left\{
  \begin{array}{ll}
    t\neq0 & \text{if  $i$ and $j$ are nearest neighbored} \\
   0 & \text{otherwise}
  \end{array}
  \right.
  \label{eq:herhop-4}
\end{equation}
with real-valued $t$, and the hoppings within the same sublattice [i.e., Eq.~\eqref{eq:tsm-2}] are zero. This, together with Eq.~\eqref{eq:herhop-2}, recovers the original eta-pairing theory for square lattice \cite{Yang1989}.

Note that the purely imaginary hoppings within the same sublattice (e.g., the next-nearest-neighbor hoppings) can be included, which does not affect the eta-pairing eigenoperators in Eq.~\eqref{eq:herhop-2}.

\textbf{Triangular lattice:}
Unlike for the square lattice, the triangular-lattice Hubbard model with Hermitian and real nearest-neighbor hoppings does not host any eta-pairing eigenoperator; and we have explained this absence of eta-pairing eigenoperator in this paper [see the Corollary in Appendix \ref{subsec:proofTheorem3} and also the discussion below Eq.~\eqref{eq:tsm-13}].

However, recall that our general two-sublattice model [e.g., Eqs.~\eqref{eq:herhop-3} and \eqref{eq:tsm-2}] with eta-pairing eigenoperators \eqref{eq:herhop-2} is defined on an arbitrary lattice $\Lambda$. Here, for the case when $\Lambda$ is a triangular lattice, the two disjoint sublattices $A$ and $B$ can be chosen as follows: any two nearest-neighbor sites that can be connected by the vector $\bm{a}_1$ belong to the same sublattice, and any two nearest-neighbor sites connected by the vector $\bm{a}_2$ or $\bm{a}_3$ belong to different sublattices; where $\bm{a}_1=(1,0)$, $\bm{a}_2=(\frac{1}{2}, \frac{\sqrt{3}}{2})$, $\bm{a}_3=(-\frac{1}{2}, \frac{\sqrt{3}}{2})$, and any two of them form the primitive translation vectors of the triangular lattice. [Thus, in this definition, both $A$ and $B$ are also regular lattices by themselves and share the same lattice translation vectors, namely $\bm{a}_1=(1,0)$ and $\bm{a}_2+ \bm{a}_3=(0, \sqrt{3})$.]

Now, let us focus on a translationally invariant $T$ with only nearest-neighbor hoppings:
\begin{equation}
\lambda_{j, j+\bm{a}_2}=t_2, \qquad \lambda_{j, j+\bm{a}_3}=t_3, \qquad t_{j, j+\bm{a}_1}=-t_{j, j-\bm{a}_1}={\rm i} t_1, \qquad (\forall j\in\Lambda),
\label{eq:herhop-5}
\end{equation}
where the real hoppings $t_2$ and $t_3$ connect nearest-neighbor sites from different sublattices (e.g., $j$ and $j+\bm{a}_{2,3}$), and the purely imaginary hoppings $\pm {\rm i} t_1$ connect nearest-neighbor sites from the same sublattice (e.g., $j$ and $j+\bm{a}_1$); see also Eqs.~\eqref{eq:herhop-3} and \eqref{eq:tsm-2}. The eta-pairing eigenoperators are given by Eq.~\eqref{eq:herhop-2}. Note that the hopping strengths $t_1$, $t_2$, and $t_3$ are not necessarily equal; when they are equal, the model \eqref{eq:herhop-5} would reduce to the one studied in Ref. \cite{huizhai}.

\textbf{Remark:}
According to Eqs.~\eqref{eq:herhop-3} and \eqref{eq:tsm-2}, the hopping ansatz for the triangular lattice in Eq.~\eqref{eq:herhop-5} is based on the definition of the two sublattices $A$ and $B$ described above. Clearly, there are many other different ways to choose $A$ and $B$ that are not necessarily regular lattices by themselves (as we mentioned in the main text, for any given lattice $\Lambda$, there are totally $2^{N_\Lambda-1}-1$ distinct partitions of $\Lambda$ into two disjoint sublattices). Thus, for triangular lattice, based on each choice of the two sublattices $A$ and $B$, one can construct models with eta-pairing eigenoperators \eqref{eq:herhop-2} via Eqs.~\eqref{eq:herhop-3} and \eqref{eq:tsm-2}.

\textbf{Honeycomb-lattice Haldane model:}
Our general two-sublattice model [e.g., Eqs.~\eqref{eq:herhop-3} and \eqref{eq:tsm-2}] with eta-pairing eigenoperators \eqref{eq:herhop-2} can also be applied to topological systems. The first example that we are going to consider is the honeycomb-lattice Haldane model (with the spin degree of freedom) \cite{Haldane1988}, which realizes the quantum Hall effect without net magnetic field (namely the quantum anomalous Hall effect or the Chern insulator phase). For honeycomb lattice, the two sublattices $A$ and $B$ are chosen such that any two nearest-neighbor sites belong to different sublattices, so both $A$ and $B$ are triangular lattices by themselves.

Now, the Haldane model can be described by Eqs.~\eqref{eq:herhop-3} and \eqref{eq:tsm-2} with the hoppings between different sublattices
\begin{equation}
  \lambda_{ij} = \left\{
  \begin{array}{ll}
    t & \text{if  $i$ and $j$ are nearest neighbored} \\
   0 & \text{otherwise}
  \end{array}
  \right.
  \label{eq:herhop-6}
\end{equation}
and the hoppings within the same sublattice
\begin{equation}
  t_{ij} =-t_{ji}= \left\{
  \begin{array}{ll}
    \pm {\rm i} t' & \text{if  $i$ and $j$ are next nearest neighbored} \\
   0 & \text{otherwise}
  \end{array}
  \right.,
  \label{eq:herhop-7}
\end{equation}
where $t$ is real and ${\rm i} t'$ is purely imaginary. The eta-pairing eigenoperators are given by Eq.~\eqref{eq:herhop-2}.

Note that the hopping strengths $t$ and $t'$ are not necessarily equal; in particular, when $t'=0$, the model \eqref{eq:herhop-6} describes the well-known Dirac-semimetal phase of graphene.

\textbf{Su-Schrieffer-Heeger (SSH) model:}
The next topological system we consider is the SSH model (with the spin degree of freedom) on one-dimensional lattice  \cite{SSH1980}. Let us denote the lattice sites as $\Lambda=\{1, 2, \ldots, N_\Lambda\}$. Then, the two sublattices are chosen to be $A=\{1, 3, \ldots\}$ and $B=\{2,4, \ldots\}$; i.e., the sublattice $A$ ($B$) consists of odd (even) sites.

Now, the SSH model can be described by Eq.~\eqref{eq:herhop-3} with the hoppings between different sublattices
\begin{equation}
  \lambda_{j,j+1} = \left\{
  \begin{array}{ll}
    w & \text{if  $j\in A$} \\
   v & \text{if  $j\in B$}
  \end{array}
  \right.
  \label{eq:herhop-8}
\end{equation}
and all other $\lambda_{ij}$ with $|i-j|>1$ are zero, where $w$ and $v$ are real. The hoppings within the same sublattice [i.e., Eq.~\eqref{eq:tsm-2}] are zero. The eta-pairing eigenoperators are given by Eq.~\eqref{eq:herhop-2}.

Thus, the SSH model describes a chain with alternating strong and weak hopping strengths. In the presence of chiral symmetry, the two gapped phases with $|w|>|v|$ and $|w|<|v|$ are topologically distinct from each other.

\textbf{Two-dimensional (2D) SSH model:}
The SSH model can be generalized to two dimensions, e.g., a square lattice \cite{2dSSH2017}. As discussed before, the two sublattices $A$ and $B$ of the square lattice are chosen such that any two nearest-neighbor sites belong to different sublattices.

Now, the 2D SSH model can be described by Eq.~\eqref{eq:herhop-3} with nearest-neighbor hoppings between different sublattices
\begin{equation}
  \lambda_{j,j+\vec{x}} = \left\{
  \begin{array}{ll}
    w_x & \text{if  $j_x$ is odd} \\
   v_x & \text{if  $j_x$ is even}
  \end{array}
  \right.,
  \qquad
  \lambda_{j,j+\vec{y}} = \left\{
  \begin{array}{ll}
    w_y & \text{if  $j_y$ is odd} \\
   v_y & \text{if  $j_y$ is even}
  \end{array}
  \right.
  \label{eq:herhop-9}
\end{equation}
and all other $\lambda_{ij}$ are zero, where $j=(j_x, j_y)$ with integers $j_{x}$ and $j_{y}$ denotes each lattice site of the square lattice, $\vec{x}=(1,0)$ and $\vec{y}=(0,1)$ are the lattice translation vectors, and the hopping parameters $w_{x,y}$ and $v_{x,y}$ are real. The hoppings within the same sublattice [i.e., Eq.~\eqref{eq:tsm-2}] are zero. The eta-pairing eigenoperators are given by Eq.~\eqref{eq:herhop-2}.

For simplicity, consider the case when $w_x=w_y=w$ and $v_x=v_y=v$. Like the above 1D case, the two phases of the 2D SSH model with $|w|>|v|$ and $|w|<|v|$ are separated by a topological phase transition at $|w|=|v|$.
[Note that when $w=v=t$, Eq.~\eqref{eq:herhop-9} would reduce to Eq.~\eqref{eq:herhop-4}.]

\textbf{2D higher-order topological insulator:}
By adding a $\pi$ flux through each square plaquette of the above 2D SSH model, we obtain a simple model for higher-order topological insulator in two dimensions \cite{HOTI2017}, which can be described by Eq.~\eqref{eq:herhop-3} with nearest-neighbor hoppings
\begin{equation}
  \lambda_{j,j+\vec{x}} = \left\{
  \begin{array}{ll}
    w_x & \text{if  $j_x$ is odd} \\
   v_x & \text{if  $j_x$ is even}
  \end{array}
  \right.,
  \qquad
  \lambda_{j,j+\vec{y}} = \left\{
  \begin{array}{ll}
    (-1)^{j_x} w_y & \text{if  $j_y$ is odd} \\
   (-1)^{j_x} v_y & \text{if  $j_y$ is even}
  \end{array}
  \right.
  \label{eq:herhop-10}
\end{equation}
and all other $\lambda_{ij}$ are zero, where the sign factors $(-1)^{j_x}$ account for the $\pi$ flux threading each plaquette [the notations in Eq.~\eqref{eq:herhop-10} are the same as those in Eq.~\eqref{eq:herhop-9}]. The hoppings within the same sublattice [i.e., Eq.~\eqref{eq:tsm-2}] are zero. The eta-pairing eigenoperators are given by Eq.~\eqref{eq:herhop-2}. (Note that the definition of the two sublattices $A$ and $B$ is the same as before.)

The model \eqref{eq:herhop-10} hosts a higher-order topological phase (with corner-localized states) as long as $|w_x|<|v_x|$ and $|w_y|<|v_y|$. When $w_x=w_y=v_x=v_y$, Eq.~\eqref{eq:herhop-10} describes a gapless phase at half-filling, namely a Dirac semimetal
(or, the so-called $\pi$-flux state \cite{Affleck1988}).

\textbf{3D higher-order topological insulator:}
The model \eqref{eq:herhop-10} can be generalized to a 3D cubic lattice \cite{HOTI2017}, realizing a higher-order topological insulator in three dimensions. Just like the case of the square lattice, the two sublattices $A$ and $B$ of the cubic lattice are chosen such that any two nearest-neighbor sites belong to different sublattices.

Now, the 3D higher-order topological insulator can be described by Eq.~\eqref{eq:herhop-3} with nearest-neighbor hoppings
\begin{equation}
  \lambda_{j,j+\vec{x}} = \left\{
  \begin{array}{ll}
    (-1)^{j_z}w_x & \text{if  $j_x$ is odd} \\
   (-1)^{j_z}v_x & \text{if  $j_x$ is even}
  \end{array}
  \right.,
  \quad
  \lambda_{j,j+\vec{y}} = \left\{
  \begin{array}{ll}
    (-1)^{j_x+j_z} w_y & \text{if  $j_y$ is odd} \\
    (-1)^{j_x+j_z}v_y & \text{if  $j_y$ is even}
  \end{array}
  \right.,
  \quad
  \lambda_{j,j+\vec{z}} = \left\{
  \begin{array}{ll}
     w_z & \text{if  $j_z$ is odd} \\
    v_z & \text{if  $j_z$ is even}
  \end{array}
  \right.
  \label{eq:herhop-11}
\end{equation}
and all other $\lambda_{ij}$ are zero, where $j=(j_x, j_y,j_z)$ with integers $j_{x,y, z}$ denotes each lattice site of the cubic lattice, $\vec{x}=(1,0,0)$, $\vec{y}=(0,1,0)$, and $\vec{z}=(0,0,1)$ are the lattice translation vectors, and the hopping parameters $w_{x,y,z}$ and $v_{x,y,z}$ are real. The hoppings within the same sublattice [i.e., Eq.~\eqref{eq:tsm-2}] are zero. The eta-pairing eigenoperators are given by Eq.~\eqref{eq:herhop-2}.

When $|w_\alpha|<|v_\alpha|$ for all $\alpha=x,y,z$, the model \eqref{eq:herhop-11} describes a 3D higher-order topological insulator (at half-filling) with corner-localized states.

\section{Off-diagonal long-range order (ODLRO)}\label{app:ODLRO}

It is known that the eta-pairing eigenstates of Hermitian systems have ODLRO \cite{Yang1989}. Here, we discuss ODLRO of eta-pairing eigenstates in non-Hermitian systems. Below, we assume that all the bonds are strong bonds.

From Eq.~\eqref{eq:m-eta-state} and Appendix \ref{subsec:RL-eta}, consider the right eta-pairing eigenstate $|R, m\rangle=(\eta^\dag)^m|0\rangle=m!\sum \omega_{k_1}\cdots \omega_{k_m}\eta_{k_1}^\dag\cdots\eta_{k_m}^\dag|0\rangle$ and the corresponding left eta-pairing eigenstate $|L, m\rangle=[(\eta')^\dag]^m|0\rangle=m!\sum \omega'_{l_1}\cdots \omega'_{l_m}\eta_{l_1}^\dag\cdots\eta_{l_m}^\dag|0\rangle$. To obtain ODLRO for non-Hermitian systems, we need to calculate the following correlation function
\begin{equation}
\langle \eta_i^\dag \eta_j \rangle_m =\frac{\bra{L,m} \eta_i^\dag \eta_j \ket{R,m}}{\langle L,m|R,m\rangle}, \quad   \text{where $i\neq j$}.
\label{eq:odlro-1}
\end{equation}
Note that $\langle \eta_i^\dag \eta_j \rangle_m =0$ if $m=0$ or $N_\Lambda$. Now, an important fact we will use is that the on-site product $\omega_k (\omega'_k)^*$ is a nonzero constant (i.e., independent of site $k$), say
\begin{equation}
\omega_k (\omega'_k)^* =C\neq0,
\label{eq:odlro-product}
\end{equation}
according to conclusion (b) in Theorem 11. By direct calculations and using Eq.~\eqref{eq:odlro-product}, we have
\begin{eqnarray}
\langle L,m|R,m\rangle &=& (m!)^2 C^{m} \binom{N_\Lambda}{m},
\nonumber\\
 \bra{L,m} \eta_i^\dag \eta_j \ket{R,m} &=& (m!)^2 (\omega'_i)^*\omega_j C^{m-1} \binom{N_\Lambda-2}{m-1},
\label{eq:odlro-2}
\end{eqnarray}
where the identities $[\eta_k^\dag,\eta_l]=\delta_{kl}(n_k-1)$ and $[\eta_k^\dag,\eta_l^\dag]=0$ have been used. Thus, from Eqs.~\eqref{eq:odlro-product} and \eqref{eq:odlro-2}, we obtain the explicit expression for Eq.~\eqref{eq:odlro-1}
\begin{equation}
\langle \eta_i^\dag \eta_j \rangle_m=\frac{m(N_\Lambda-m)}{N_\Lambda(N_\Lambda-1)}\cdot\frac{\omega_j}{\omega_i}.
\label{eq:odlro-01}
\end{equation}
Note that Eq.~\eqref{eq:odlro-01} holds for $0\leq m\leq N_\Lambda$. For two given sites $i,j$ and taking the limit $N_\Lambda\rightarrow\infty$ (e.g., the thermodynamic limit) at a fixed filling fraction $\nu= m/N_\Lambda$ (or, more precisely, $m/N_\Lambda\rightarrow\nu\in[0,1]$), the magnitude of Eq.~\eqref{eq:odlro-01} reads
\begin{equation}
|\langle \eta_i^\dag \eta_j \rangle_m| \rightarrow \nu(1-\nu)\frac{|\omega_j|}{|\omega_i|}.
\label{eq:odlro-mag}
\end{equation}

Now, we discuss the behavior of the correlation function in Eq.~\eqref{eq:odlro-mag}. For Hermitian systems (or more generally, the case when the hopping amplitudes are symmetric for every bond), we have $|L, m\rangle=|R, m\rangle$ (i.e., $\eta'=\eta$ or $\omega'_k=\omega_k$) and $|\omega_k|$ is site independent; see also the discussion below Theorem 4. Thus, the correlation function \eqref{eq:odlro-mag} reduces to $\nu(1-\nu)$, which does not depend on the sites $i$ and $j$, and the eta-pairing eigenstates exhibit ODLRO.

On the other hand, for non-Hermitian systems, $|\omega_k|$ is generally site dependent and hence the correlation function \eqref{eq:odlro-mag} generally depends on the sites $i$ and $j$.
To be specific, let us consider the Hatano-Nelson-Hubbard model and the general two-sublattice model studied in this paper: (1) For the one-dimensional Hatano-Nelson-Hubbard model, using Eq.~\eqref{eq:1d-eta-pairing}, we see that Eq.~\eqref{eq:odlro-mag} becomes $\nu(1-\nu)|q|^{i-j}$, where $i,j=1,2,\ldots$; thus, assuming $|q|>1$, the correlation decays (grows) exponentially with the distance $|i-j|$ if $i<j$ ($i>j$). (2) For the general two-sublattice model, using Eq.~\eqref{eq:twosublattice-3}, we see that Eq.~\eqref{eq:odlro-mag} becomes $\nu(1-\nu)$ if $i$ and $j$ belong to the same sublattice and becomes $\nu(1-\nu)\frac{|\omega_A|}{|\omega_B|}$ [or $\nu(1-\nu)\frac{|\omega_B|}{|\omega_A|}$] if $i$ and $j$ belong to different sublattices; thus, the eta-pairing eigenstates exhibit ODLRO.

\end{document}